\documentclass[11pt]{article}
\usepackage{latexsym,array,theorem,mathrsfs,bm,float}
\usepackage{psfrag}
\usepackage{amsfonts,amsmath,amssymb,latexsym,array,afterpage,
theorem,mathrsfs,bm,float,epsfig,color,graphicx,tabularx,
here,multirow,mediabb}
\usepackage[dvipdfmx]{hyperref}
\usepackage{fancybox}

  \makeatletter
    
    \@addtoreset{equation}{section}
  \makeatother

\usepackage[top=30mm,bottom=35mm,left=25mm,right=25mm]{geometry}

\newcommand{\bea}{\begin{eqnarray}}
\newcommand{\ena}{\end{eqnarray}}
\newcommand{\beann}{\begin{eqnarray*}}
\newcommand{\enann}{\end{eqnarray*}}

\newcommand{\ma}[1]{\mbox{$\mathcal{#1}$}}

\newcommand{\ti}{\tilde}

\newcommand{\calhR}[1]{\raisebox{2ex}{\tiny ({\em h})}\hspace{-0.8em}{\ma R}}

\newcommand{\BS}{\boldsymbol}
\newcommand{\MC}{\mathcal}

\newcommand{\p}{\prime}
\newcommand{\DS}{\displaystyle}

\newif\iffigure
\figurefalse
\figuretrue

\begin{document}

\title{\LARGE{\bf{
Thermodynamics of the Einstein-Maxwell system }}
}

\author{
\\
Shoichiro Miyashita\thanks{e-mail address : s-miyashita"at"gms.ndhu.edu.tw} 
\\ \\ 
{\it Department of Physics, National Dong Hwa University, Hualien, Taiwan, R.O.C.} 
\\ \\
}

\date{~}


\maketitle
\begin{abstract}
At first glance, thermodynamic properties of gravity with asymptotically AdS conditions and those with box boundary conditions, where the spatial section of the boundary is a sphere of finite radius, appear similar. Both exhibit a similar phase structure and Hawking-Page phase transition. However, when we introduce a U(1) gauge field to the system, discrepancies in thermodynamic properties between the two cases has been reported in \cite{BasuKrishnanSubramanian} (JHEP \textbf{11} (2016) 041). In this paper, by accepting the assumption that all Euclidean saddles contribute to the partition function, I found that these discrepancies are resolved due to the contribution from the "bag of gold (BG)," which is the class of Euclidean geometries whose area of bolt is bigger than that of the boundary. As a result, the Hawking-Page phase structure is restored, with the unexpected properties that the upper bound of thermodynamic entropy is always larger than the boundary area divided by 4G when the chemical potential is non-zero, and that such high entropy states are realized at sufficiently high temperature.
\end{abstract}

\clearpage

\tableofcontents

\clearpage

\section{Introduction}
It is often said that BH is thermodynamically unstable, but becomes stable when enclosed by a ``box''. There are two ways of realizing this; one is to impose an asymptotic AdS condition and the other is to impose a box boundary condition, i.e. we actually place a sphere boundary of finite radius. In terms of gravitational thermodynamics, the BH itself is not a thermodynamic system, but is realized as a thermodynamically stable phase in the gravitational system enclosed by the ``box''. The thermodynamic properties are studied in \cite{HawkingPage} for the AdS boundary and in \cite{York} for the box boundary (for the case with cosmological constant see \cite{BrownCreightonMann, Miyashita, DraperFarkas, BanihashemiJacobson}), and from their works we know that both cases have similar thermodynamic properties; the empty (or thermal gas) phase is realized at low temperature, the BH phase is realized at high temperature, and there is a phase transition called Hawking-Page phase transition (Fig. \ref{1}). Since both cases have a ``box'' boundary and exhibit similar thermodynamic properties for the case of pure gravity, one might expect this kind of similarity to hold when we add some other fields to the system. However, it has been reported that in the Einstein-Maxwell system there is some discrepancy between them \cite{BasuKrishnanSubramanian}. Let's focus on the case of the grand canonical ensemble. For the AdS boundary 
 condition, the system exhibits a phase structure similar to the pure gravity case, i.e. the behavior of the free energy is qualitatively the same as in pure gravity (for sufficiently low chemical potential $\mu$) \cite{ChamblinEmparanJohnsonMyers}. See Fig. \ref{1}. On the other hand, for the box boundary condition, the dominant BH branch does not extend to infinity and terminates at some finite temperature (Fig. \ref{2}). The termination point is controlled by the value of the chemical potential $\mu$ (and the cosmological constant $\Lambda$ and the radius of the box $r_{b}$). The only branch that exists beyond the temperature is the empty phase, and apparently the free energy has a discontinuity there. In this sense, thermodynamics itself is ill-defined for this system, since we cannot define thermodynamic quantities such as energy or electric charge there.
                                                 %
\iffigure
\begin{figure}[h]
\begin{center}
	\includegraphics[width=8.cm]{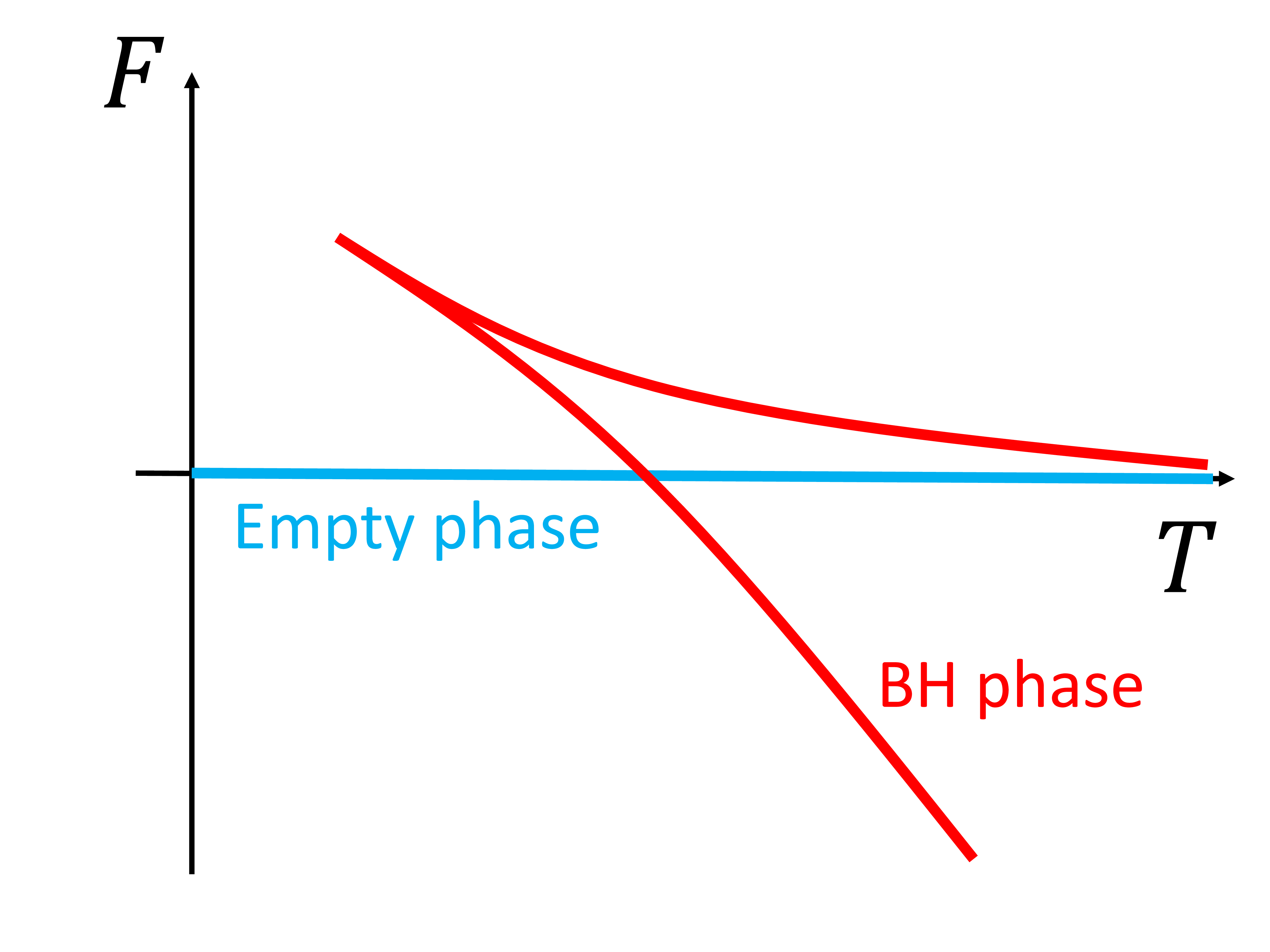} 
	\caption{Qualitative behavior of free energies $F$ for boxed gravity systems. $F$ is defined by the on-shell action times the temperature. This behavior can be seen in the pure gravity system with AdS boundary, that with box boundary (for $\Lambda \leq 0$), and the Einstein-Maxwell system with AdS boundary (for sufficiently low $\mu$). A previous study \cite{BasuKrishnanSubramanian} claimed that this cannot be seen in the Einstein-Maxwell system with box boundary and it behaves like Fig. \ref{2}.}
\label{1}
\end{center}
\end{figure}
\fi
                                                 %

                                                 %
\iffigure
\begin{figure}[h]
\begin{center}
	\includegraphics[width=7.cm]{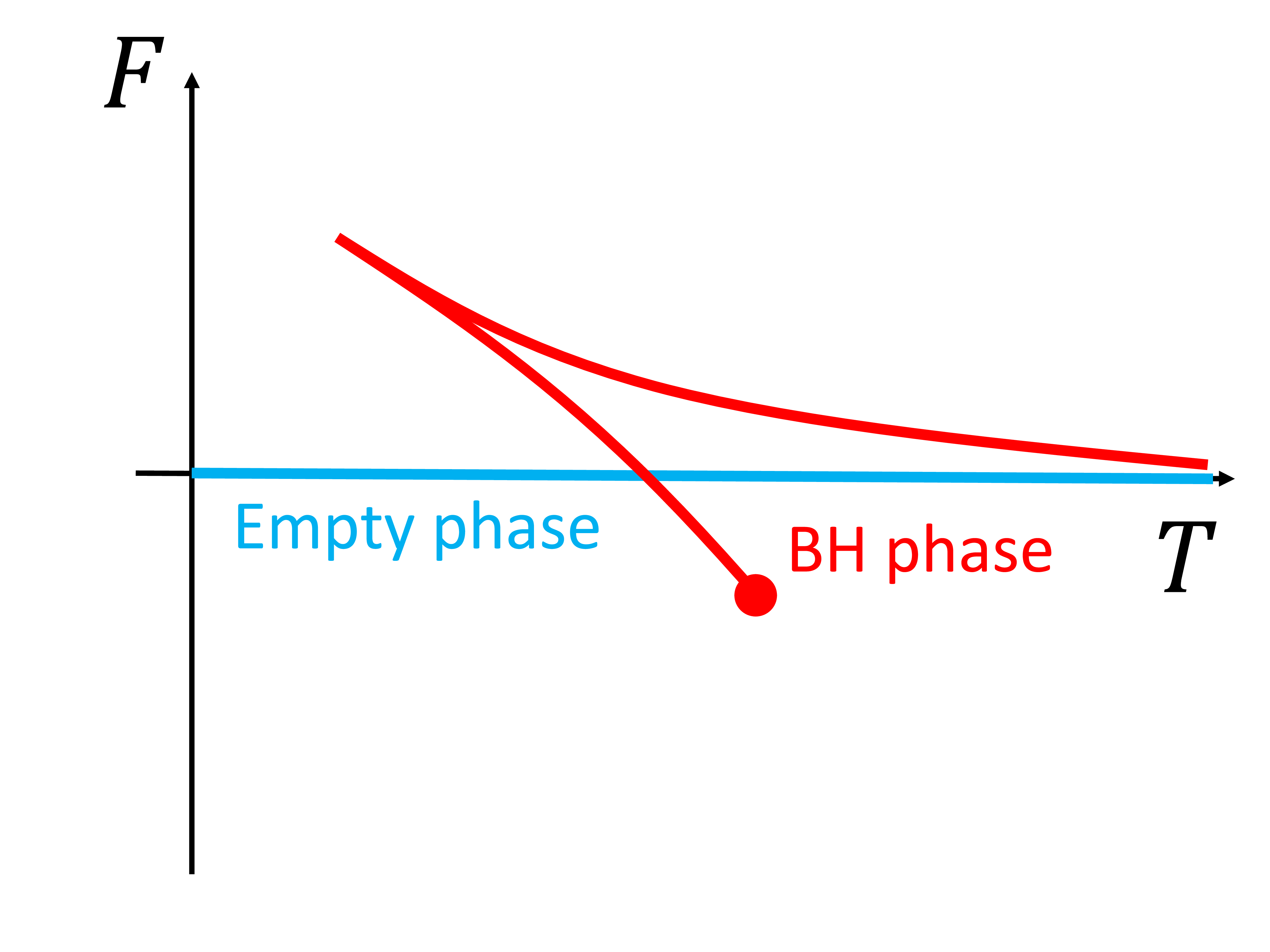} 
	\caption{Qualitative behavior of the free energy of the empty phase and the free energy of the BH phase in the Einstein-Maxwell system with box boundary. The BH branch terminates at some finite temperature depending on $\mu$ (and $\Lambda$ and $r_{b}$). If there are no other dominant saddle points, the partition function would be a discontinuous function and one might say that the thermodynamics of the system is ill-defined. My claim in this paper is very simple: the bag of gold saves the day. (See Fig. \ref{7}). }
\label{2}
\end{center}
\end{figure}
\fi
                                                 %
 
In this paper, I want to resolve the problem of this ill-definedness of the Einstein-Maxwell system by considering ``bag of gold (BG)'' saddles, the class of Euclidean saddle point geometries whose area of a bolt is larger than that of the boundary \cite{Miyashita}.
\footnote{
Independently, their importance in gravitational thermodynamics was discussed in \cite{DraperFarkas, BanihashemiJacobson}. In \cite{DraperFarkas}, they called the class of geometries ``cosmological side solution.''
}

In the previous paper \cite{Miyashita} I discussed the role of BG saddles in the thermodynamics of pure gravity with positive $\Lambda$.  Including them in the path integral leads to the {\it lack} of thermodynamic stability and the entropy bound, which are the universal properties of pure gravity with $\Lambda\leq 0$. In a sense, I have shown a {\it bad} aspect of the BG saddle. It breaks the seemingly universal properties of gravitational thermodynamics. If one believes that thermodynamic stability or the entropy bound must be universal, one may argue that the BG saddles should be excluded from the path integral, since we do not know which saddle points contribute to the Euclidean path integral of gravity \cite{GibbonsHawkingPerry}.
\footnote{
There are some attempts to find the definition of the Euclidean path integral of gravity or the partition function of gravity.
}
 
 Here I will show a {\it good} aspect of the BG saddle by showing that it leads to the well-defined thermodynamic description of the Einstein-Maxwell system with box boundary condition and similarity to that with AdS boundary condition. The reason is simple. The BG branch appears and is smoothly connected to the BH branch and it extends to infinity (see Fig. \ref{7} and compare with Fig. \ref{2}). And in fact, the BG branch is always thermodynamically stable, in contrast to the case of positive $\Lambda$.
 
 Throughout this paper I use the standard path integral ``definition'' of the gravitational partition function and assume that the dominant saddle points are in the following class of metrics and U(1) gauge fields on a manifold of $B\times S^1$ topology or of $S^{2} \times D$ topology:
\footnote{
For the former, $B$ means 3-dimensional ball and $S^1$ represents the Euclidean time circle. For the latter, $D$ means 2-dimensional disk and the angular direction of $D$ is the Euclidean time direction. Their boundary topologies are the same, $S^2 \times S^1$.
}
\bea
&& \BS{g}= f(r)dt^2 + \frac{1}{f(r)}dr^2 + r^2 d\Omega^2 \\
&& ~~~~~~~~ f(r)= 
\begin{cases}
\displaystyle \hspace{1.3cm} 1  - \frac{\Lambda}{3}r^2    \hspace{2.42cm} {\rm for ~ } B\times S^1 \\
\displaystyle 1 - \frac{2G\ti{M}}{r}+ \frac{G\ti{Q}^2}{r^2} - \frac{\Lambda}{3}r^2 ~~~~~~~ {\rm for ~ } S^{2} \times D
\end{cases}  \\
&&  A_{t}(r)= 
\begin{cases}
 \hspace{0.7cm} -i\ti{\mu}    \hspace{2.42cm} {\rm for ~ } B\times S^1 \\
  -i\ti{Q} \left( \frac{1}{r_{H}}-\frac{1}{r} \right) ~~~~~~~~~~ {\rm for ~ } S^{2} \times D
\end{cases} 
\ena
These are Euclidean solutions of the Einstein-Maxwell system.
\footnote{
Euclidean here means that the geometry is purely Euclidean. The gauge field is allowed to be complex.
}
The coordinate ranges and parameter values depend on the boundary condition. The fact that $\ti{M}$ and $\ti{\mu}$ are identified with the total energy $E$ and the chemical potential $\mu$ for the AdS (and asymptotically flat) boundary condition but not for the box boundary condition may be familiar. Less familiar is the relation between $\tilde{Q}$ and the total charge $Q$. Normally $\ti{Q}$ can be identified with $Q$ ($\ti{Q} = Q$). However, for the BG saddle, $\ti{Q}$ is not $Q$, but there is a minus sign ($-\ti{Q}=Q$), as I will explain in Section 3. 
 
 The organization of this paper is as follows. In Section 2, I review the thermodynamics of the Einstein-Maxwell system with AdS boundary \cite{ChamblinEmparanJohnsonMyers}. In subsection 3.1, I review the properties of the BH phase branch in the Einstein-Maxwell system with box boundary without $\Lambda$, which are systematically studied in \cite{BasuKrishnanSubramanian}. 
\footnote{
The study of this system was initiated long ago by \cite{BradenBrownWhitingYork}. The higher dimensional generalization was recently studied by \cite{FernandesLemos}.
}
I then study the properties of the BG phase in subsection 3.2 and claim that this leads to the well-defined thermodynamics in the system in subsection 3.3. In Section 4, I generalize the analysis of Section 3 to the case with a cosmological constant $\Lambda$. The $\Lambda < 0$ case is qualitatively the same as the $\Lambda=0$ case. In the $\Lambda>0$ case, since not only {\it good} BGs but also {\it bad} BGs seem to appear, one might expect the system to become thermodynamically unstable, as in the pure gravity case. This is certainly true for $0<\sqrt{G} \mu<1$. However, for the case of $\mu>1$, {\it bad} BGs no longer exist, but BHs do. As a result, the system is thermodynamically stable in this case. I summarize all the material in this paper in Section 5.
 
In the main part of this paper I will focus only on grand canonical ensembles, where the corresponding Euclidean action is given by \cite{BradenBrownWhitingYork, HawkingRoss}.
\footnote{
The last term is the Mann counterterm \cite{Mann}, which has been shown to be a suitable boundary counterterm for spacetime with a spherical boundary \cite{Miyashita2}, in the sense that it can not only subtract the divergence of the action, but also set the ground state energy to zero. 
}
\bea
I^{E}[\BS{g}, \BS{A}]= \frac{-1}{16\pi G} \int_{\MC{M}} d^4 x \sqrt{g} (\MC{R}- 2\Lambda) + \frac{1}{16 \pi}\int d^4 x \sqrt{g} F_{\mu\nu} F^{\mu\nu} \hspace{2.5cm} \notag \\
+ \frac{-1}{8\pi G} \int d^3 y \sqrt{\gamma} \Theta + \frac{1}{8\pi G} \int d^3 y \sqrt{\gamma} \sqrt{2\MC{R}^{(3)}- \frac{4\Lambda}{3} }
\ena
However, since canonical and microcanonical ensembles are equally important, I investigate their properties in Appendix A and some of the results are also listed in Section 5.

\section{AdS Boundary Condition}
In this section, I briefly review the result of \cite{ChamblinEmparanJohnsonMyers}.  The upshot is that there exists a critical value of chemical potential $\mu_{cr}$; for $\mu<\mu_{cr}$, there exist the AdS phase and the BH phase, and the system exhibits Hawking-Page phase transition, and for $\mu \geq\mu_{cr}$, there is only the BH phase. (Fig. \ref{3})
                                                 %
\iffigure
\begin{figure}[h]
\begin{center}
	\includegraphics[width=7.cm]{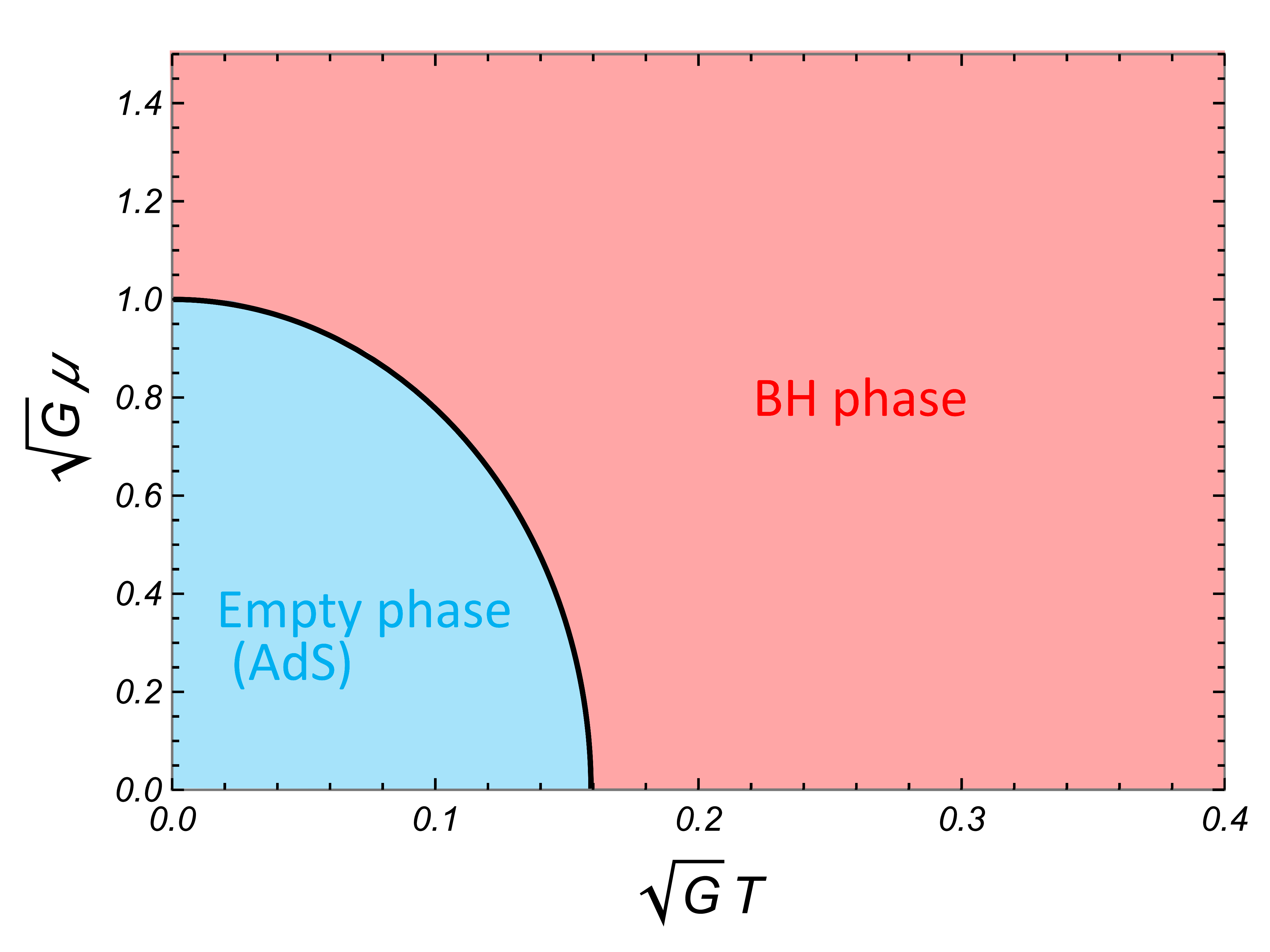} 
	\caption{Phase diagram of the Einstein-Maxwell system with AdS boundary condition. For four dimensions, the critical value of the chemical potential is given by $\mu_{cr}= 1/ \sqrt{G}$ as shown in the figure. The phase boundary is given by the function $T=\frac{1}{\pi l_{-}}\sqrt{1-G\mu^2}$. ($l_{-}$ is the AdS radius and I set $l_{-}=2\sqrt{G}$ here.)}
\label{3}
\end{center}
\end{figure}
\fi
                                                 %

For the AdS boundary case, the saddle points are Euclidean AdS ($B \times S^{1}$ topology) and Euclidean Reissner-Nordstr$\ddot{{\rm o}}$m(RN) AdS BH ($S^2 \times D$ topology). The coordinate range of $r$ is $[0, \infty)$ for the former and $[r_{H}, \infty)$ for the latter. The inverse temperature $\beta$ is equal to the circumference of $t$. The free energy $F(\beta, \mu)$ is given by
\bea
F(\beta, \mu) = \frac{I^{E}[{\rm saddle}_{(\beta, \mu)}]}{\beta} \label{free}
\ena
where ${\rm saddle}_{(\beta, \mu)}$ represents a saddle point which satisfies the following conditions;
\bea
 \ti{\mu}= \mu ~~~~~~~~~~~~~~~  {\rm for ~ } B\times S^1\\
 \frac{4\pi}{f^{\p}(r_{H})}  = \beta,  ~~~ \frac{\ti{Q}}{r_{H}} = \mu ~~~~~~   {\rm for ~ } S^2 \times D
\ena
From these conditions, together with $f(r_{H})=0$, the parameters $r_{H}, \ti{M}, \ti{Q}, \ti{\mu}$ are determined. For some $\beta, \mu$ they are not uniquely determined. In this case we get many ``free energies'' for given $\beta, \mu$. For example, for sufficiently high temperatures in Fig. \ref{4} (Left) (or Fig. \ref{1}), there are three saddles; one AdS saddle and two BH saddles. 
\footnote{
Of course, the free energy must be uniquely determined and it is given by the dominant saddle. However, I will abuse the term ``free energy'' in this paper. 
}

Explicitly, the free energy for each saddle is given by
\bea
F_{AdS}(\beta, \mu)=0 \hspace{8.05cm} \\
F_{BH}(\beta,\mu) = \frac{r_{H}(\beta, \mu)}{4G}\left( 1-G\mu^2 - \frac{r_{H}(\beta, \mu)^2}{l_{-}^2} \right), \label{FBHAdS} \hspace{2.4cm}\\
 r_{H}(\beta, \mu) = \frac{1}{3} \left( 2\pi T l_{-}^2 \pm \sqrt{ ( 2\pi T l_{-}^2)^2 -3l_{-}^2 (1-G\mu^2)}\right) \label{EqrHAdS}
\ena
where $l_{-} \equiv \sqrt{-3/\Lambda} $ is the AdS radius. In order for the corresponding real Euclidean geometry to exist, $r_{H}>0$. Therefore, from eq. $(\ref{EqrHAdS})$, we could know the followings;\\
~~~~ $\cdot$ For $G \mu^2 <1$, $T_{min} = \frac{\sqrt{3}}{2\pi l_{-}} \sqrt{1-G\mu^2}$ is the minimum temperature. There are two BH saddles\\ ~~~~~~ when $T>T_{min}$ and one BH saddle when $T=T_{min}$. Below $T_{min}$, there are no BH saddles. \\   
~\\ 
~~~~ $\cdot$ For $G \mu^2 \geq 1$, there is only one BH saddle for each temperature.  \\
~\\ 
And with a little algebra, we also know\\
~~~~ $\cdot$ for $G \mu^2 <1$, the BH saddle with the larger horizon radius always thermodynamically stable.\\
~~~~~~  Above $T_{tr}=\frac{1}{\pi l_{-}}\sqrt{1-G\mu^2}$, it becomes the dominant saddle, and below $T_{tr}$, the dominant\\
~~~~~~  saddle is the AdS saddle.\\
~\\
~~~~ $\cdot$ for $G \mu^2 \geq 1$, the BH saddle is always thermodynamically stable and the dominant one.\\
~\\
Therefore, the behavior of the system is qualitatively different depending on whether $\mu$ is less than $\mu_{cr}\equiv 1/\sqrt{G}$ or not.
\footnote{
Since we can take $\mu \geq 0$ without loss of generality, I will restrict the analysis to this case. 
}
 The qualitative behavior of the free energies in each case is shown in Fig. \ref{4}, and the phase diagram is shown in Fig. \ref{3}.

                                                 %
\iffigure
\begin{figure}[h]
\begin{center}
	\includegraphics[width=7.cm]{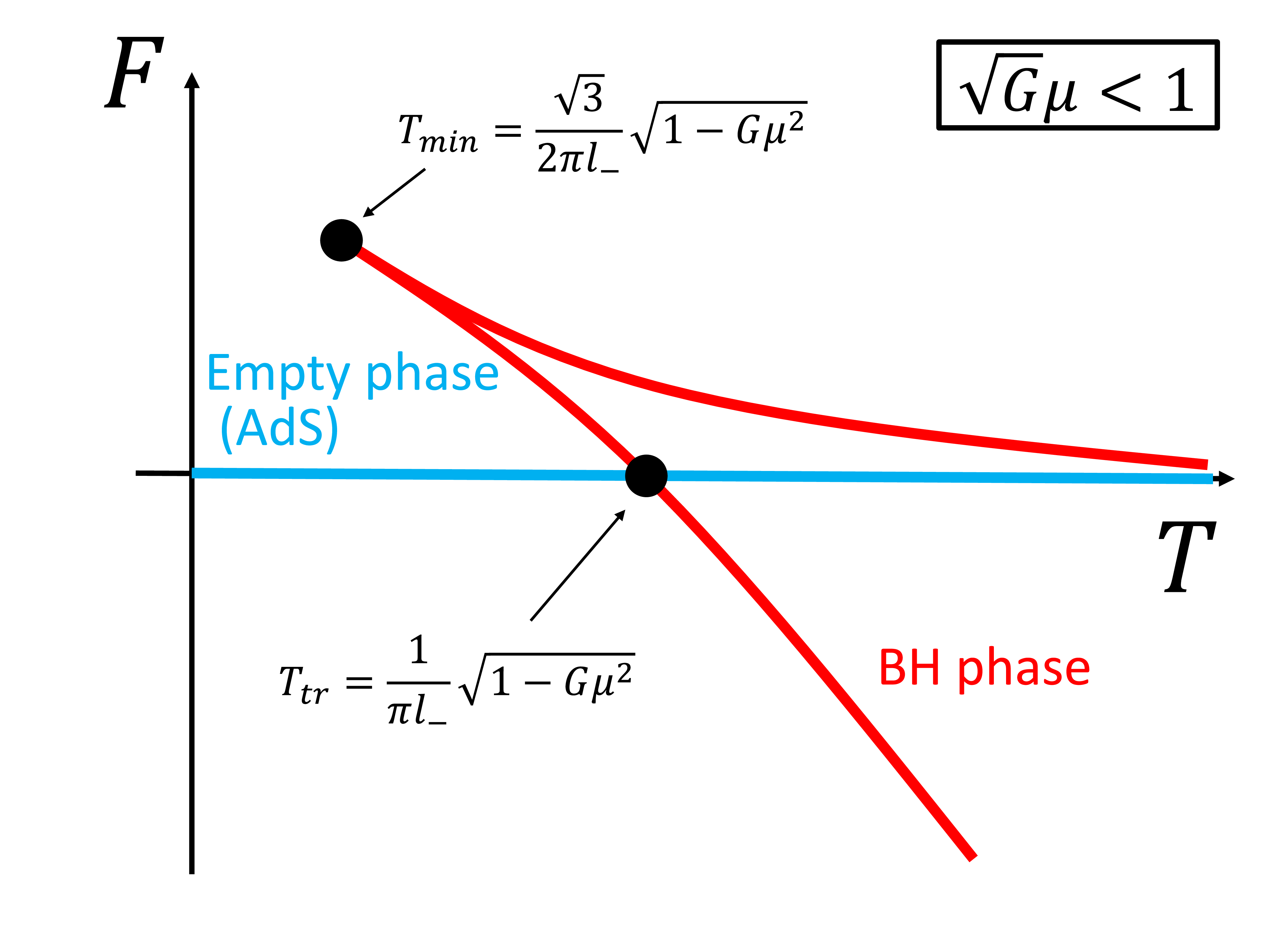}  ~~~~ 	\includegraphics[width=7.cm]{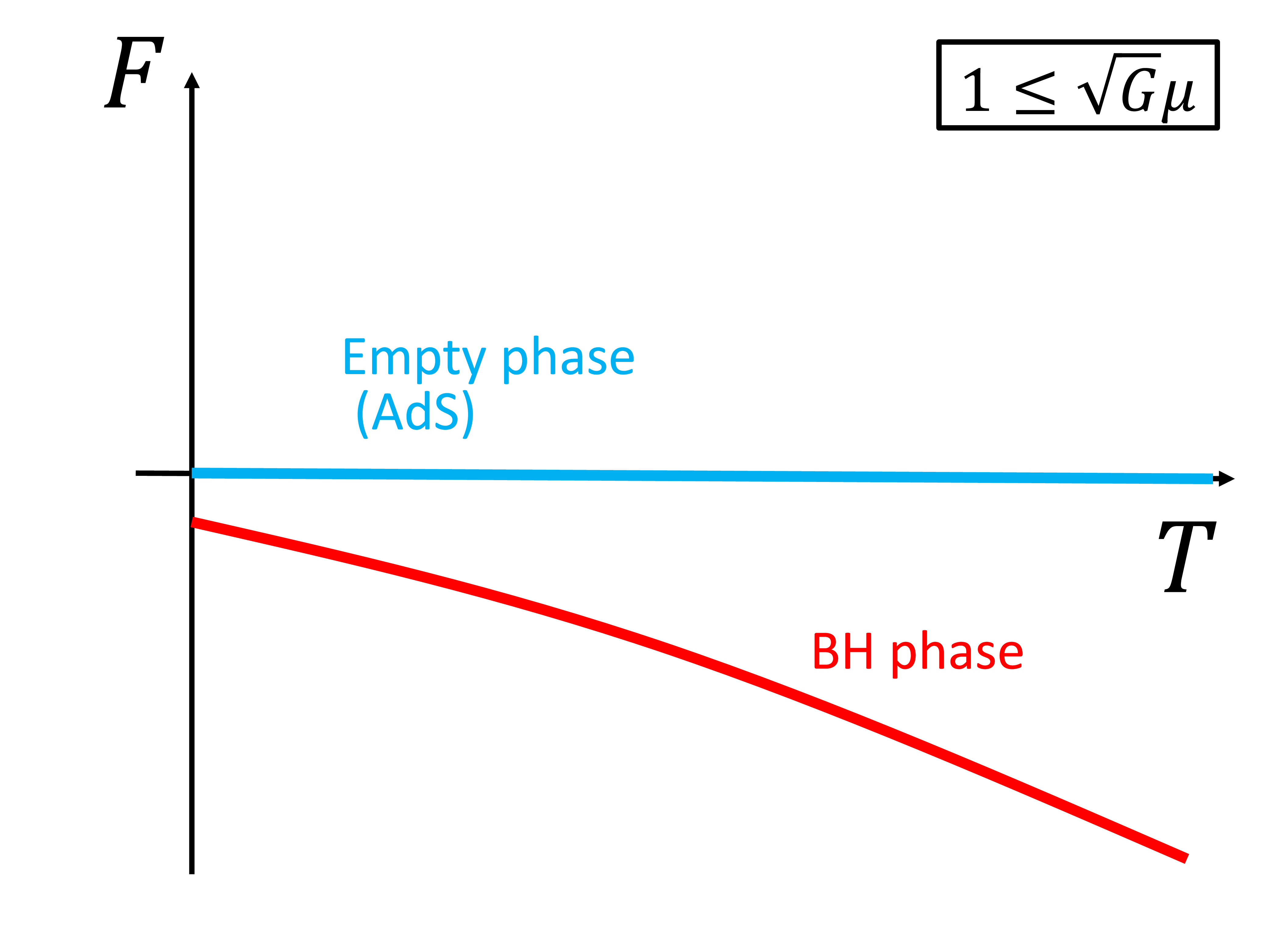} 
	\caption{Behavior of free energies. (Left) $G\mu^2 <1$. There are two phases; AdS phase and BH phase. Hawking-Page phase transition occurs at $T_{tr}=\frac{1}{\pi l_{-}}\sqrt{1-G\mu^2}$. (Right) $G\mu^2 \geq 1$. Only BH phase exists since the corresponding BH saddle is always the dominant one. The zero temperature state corresponds to the extremal BH with the horizon radius $r_{H}= l_{-}\sqrt{\frac{G\mu^2 -1}{3}}$.}
\label{4}
\end{center}
\end{figure}
\fi
                                                 %

\section{Box Boundary Condition}
In this section I consider Einstein-Maxwell thermodynamics in the case of a box boundary without $\Lambda$. Considering the box boundary, we can find empty saddles and BH saddles, similar to the previous section. Although one might expect the qualitatively similar behavior to the AdS boundary case, the result is somewhat strange; stable BH saddles cease to exist above a certain temperature (Fig. \ref{2}). This is the result obtained by Basu, Krishnan, and Subramanian in \cite{BasuKrishnanSubramanian}. 
\footnote{
Note that the main purpose of their paper is to study the thermodynamics of the Einstein-Maxwell-scalar system. The analysis of Einstein-Maxwell system is a kind of warm-up, but clearly summarizes the thermodynamic properties of BH saddles and empty saddles. I recommend the readers to refer the paper. Note also that in the analysis of the Einstein-Maxwell-scalar system, they did not take into account the contributions of BG saddles. Investigating the possibility of modifying their result by including them is left for future study.
}
I briefly review their analysis in subsection 3.1. Then, in subsection 3.2, I show that there are other types of saddles, which I call BG saddles \cite{Miyashita}. And in subsection 3.3, by including their contribution to the path integral, I show that the thermodynamic behavior becomes regular, similar to the AdS boundary case.

\subsection{empty saddle and BH saddle}
Similar to the AdS boundary case, there are two types of saddle points, empty saddle ($B \times S^{1}$ topology) and Euclidean RN BH saddle ($S^2 \times D$ topology);
\bea
{\rm \underline{field ~ configuration}} \hspace{11.5cm} \notag \\
{\rm Empty ~ saddle:} ~~ r\in[0, r_{b}], ~ f(r)=1, ~ A_{t}(r)=-i \ti{\mu} \hspace{5.1cm} \\
{\rm Euclidean ~  RN ~ BH ~ saddle:} ~~ r\in[r_{H}, r_{b}], ~ f(r)= 1- \frac{2GM}{r} + \frac{G\ti{Q}^2}{r^2}, ~ A_{t}(r)=-i \ti{Q} \left( \frac{1}{r_{H}} - \frac{1}{r} \right) \label{configBH} 
\ena
The free energy $F(\beta, r_{b}, \mu)$  of each saddle is defined as eq. (\ref{free}) and the parameters of a saddle are determined by boundary conditions (and $f(r_{H})=0$ for BH saddles);
\bea
{\rm \underline{boundary ~ condition}} \hspace{10.7cm} \notag \\
{\rm Empty ~ saddle:} ~~ \frac{\ti{\mu}}{\sqrt{f(r_{b})}} = \mu    \hspace{5.6cm} \\
{\rm Euclidean ~  RN ~ BH ~ saddle:} ~~ \frac{4 \pi}{f^{\p}(r_{H})}\sqrt{f(r_{b})}= \beta, ~~ \frac{\ti{Q}}{\sqrt{f(r_{b})}} \left( \frac{1}{r_{H}} - \frac{1}{r_{b}} \right)=\mu 
\ena
and their circumferences of $t$ are given by $\beta$ and $\frac{4\pi}{f^{\p}(r_{H})}$ respectively. The corresponding energy $E$ and charge $Q$ are given by
\footnote{
Using boundary currents $\displaystyle \tau_{ij}\equiv \frac{2}{\sqrt{\gamma}} \frac{\delta I^E}{\delta \gamma^{ij}}, ~~ j^{i}\equiv \frac{-1}{\sqrt{\gamma}} \frac{\delta I^E}{\delta A_{i}} $, energy and charge are defined by\\
 $\displaystyle E\equiv \int d^2 z \sqrt{\sigma} u_{i} \tau^{ij} \xi_{j} ,$ $Q \equiv \frac{1}{i} \int d^2 z \sqrt{\sigma} j^{i}u_{i} $, where $\xi^{i}$ is the normalized Killing vector of the $U(1)$ isometry on the boundary and $u^i$ is the normal vector of the integration surface. The integrals should be over a surface homologous to the sphere in order to identify them with the conserved charges in Lorentzian theory. If we simply take the sphere as the integration surface, then $\BS{u} = u_{t}dt= \sqrt{f(r_{b})}dt$. 
}
\bea
{\rm \underline{thermodynamical ~ quantity}} \hspace{9.7cm} \notag \\
{\rm Empty ~ saddle:} ~~ E=0, ~~ Q=0 \hspace{7.1cm} \\
{\rm Euclidean ~  RN ~ BH ~ saddle:} ~~ E=\frac{r_{b}}{G}\left(1-(r_{b}-r_{H}) \sqrt{\frac{1}{r_{b}(r_{b}-r_{H}(1-G\mu^2)) }}  \right), ~~ Q=\ti{Q}
\ena

The detailed analysis of thermodynamic properties of this system has been done in \cite{BasuKrishnanSubramanian}. Probably, an easy way to see the peculiarity found in \cite{BasuKrishnanSubramanian} is to see the relationship between temperature and horizon radius;
\bea
T=\frac{1-G\mu^2}{4\pi r_{H}} \sqrt{\frac{r_{b}}{r_{b}-r_{H}(1-G\mu^2)}} ~~~~~~~  r_{H}\in (0, r_{b})  \label{temp}
\ena 
When the chemical potential is turned off ($\mu=0$), it reduces to $T=\frac{1}{4\pi r_{H}} \sqrt{\frac{r_{b}}{r_{b}-r_{H}}}$ and will diverge as the horizon radius $r_{H}$ approaches the box radius $r_{b}$. Therefore, the BH branch extends to infinity in the $F$-$T$ diagram and the behavior of the free energy is similar to the AdS boundary case as shown in Fig. \ref{1} or Fig. \ref{4} (Left) \cite{York}. However, once the chemical potential is turned on ($\mu >0$), the denominator in the square root of eq. (\ref{temp}) never reaches zero for $r_{H}\leq r_{b}$, i.e. the temperature remains finite when we take the limit $r_{H}\to r_{b}$. This leads to the peculiar $F$-$T$ diagram and phase diagram as in Fig. \ref{5}.
                                                 %
\iffigure
\begin{figure}[h]
\begin{center}
	\includegraphics[width=7.cm]{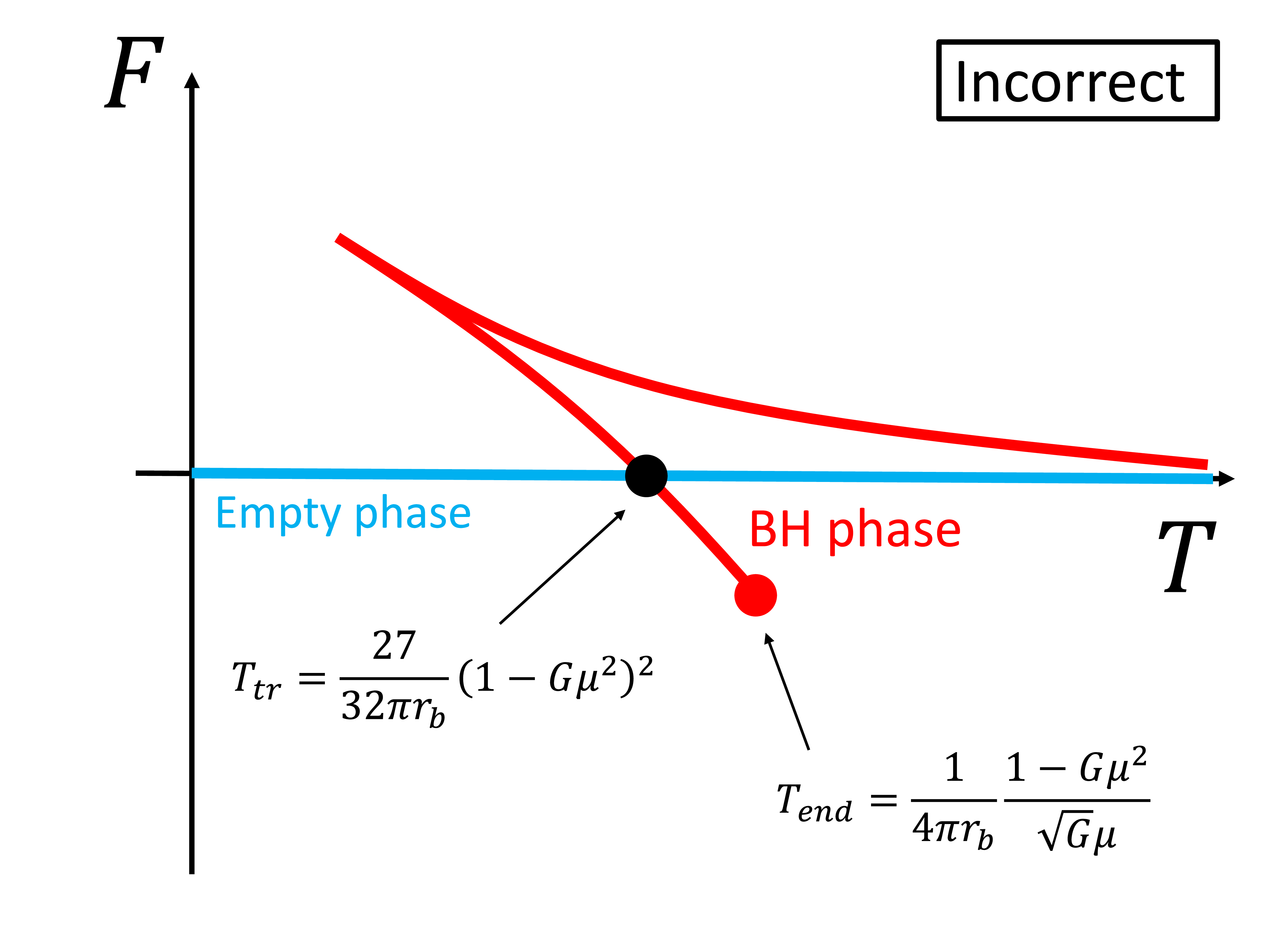}  ~~~~ 	\includegraphics[width=7.cm]{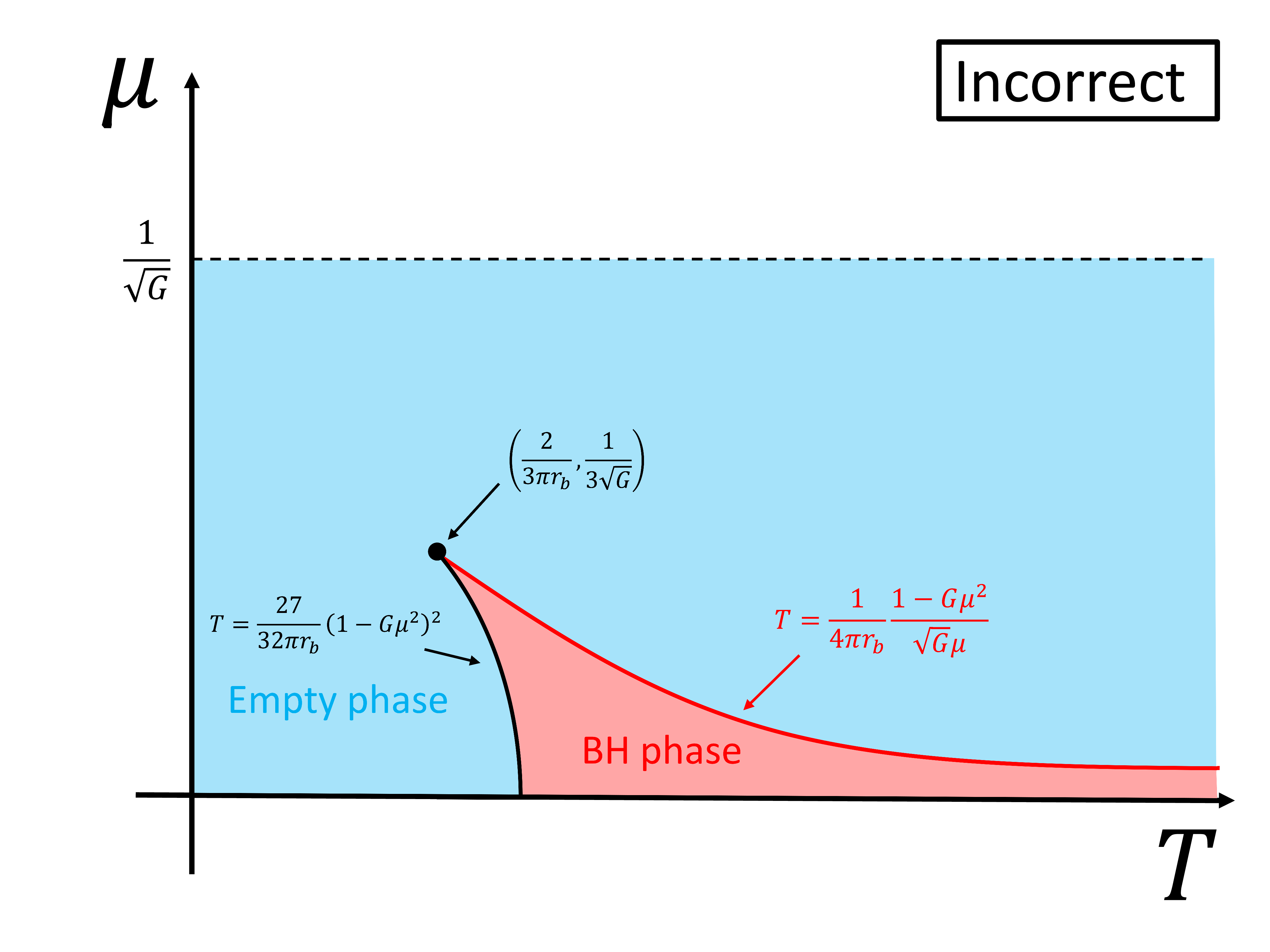} 
	\caption{{\it Incorrect} $F-T$ diagram and phase diagram.}
\label{5}
\end{center}
\end{figure}
\fi
                                                 %
At $T= \frac{1-G\mu^2}{4\pi \sqrt{G}r_{b} \mu }$, the BH branch ceases to exist and the empty saddle becomes the dominant one above the temperature. In the sense that thermodynamic quantities, such as energy and charge, are ill-defined there because of the discontinuity, one might  conclude that thermodynamics is ill-defined in this system. However, I would like to argue that this is not true.

\subsection{{\it good} BG saddle}
When evaluating a partition function of gravity with a spherical boundary, we usually approximate it with saddles with $U(1)\times SO(3)$ isometry. The saddles that dominantly contribute when the phase has finite entropy at zero-loop order can be obtained by Euclideanizing an appropriate part of some static and spherical symmetric spacetime including a horizon. It can be a BH horizon but not necessarily be. The other type of horizon is the inner horizon, which appears inside the BH horizon when a BH has conserved charges other than energy, such as electric charge or angular momentum. 
\footnote{
Here, what I have in mind are those in the Kerr-Newmann family. In general, hairy BHs do not have inner horizons. (However, I suspect that BG instantons {\it do} exist even for such systems. This will be left for future study.) 
}
 In this case, we can obtain an Euclidean saddle by Euclideanizing a suitable part between the inner horizon and the time-like singularity. (Fig. \ref{6}) 
                                                 %
\iffigure
\begin{figure}[h]
\begin{center}
	\includegraphics[width=13cm]{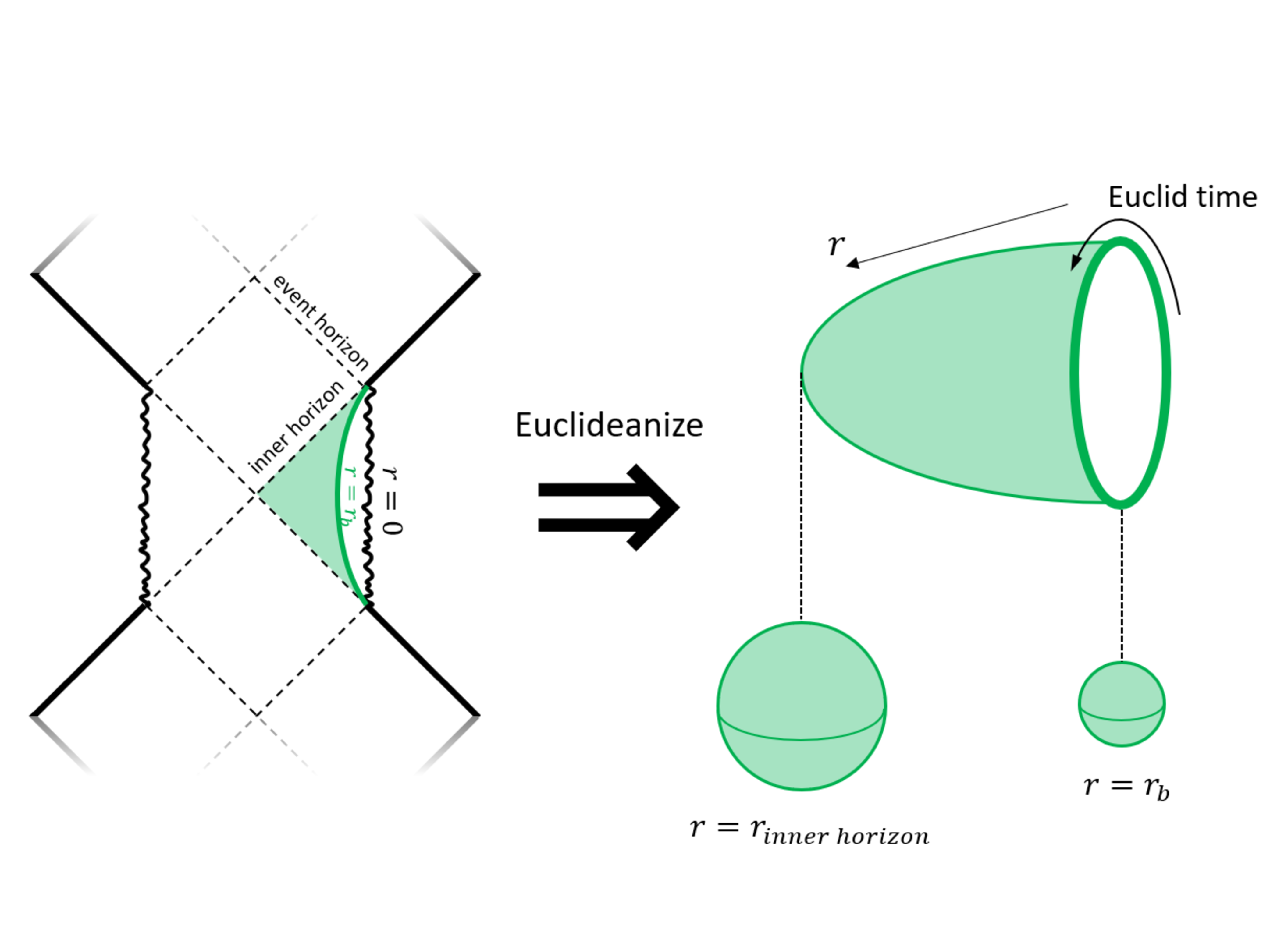}  ~~~~ 
	\caption{(Left) Penrose diagram of the Reissner-Nordstr$\ddot{{\rm o}}$m BH. By Euclideanizing the colored region inside the inner horizon, we obtain a BG geometry. (Right) Schematic picture of the BG geometry. The thick closed curve on the right represents the boundary Euclidean time circle. The area of the bolt is larger than that of the boundary sphere.}
\label{6}
\end{center}
\end{figure}
\fi
                                                 %
This saddle is slightly different from the BH saddle in the sense that the area of the bolt is larger than the area of the boundary sphere. This means that the corresponding entropy exceeds the ``boundary Bekenstein-Hawking entropy,'' which is given by the boundary area divided by $4G$. In the previous paper \cite{Miyashita}, I call this type of saddles ``bag of gold(BG)''.
\footnote{
This is because the system can contain an arbitrary amount of entropy due to the saddle (similar to the Wheeler's bag of gold spacetime), or due to its shape.
}

In the Einstein-Maxwell system, BG saddles and the corresponding thermodynamical quantities are simply given by
\footnote{
$r_G$ represents the position of the bolt (,or the position of ``gold.''  Since the bulk is empty at zero-loop order, the golds should be at the bottom of the ``bag'').
}
\bea
{\rm \underline{field ~ configuration}} \hspace{11cm} \notag \\
{\rm BG ~ saddle:} ~~ r\in[r_{b}, r_{G}], ~ f(r)= 1- \frac{2GM}{r} + \frac{G\ti{Q}^2}{r^2}, ~ A_{t}(r)= -i\ti{Q} \left( \frac{1}{r_{G}} - \frac{1}{r} \right) \label{configBG} \\
{\rm \underline{boundary ~ condition}} \hspace{10.7cm} \notag \\
{\rm BG ~ saddle:} ~~ \frac{4 \pi}{f^{\p}(r_{G})}\sqrt{f(r_{b})}= -\beta, ~~ \frac{\ti{Q}}{\sqrt{f(r_{b})}} \left( \frac{1}{r_{G}} - \frac{1}{r_{b}} \right)=\mu \hspace{2.3cm} \label{bcBG} \\
{\rm \underline{thermodynamical ~ quantity}} \hspace{9.6cm} \notag \\
{\rm BG ~ saddle:} ~~ E=\frac{r_{b}}{G}\left(1-(r_{b}-r_{G}) \sqrt{\frac{1}{r_{b}(r_{b}-r_{G}(1-G\mu^2)) }}  \right), ~~ Q=-\ti{Q} \hspace{0.4cm} \label{tqBG}
\ena
The form of the metric and gauge field (\ref{configBG}) is exactly the same as that of the BH (\ref{configBH}). The only difference is that the radial coordinate $r$ increases from the boundary towards the bolt, as opposed to BH saddles. This fact leads to the following differences from the BH saddle case;
\begin{itemize}
\item The minus sign appears in front of $\beta$ in (\ref{bcBG}) because $f^\p(r_G)$ is now negative. 
\item The minus sign appears in front of $\ti{Q}$ in (\ref{tqBG}) since $n_{r}$, which appears in the definition of charge $Q\equiv\int d^2 z \sqrt{\sigma}\left( \frac{1}{4\pi} n_{\nu}F^{\nu \mu} \right) u_{\mu} $, is now negative, while the other parts are unchanged. 
\end{itemize}

\subsection{thermodynamics}
Here, I will present the thermodynamic properties of the system including the contribution of BG saddles. Before doing so, I want to give a simple check that BGs lead to well-defined thermodynamics. In 3.1, I claimed that the BH branch ceases to exist at some finite temperature because of the relation between $T$ and $r_{H}$ (\ref{temp}).  From the equations in the last subsection, we know the following relation for BG:
\bea
T=\frac{1-G\mu^2}{4\pi r_{G} } \sqrt{\frac{r_{b}}{r_{b}-r_{G}(1-G\mu^2) }} ~~~~~~~  r_{G}\in \left(r_{b}, \frac{r_{b}}{(1-G\mu^2)} \right) \label{tempBG}
\ena
This function form is exactly same as  (\ref{temp}). Combining (\ref{temp}) and (\ref{tempBG}), we may be able to regard it as a single (differentiable) function defined on the range $\left(0, \frac{r_{b}}{(1-G\mu^2)} \right)$. Nothing special happens at $r_{b}$ and the temperature goes to infinity when we take the limit $r_{G} \to \frac{r_{b}}{(1-G\mu^2)}$, similar to the pure gravity case. One can show that the same thing happens for other thermodynamic quantities, including free energy. Therefore, the BH branch and the BG branch can be smoothly connected and we can regard them as a single phase. One might worry that although the thermodynamic quantities are regular at $r_{H}=r_{b}$ , the corresponding geometry may become zero size and singular. But actually this does not happen and the $r_{H} ({\rm or~} r_{G}) \to r_{b}$ limit of BH/BG is (Euclidean) Bertotti-Robinson(BR) geometry, which is a direct product of $AdS_{2}$ and $S^{2}$. See Appendix C for more details.

Let me move on to the thermodynamics. Explicitly, the free energy of the BH/BG phase is 
\bea
F_{BH/BG}(\beta, r_{b}, \mu)=E(\beta, r_{b}, \mu)-\frac{1}{\beta}S(\beta, r_{b}, \mu)-\mu Q(\beta, r_{b}, \mu) \hspace{3.1cm}\\
E(\beta, r_{b}, \mu)=\frac{r_{b}}{G}\left(1-(r_{b}-R) \sqrt{\frac{1}{r_{b}(r_{b}-R(1-G\mu^2)) }}  \right) \hspace{1.45cm} \label{HGene} \\
S(\beta, r_{b}, \mu)= \frac{\pi R^2}{G} \hspace{7.5cm} \\
Q(\beta, r_{b}, \mu)= \frac{\mu^2 r_{b} R^2}{r_{b}-R(1-G\mu^2)} \hspace{5.45cm}
\ena
where $R=R(\beta, r_{b}, \mu)$ is determined by the relation
\bea
T=\frac{1-G\mu^2}{4\pi R } \sqrt{\frac{r_{b}}{r_{b}-R(1-G\mu^2) }} 
\ena
The resulting $F$-$T$ diagram and phase diagram are shown in Fig. \ref{7}. As noted, the BG phase appears and extends to infinity. (Fig. \ref{7} (Left)) Therefore, the system will be empty phase at low temperature and be BH/BG phase at high temperature.  This is the main claim of this paper; this kind of ``Hawking-Page phase structure'' may be common for gravity systems with a sufficiently small magnitude of an external field and I showed that it is true for the Einstein-Maxwell system. What is non-trivial may not be that it is true, since after all we are only doing a small deformation of pure gravity. What is non-trivial is how it is realized, i.e. that it is realized by the appearance of BG phases (or saddles).
                                                 %
\iffigure
\begin{figure}[h]
\begin{center}
	\includegraphics[width=7cm]{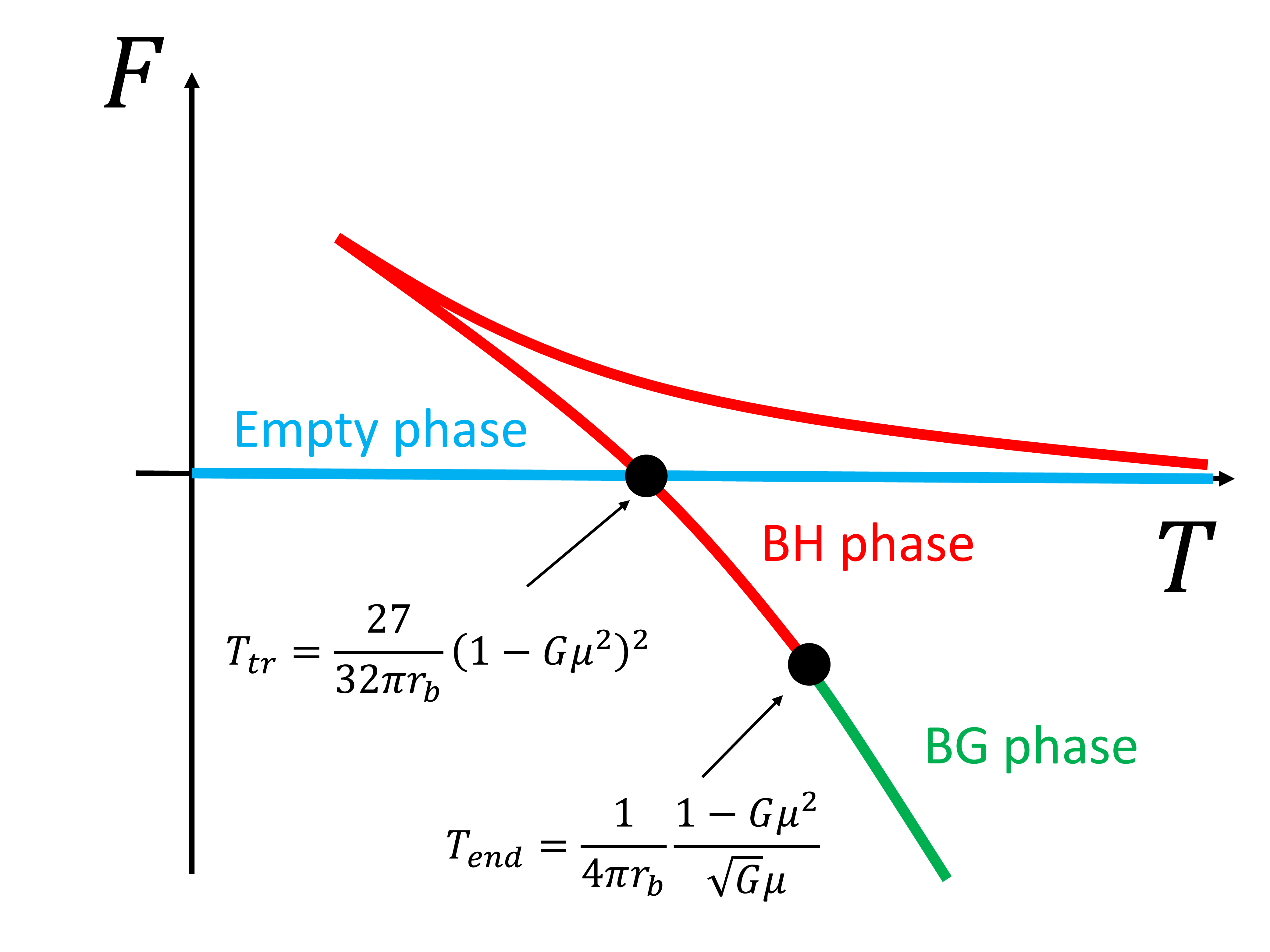}  ~~~~~
	\includegraphics[width=7cm]{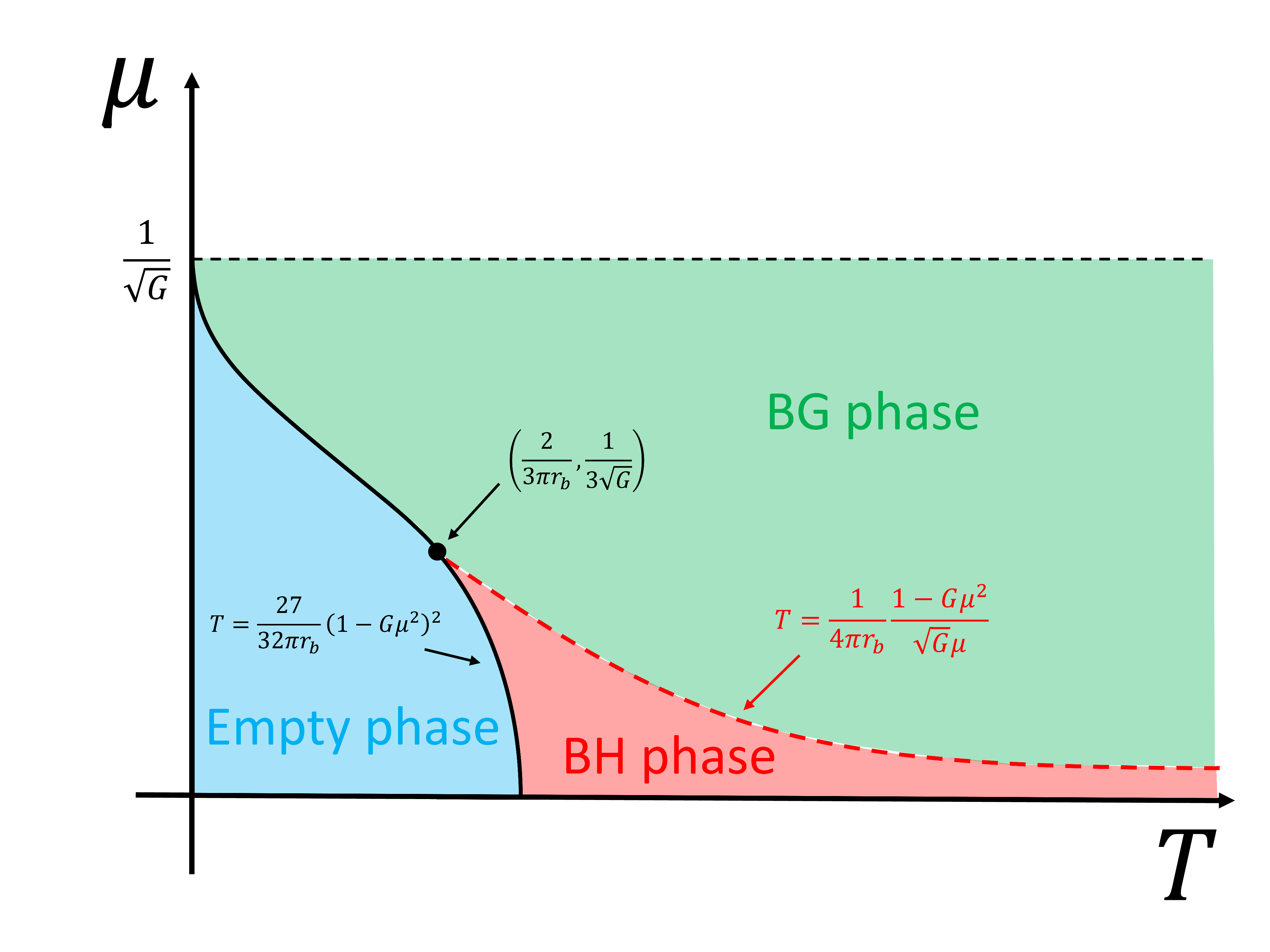}
	\caption{(Left) Qualitative behavior of the free energy versus  temperature when $\Lambda=0$. The BH branch can be extended beyond $T_{end}=\frac{1}{4\pi r_{b}} \frac{1-G\mu^2}{\sqrt{G}\mu}$ and this branch comes from the contribution of BG saddles. (Right) Phase diagram. Although I do not know whether the BH phase and the BG phase should be considered as different phases, I will tentatively treat them as different phases in this paper.  }
\label{7}
\end{center}
\end{figure}
\fi
                                                 %

Of course, there are a number of differences from the pure gravity system or the Einstein-Maxwell system with AdS boundary conditions. So I list them below. 
\begin{itemize}
\item New entropy bound and no energy bound \\
For the case of pure gravity with box boundary, it is known that there is the entropy bound and the energy bound. This is because the area of the bolt of BH saddles is bounded by the area of the boundary $4\pi r_{b}^2$ and, when $\mu=0$ and taking the limit $R \to r_{b}$ of eq. (\ref{HGene}), $E = \frac{r_{b}}{G}$. 

In the Einstein-Maxwell system, however, the maximum area of the bolt is $4 \pi \frac{r_{b}^2}{(1-G\mu^2)^2}$ for a given $\mu < 1$. From eq. (\ref{HGene}),  we know that the energy can be arbitrarily large. Therefore, for a fixed $\mu<1$, there is a new entropy bound $\frac{\pi r_{b}^2}{G} \frac{1}{(1-G\mu^2)^2}$, which is always larger than that of pure gravity $\frac{\pi r_{b}^2}{G} $, and no energy bound.
\item Transition temperature \\
In \cite{Miyashita}, it was shown that the transition temperature depends strongly on the radius of the boundary and weakly depends on other parameters for the case of box boundary: for a given $r_{b}$, the allowed range of the transition temperature is $\frac{27}{32 \pi r_{b}} \leq T_{tr} \leq \frac{1}{\pi r_{b}}$ for pure gravity with negative $\Lambda$, and $\sqrt{\frac{2}{3}} \frac{1}{\pi r_{b}} \leq T_{tr} \leq \frac{1}{\pi r_{b}}$ for the gravity-scalar system. 

As shown in  Fig. \ref{7}, the transition temperature of the Einstein-Maxwell system can also depend strongly on $\mu$ and can be taken to be $0$. 
\footnote{
If we focus only on the transition temperature between the empty phase and the BH phase, it is still confined around $\frac{1}{\pi r_{b}}$, i.e. the range is $\frac{2}{3} \frac{1}{\pi r_{b}} \leq T_{tr} \leq \frac{1}{\pi r_{b}}$. However, in the next section, we will see that even the transition temperature between the empty phase and the BH phase can be taken to be zero.
}
\item Existence of solutions above $\mu=1$ \\
For the case of the AdS boundary condition, the BH phase exists above $\mu_{cr} = 1/\sqrt{G}$ (Fig \ref{3}), i.e. BH saddles exist for all $\mu$. However, in the case of the box boundary condition, BH saddles and BG saddles exist only below $\mu_{cr}$ and we cannot find any non-trivial Euclidean saddles above $\mu_{cr}$ (Fig. \ref{7}). There may exist complex saddles above it and they may contribute to the partition function. However, I will not pursue this possibility here and leave it for future work. 
\end{itemize}

\section{Box Boundary Condition with $\Lambda$}
In this section, I introduce a cosmological constant $\Lambda$ and examine how it changes the properties shown in the previous section. Naively, a negative $\Lambda$ does not change them qualitatively and a positive $\Lambda$ does due to the {\it bad} BG saddle. (Although there is an unexpected behavior when $\sqrt{G}\mu>1$ in the positive $\Lambda$ case.) If you are familiar with gravitational thermodynamics and the BG saddle (i.e. one of the references \cite{Miyashita, DraperFarkas, BanihashemiJacobson}), you can skip this section and go to the final section.  Subsection 4.1 is devoted to the case of negative $\Lambda$ and subsection 4.2 is to the case of positive $\Lambda$. At the beginning of subsection 4.2, I give a very brief review of the thermodynamical properties of pure gravity with positive $\Lambda$. In both cases, the field configuration, boundary condition, and the  thermodynamical quantity are given by
\bea
{\rm \underline{field ~ configuration}} \hspace{11.5cm} \notag \\
{\rm Empty ~ saddle:} ~~ r\in[0, r_{b}], ~ f(r)=1, ~ A_{t}(r)=-i\ti{\mu} \hspace{5cm} \\
{\rm BH ~ saddle:} ~~ r\in[r_{H}, r_{b}], ~ f(r)= 1- \frac{2GM}{r} + \frac{G\ti{Q}^2}{r^2}- \frac{\Lambda}{3}r^2 ,  \hspace{2.65cm} \notag \\
~ A_{t}(r)= -i\ti{Q} \left( \frac{1}{r_{H}} - \frac{1}{r} \right)  \hspace{4.7cm}  \\
{\rm BG ~ saddle:} ~~ r\in[r_{b}, r_{G}], ~ f(r)= 1- \frac{2GM}{r} + \frac{G\ti{Q}^2}{r^2}- \frac{\Lambda}{3}r^2 ,  \hspace{2.65cm} \notag \\
 ~ A_{t}(r)= -i\ti{Q} \left( \frac{1}{r_{G}} - \frac{1}{r} \right) \hspace{4.7cm} \\
{\rm \underline{boundary ~ condition}} \hspace{11.2cm} \notag \\
{\rm Empty ~ saddle:} ~~ \frac{\ti{\mu}}{\sqrt{f(r_{b})}} = \mu    \hspace{8.2cm} \\
{\rm BH ~ saddle:} ~~ \frac{4 \pi}{f^{\p}(r_{H})}\sqrt{f(r_{b})}= \beta, ~~ \frac{\ti{Q}}{\sqrt{f(r_{b})}} \left( \frac{1}{r_{H}} - \frac{1}{r_{b}} \right)=\mu \label{bcBHnL} \hspace{2.55cm} \\
{\rm BG ~ saddle:} ~~ \frac{4 \pi}{f^{\p}(r_{G})}\sqrt{f(r_{b})}= -\beta, ~~ \frac{\ti{Q}}{\sqrt{f(r_{b})}} \left( \frac{1}{r_{G}} - \frac{1}{r_{b}} \right)=\mu \hspace{2.3cm}  \label{bcBGnL} \\
{\rm \underline{thermodynamical ~ quantity}} \hspace{10cm} \notag \\
{\rm Empty ~ saddle:} ~~ E=0, ~~ Q=0 \hspace{7.9cm} \\
{\rm BH ~ saddle:} ~~ E=\frac{r_{b}}{G}\left(\sqrt{1- \frac{\Lambda}{3}r_{b}^2} - (r_{b}-r_{H}) \sqrt{ \frac{1- \frac{\Lambda}{3}\left( r_{b}^2 + r_{b}r_{H} + r_{H}^2 \right) }{r_{b}(r_{b}-r_{H}(1-G\mu^2)) }}  \right), \notag \\
 Q=\ti{Q} \hspace{9.2cm} \\
{\rm BG ~ saddle:} ~~ E=\frac{r_{b}}{G}\left(\sqrt{1- \frac{\Lambda}{3}r_{b}^2} - (r_{b}-r_{G}) \sqrt{ \frac{1- \frac{\Lambda}{3}\left( r_{b}^2 + r_{b}r_{G} + r_{G}^2 \right) }{r_{b}(r_{b}-r_{G}(1-G\mu^2)) }}  \right), \notag \hspace{0.1cm} \\
Q=-\ti{Q} \hspace{8.9cm} 
\ena

\subsection{negative $\Lambda$}
When $\Lambda=0$, there always exist the cusp structure as in Fig. \ref{7} and there are no extremal BH/BG states, i.e. the system does not show the behavior as in the right panel of Fig. \ref{4}, which can be seen in the case of the AdS boundary. When a negative $\Lambda$ is turned on, a new type of behavior and the behavior like the right panel of Fig. \ref{4} appear depending on the parameters $\mu, \Lambda, r_{b}$ and, as a result, extremal BH/BG states appear. To see why they do (not) appear in the case of $\Lambda\neq0 (\Lambda=0)$, let us look again at the relation between $T$ and $r_{H}$. From (\ref{bcBHnL}) or (\ref{bcBGnL}) we get
\bea
Q^2 = \mu^2 \frac{r_{H}^2 r_{b}\left( 1-\frac{\Lambda}{3}(r_{b}^2 +r_{b}r_{H} + r_{H}^2  )  \right) }{r_{b}-r_{H}(1-G\mu^2)  }
\ena
So $T$ can be written as
\footnote{
Here, $\epsilon=1$ for $R<r_{b}$ and $\epsilon=-1$ for $R>r_{b}$.
}
\bea
T=\frac{\epsilon}{4\pi \sqrt{f(r_{b})}}\left( \frac{1}{R} - \frac{Q^2}{R^3} -\Lambda R \right) ~~~~~~~~~~~ \notag \\
 = \frac{\epsilon}{4\pi \sqrt{f(r_{b})}}\frac{r_{b}-R}{3R(r_{b}-R(1-G\mu^2))}\MC{T}(R) \notag\\
  = \frac{1}{4\pi \sqrt{f(r_{b})}}\frac{|r_{b}-R|}{3R(r_{b}-R(1-G\mu^2))}\MC{T}(R)
\ena 
where $\MC{T}$ is defined by
\bea
\MC{T}(R)\equiv -3\Lambda (1-G\mu^2) R^2 + 2\Lambda r_{b} G\mu^2 R + 3-(3-\Lambda r_{b}^2) G\mu^2
\ena 
The plot of $\MC{T}$ for sufficiently small $\mu$ is shown in Fig. \ref{8}.  
                                                 %
\iffigure
\begin{figure}[h]
\begin{center}
	\includegraphics[width=10cm]{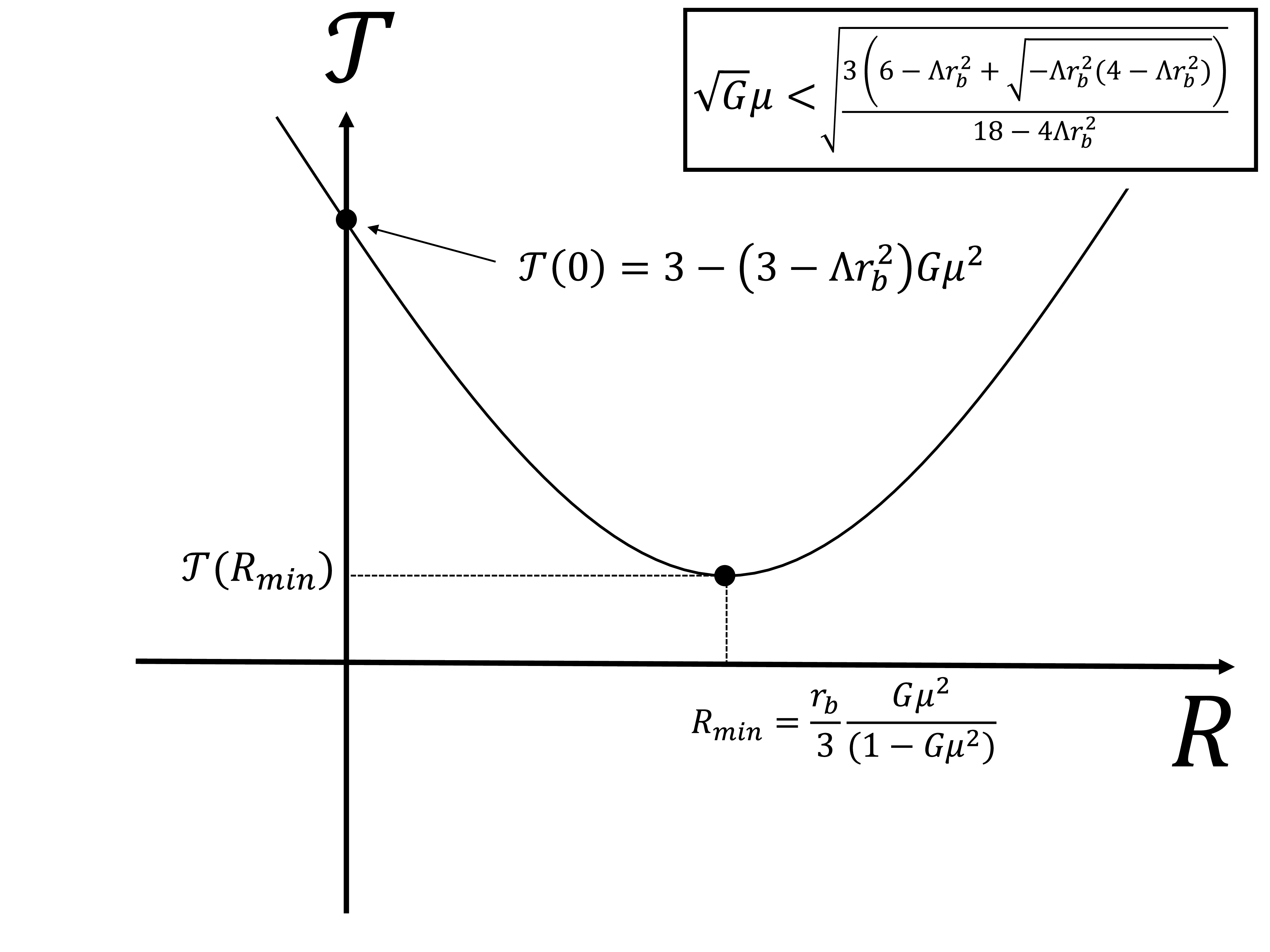} 
	\caption{The plot of $\MC{T}$ for $\sqrt{G}\mu<\sqrt{\frac{3 \left( 6- r_{b}^2 \Lambda + \sqrt{-r_{b}^2 \Lambda (4-r_{b}^2 \Lambda)} \right) }{18-4r_{b}^2 \Lambda} } $. When $\sqrt{\frac{3 \left( 6- r_{b}^2 \Lambda - \sqrt{-r_{b}^2 \Lambda (4-r_{b}^2 \Lambda)} \right) }{18-4r_{b}^2 \Lambda} } < \sqrt{G}\mu< \sqrt{ \frac{3}{(3-\Lambda r_{b}^2)} }$, the minimum of $\MC{T}$ becomes negative and there are two positive roots corresponding to extremal BH or BG states. When $ \sqrt{\frac{3}{(3-\Lambda r_{b}^2)}}< \sqrt{G} \mu <1$, $\MC{T}(0)$ also becomes negative and there is only one positive root. Explicitly, the two roots are $R_{ex}=\frac{1}{-3\Lambda (1-G\mu^2)}\left( -\Lambda r_{b} G \mu^2 \pm \sqrt{-\Lambda \left( -9-3(-6+r_{b}^2 \Lambda )G \mu^2 + (-9+2 r_{b}^2 \Lambda) G^2\mu^4 \right)} \right)$. }
\label{8}
\end{center}
\end{figure}
\fi
                                                 %
Because the other factors of temperature are always positive,
\footnote{
For $R<\frac{r_{b}}{1-G\mu^2}$ and $\sqrt{G}\mu<1$.
}
the positivity of $\MC{T}$ can be used to judge the existence of solution for a given $R$. Let the radius of the minimum $R_{min}$, which is given by 
$R_{min}= \frac{r_{b}}{3}\frac{G\mu^2}{1-G\mu^2}$. As shown in the figure, the positivity of $\MC{T}(R_{min})$ and $\MC{T}(0)$ depends on the parameters  $\mu, r_{b}, \Lambda$;
\begin{itemize}
\item $0< \sqrt{G}\mu < \sqrt{\frac{3 \left( 6- r_{b}^2 \Lambda - \sqrt{-r_{b}^2 \Lambda (4-r_{b}^2 \Lambda)} \right) }{18-4r_{b}^2 \Lambda} }$ \\
In this range, $\MC{T}(R)$ is always positive, as shown in Fig. \ref{8}. So there are no extremal states and there are always solutions for $0<R<\frac{r_{b}}{(1-G\mu^2)}$.
\item  $\sqrt{\frac{3 \left( 6- r_{b}^2 \Lambda - \sqrt{-r_{b}^2 \Lambda (4-r_{b}^2 \Lambda)} \right) }{18-4r_{b}^2 \Lambda} } < \sqrt{G}\mu < \sqrt{\frac{3}{(3-\Lambda r_{b}^2)}}$ \\
The minimum $\MC{T}(R_{min})$ becomes negative but $T(0)$ is still positive. So there are two positive roots, which are given by $R_{ex}= \frac{1}{-3\Lambda (1-G\mu^2)} \left(-\Lambda r_{b} G\mu^2 \pm \sqrt{3\Lambda(1-G\mu^2)\MC{T}(R_{min})}\right)$. These two roots represent extremal BHs (or BGs). The one with the larger radius is thermodynamically (locally) stable and the other is unstable.
\item $\sqrt{\frac{3}{(3-\Lambda r_{b}^2)}} < \sqrt{G}\mu <1$ \\
Since both $\MC{T}(0)$ and $\MC{T}(R_{min})$ are negative, there is only one positive root. 
\end{itemize}
The qualitative behaviors of the free energies of these cases are shown in Fig. \ref{9}. Note that this time, unlike the $\Lambda=0$ case, we may not have the exact expression of the transition temperature between the empty phase and the BH/BG phase. However, we still have that of $T_{end}$, which is the "transition" temperature between the BH and BG phases:
\bea
T_{end} = \frac{1-G\mu^2 - \Lambda r_{b}^2 (1-2G\mu^2) }{4\pi \sqrt{G} \mu r_{b} \sqrt{1-\Lambda r_{b}^2}}  \label{negativeend}
\ena
                                                 %
\iffigure
\begin{figure}[h]
\begin{center}
	\includegraphics[width=5.2cm]{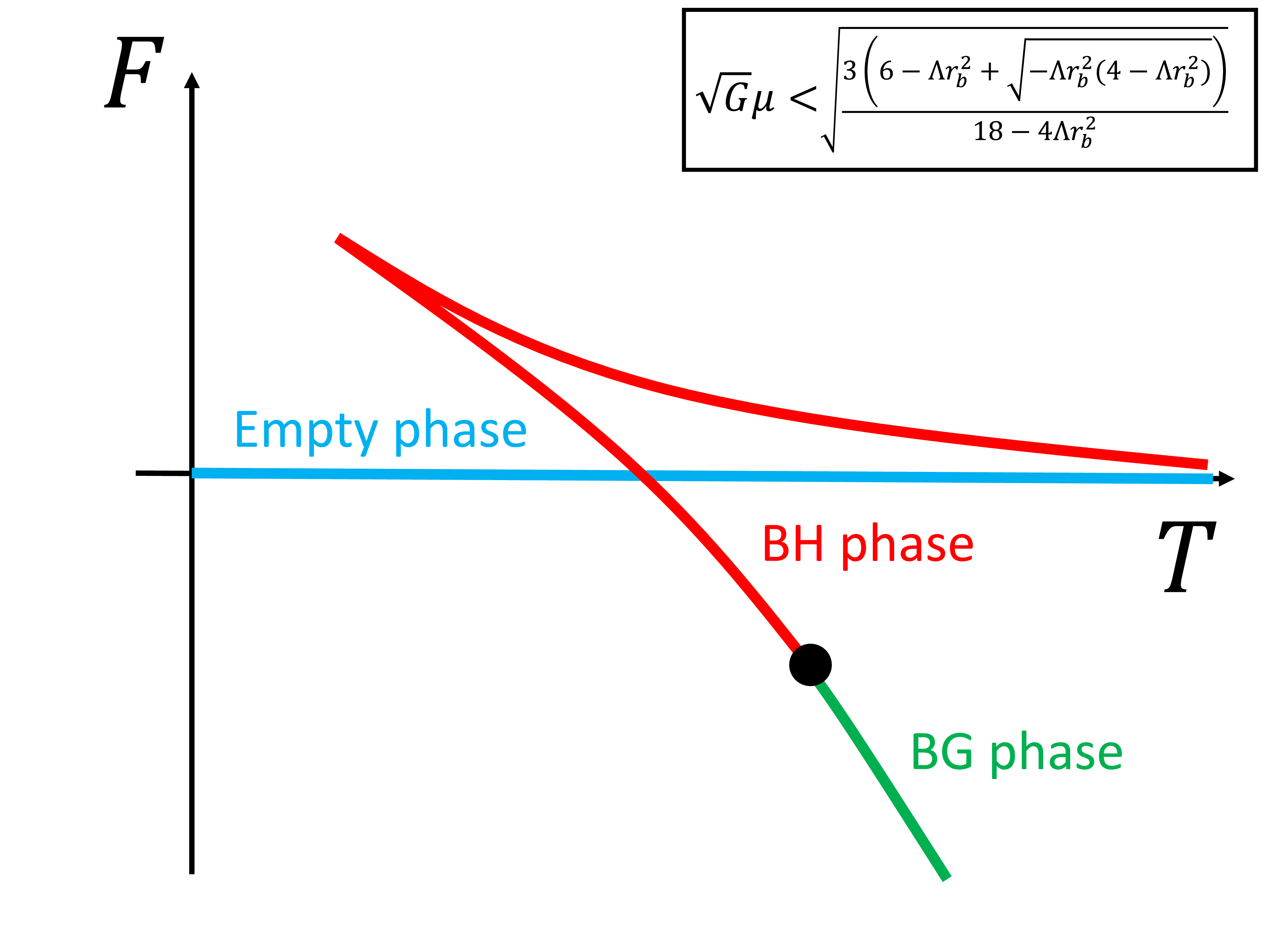} ~
	\includegraphics[width=5.2cm]{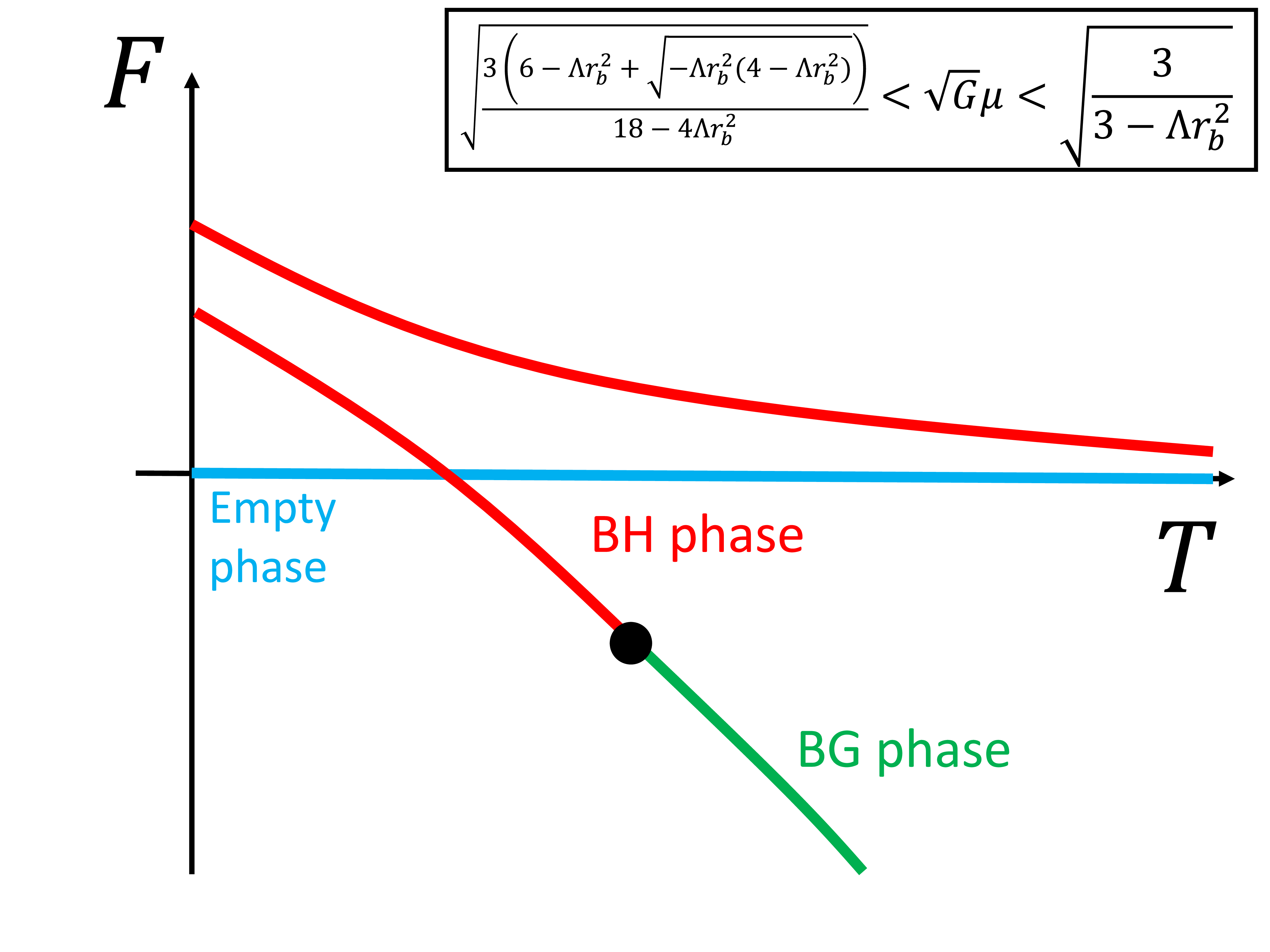} ~
	\includegraphics[width=5.2cm]{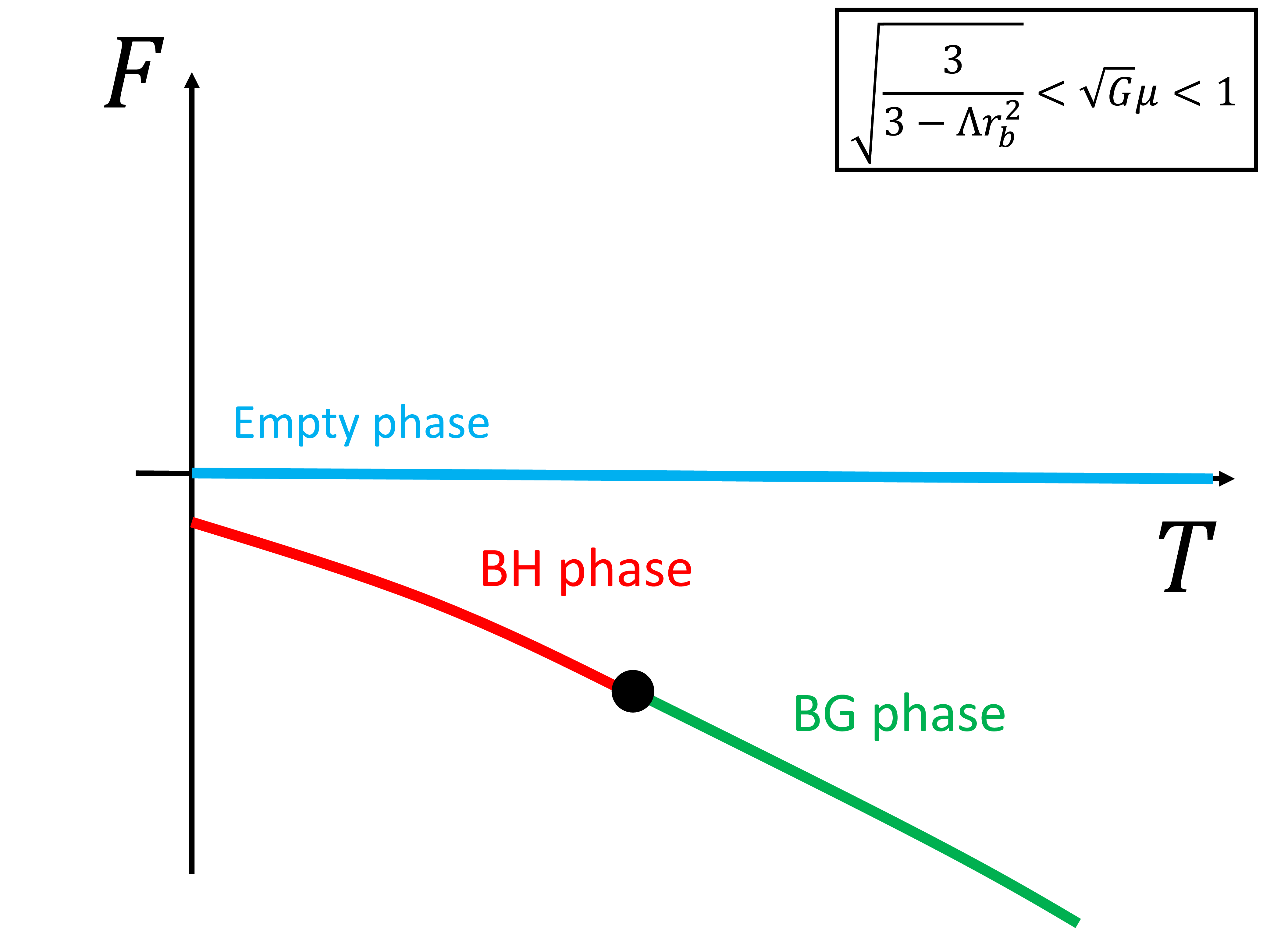}  
	\caption{Qualitative behaviors of the free energies when $\Lambda<0$. In the case of $\Lambda<0$, there are qualitatively different behaviors depending on the chemical potential.  For each type of behavior, the position of the boundary between BH and BG also depends on the chemical potential. For the phase boundary behavior, see Fig. \ref{11}. Note that although in the middle figure I show the case where the empty phase has the lowest free energy at low temperature, the global stability of the locally stable BH/BG branch in the figure also depends on the chemical potential. There exists the critical chemical potential, which depends on the combination $-\Lambda r_{b}^2$, and empty phase is realized when the chemical potential below it and BH/BG phase is realized when the chemical potential above it at low temperature in the middle case. (See Fig. \ref{10}.)}
\label{9}
\end{center}
\end{figure}
\fi
                                                 %
In the second range of $\mu$, the locally stable extremal BH/BG appears but it is not always the dominant saddle. Since the extremal BH/BG in the third range is always the dominant saddle, the exchange of the dominance between the empty phase and the extremal BH/BG phase would occur in the second range. This critical chemical potential, above which extreme BH/BGs become the dominant saddles, is plotted in Fig. \ref{10}. This depends only on the combination $-\Lambda r_{b}^2$. In the left panel, I also show the upper and lower boundary values of the second range. In the right panel, I have plotted the relative values, relative to the lower boundary value $\sqrt{\frac{3 \left( 6- r_{b}^2 \Lambda - \sqrt{-r_{b}^2 \Lambda (4-r_{b}^2 \Lambda)} \right) }{18-4r_{b}^2 \Lambda} }$.
                                                 %
\iffigure
\begin{figure}[h]
\begin{center}
	\includegraphics[width=8cm]{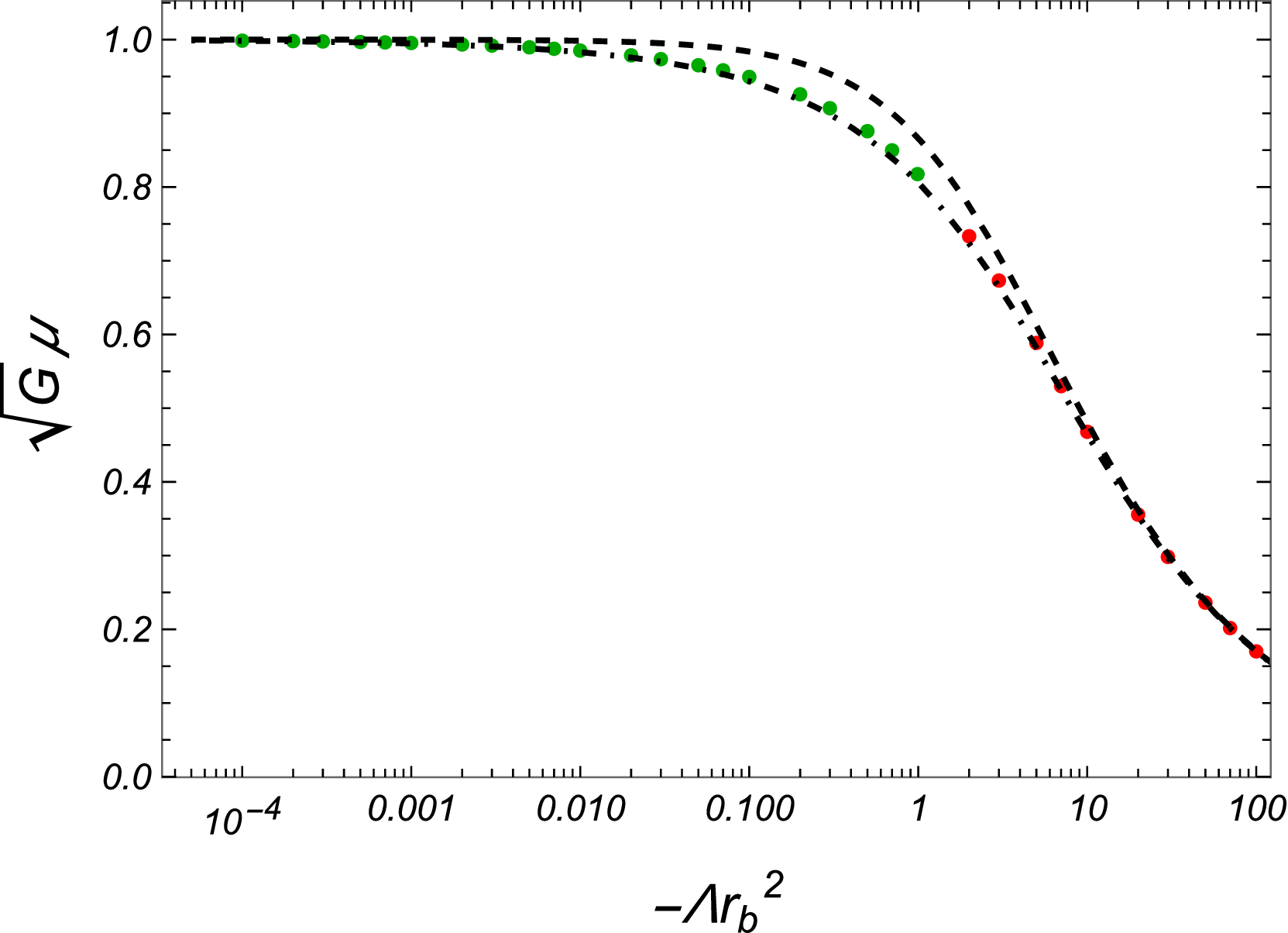} ~~~
	\includegraphics[width=8cm]{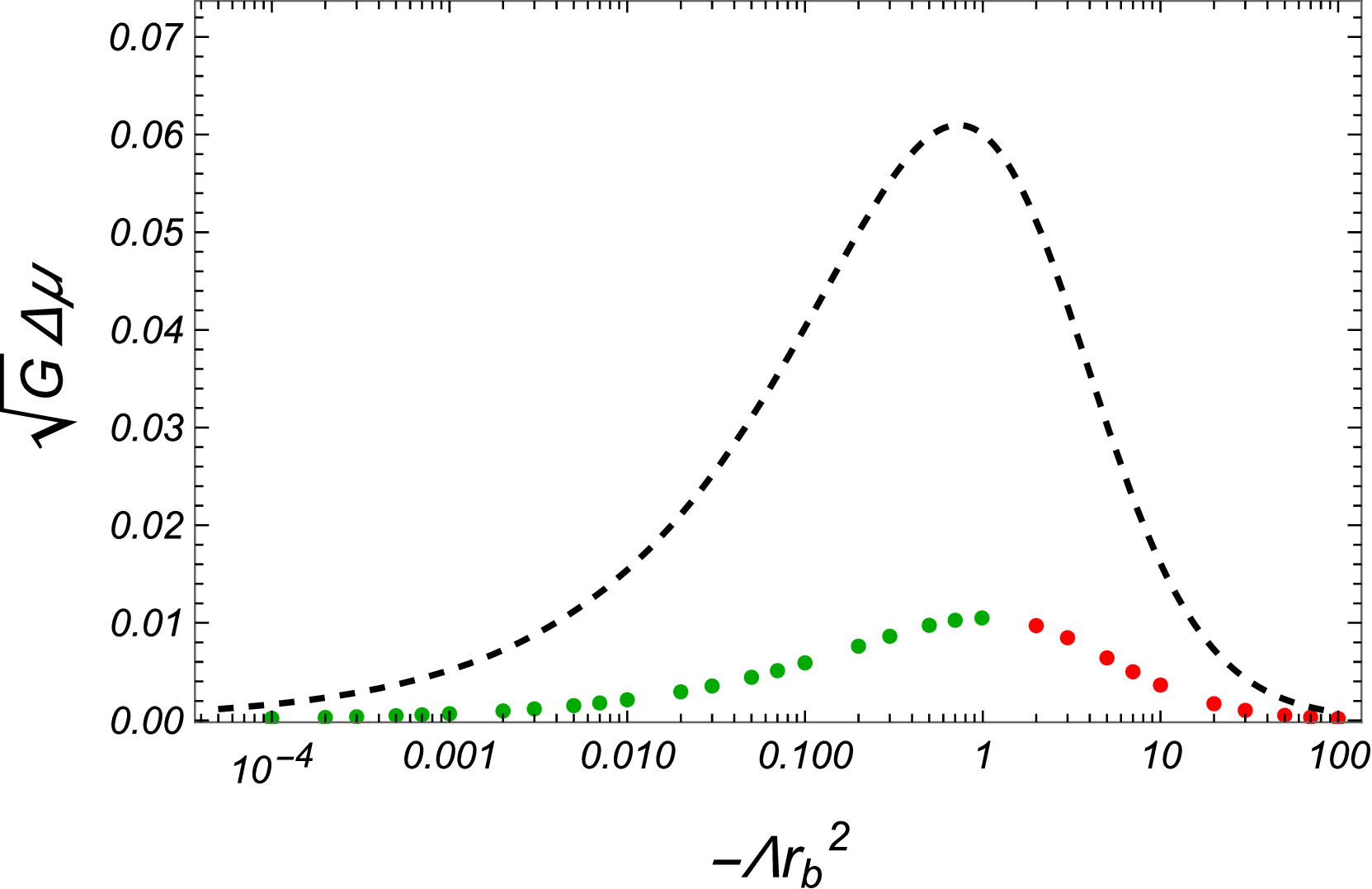}  
	\caption{(Left) The plot of the critical chemical potential above which the extremal BH/BGs become the dominant saddles. Green and red dots represent the values. It depends on the combination $-\Lambda r_{b}^2$. I also show the upper and lower boundary values of the second range, that is, the dashed curve (upper) represents $\sqrt{G} \mu= \sqrt{\frac{3}{3-\Lambda r_{b}^2}} $ and the dot-dashed curve (lower) represents $\sqrt{G}\mu = \sqrt{\frac{3 \left( 6- r_{b}^2 \Lambda + \sqrt{-r_{b}^2 \Lambda (4-r_{b}^2 \Lambda)} \right) }{18-4r_{b}^2 \Lambda} } $. Green dots represent the transition is to BG phase and red ones represent the transition is to BH phase. The exchange occurs at $-\Lambda r_{b}^2 =1$ (i.e. $\sqrt{G}\mu=\frac{2}{3}$). (Right) Relative critical values and the upper boundary values, relative to the lower boundary value.  }
\label{10}
\end{center}
\end{figure}
\fi
                                                 %
As shown in the figure, the second range shrinks for large and small values of $-\Lambda r_{b}^2$ and we can approximate the critical chemical potential as, for example, $\sqrt{\frac{3}{(3-\Lambda r_{b}^2)}}$. Only when the boundary radius $r_{b}$ and the AdS length $l_{-} \sim 1/\sqrt{-\Lambda}$ are comparable, the critical chemical potential has non-trivial dependence.

                                                 %
\iffigure
\begin{figure}[h]
\begin{center}
	\includegraphics[width=5.2cm]{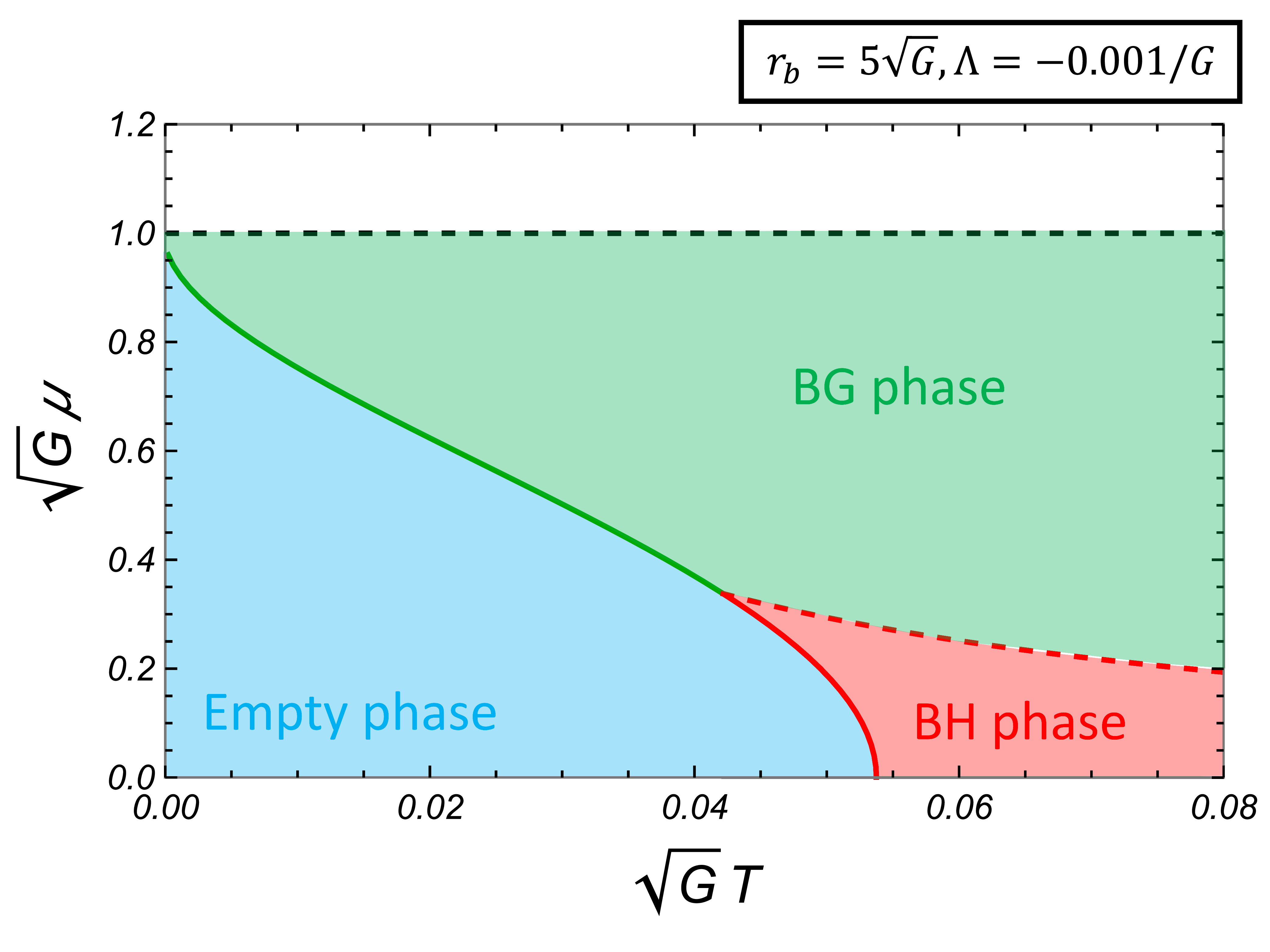} ~
	\includegraphics[width=5.2cm]{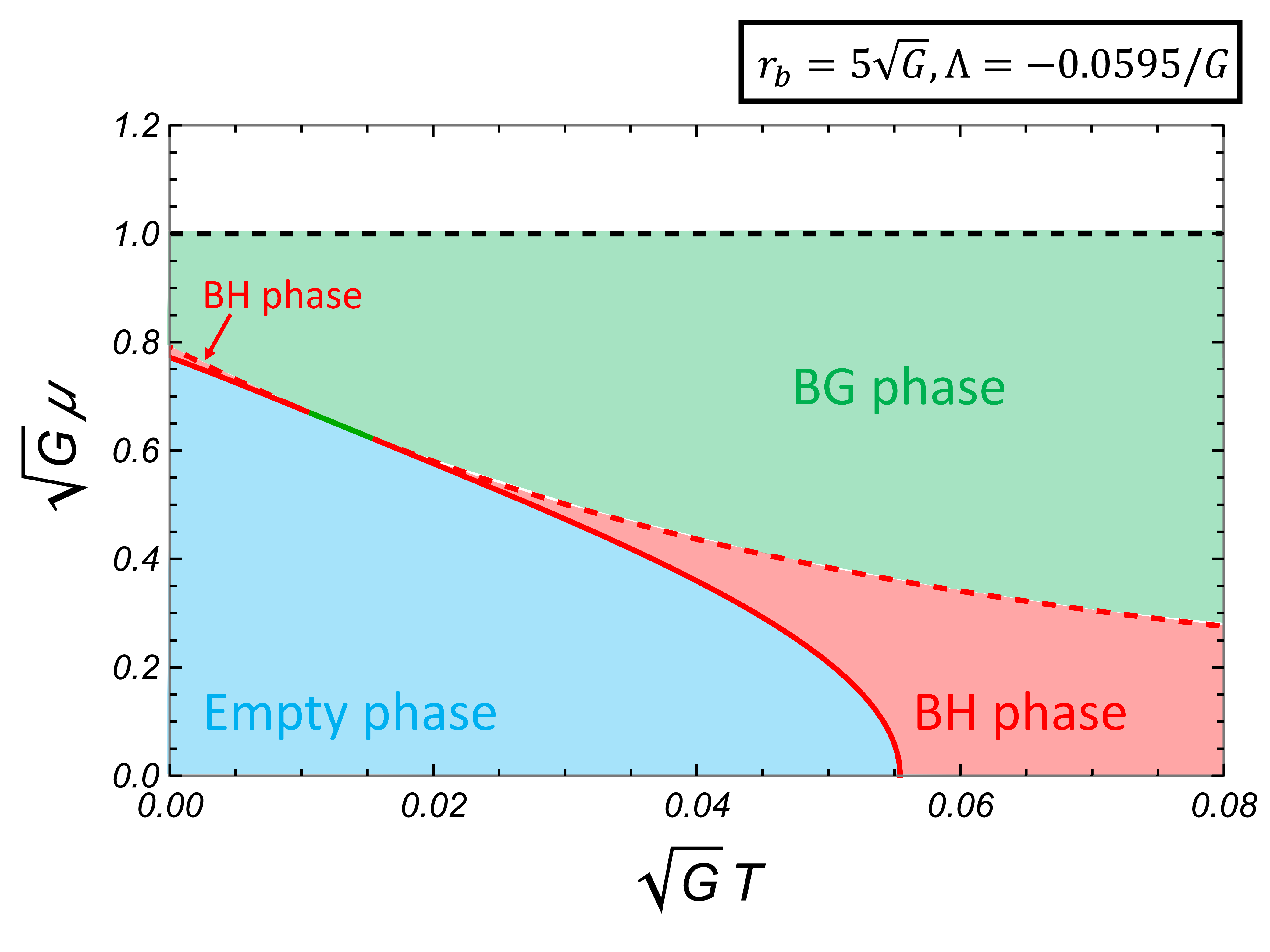} ~
	\includegraphics[width=5.2cm]{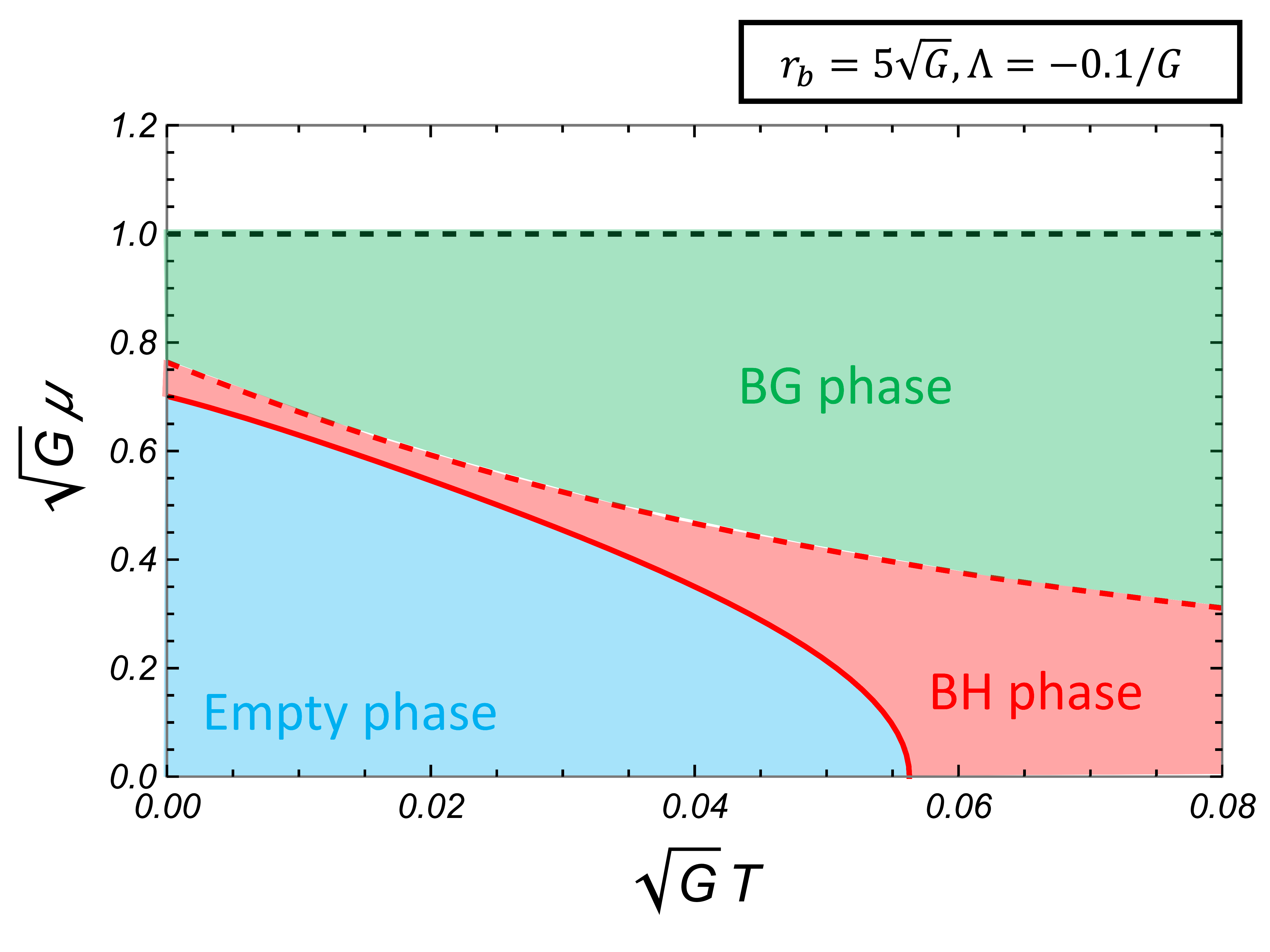}  
	\caption{When $\Lambda<0$, phase diagrams are classified into three types depending on the value of $-\Lambda r_{b}^2$. These are examples of each type. (Left)  $0<-\Lambda r_{b}^2< 1$. (Middle)  $1<-\Lambda r_{b}^2<\frac{3}{2}$, (Right) $\frac{3}{2}<-\Lambda r_{b}^2$. }
\label{11}
\end{center}
\end{figure}
\fi
                                                 %
The phase diagrams also exhibit qualitatively different behavior, as shown in Fig. \ref{11}. When $-\Lambda r_{b}^2$ is sufficiently small, it is qualitatively same as the $\Lambda=0$ case. When $1<-\Lambda r_{b}^2<\frac{3}{2}$, another BH phase appears, which is separated from the one already existing for small $-\Lambda r_{b}^2$. (Fig. \ref{11} (Middle) and Fig. \ref{12})
                                                 %
\iffigure
\begin{figure}[h]
\begin{center}
	\includegraphics[width=7cm]{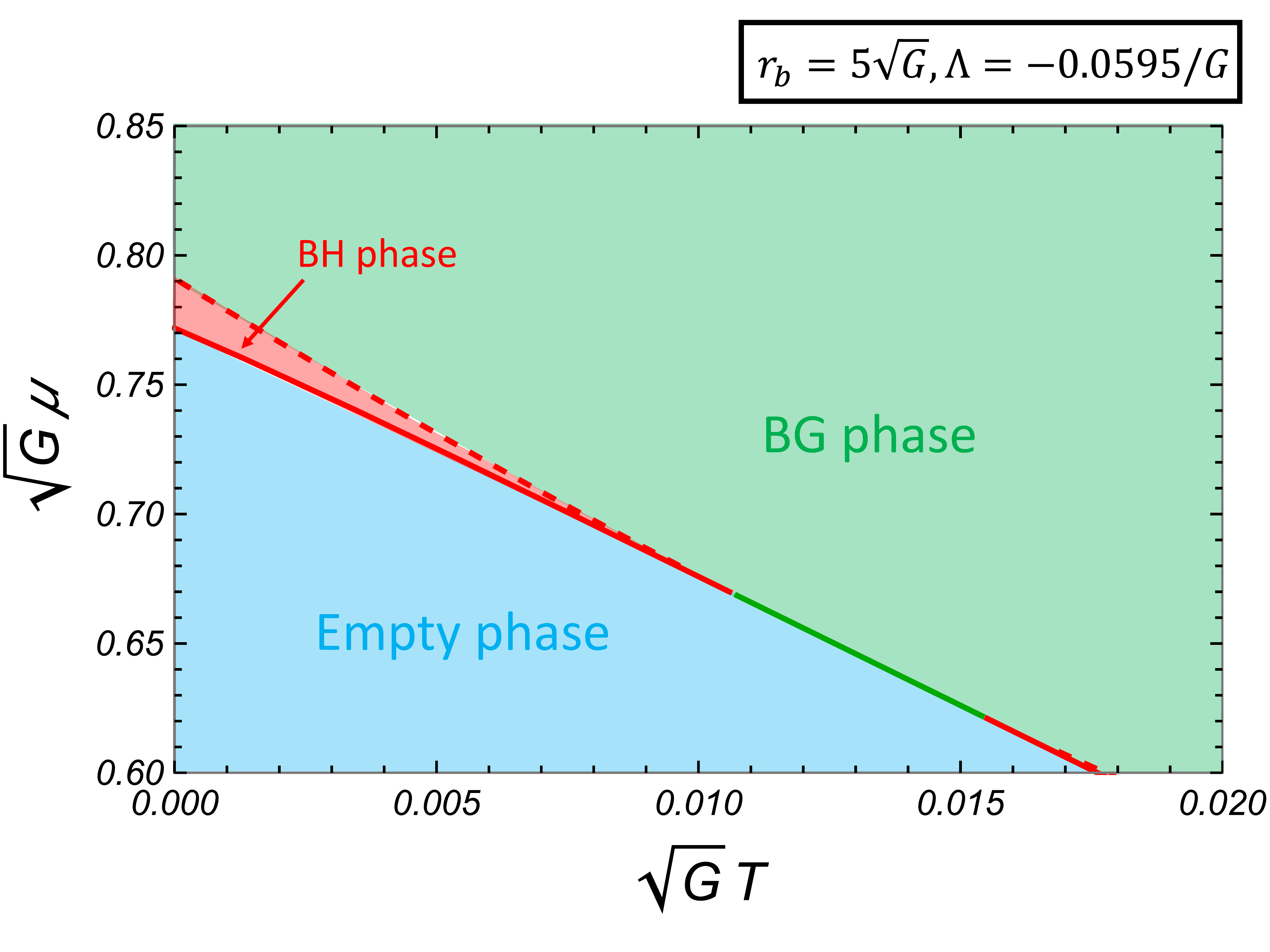} 
	\caption{The enlarged figure near the disconnected BH phase of the Middle panel of Fig \ref{11}.}
\label{12}
\end{center}
\end{figure}
\fi
                                                 %
 When $\frac{3}{2}<-\Lambda r_{b}^2$, the BG phase is completely separated from the empty phase and the BH phases are connected. 

Note that, since this system has a negative $\Lambda$, $r_{b}/l_{-}\to \infty$ limit, with a suitable redefinition of the variables, leads to the system of AdS boundary conditions. More precisely, by defining
\bea
T_{AdS}\equiv \left(\frac{r_{b}}{l_{-}} \right) T, ~~~ \mu_{AdS} \equiv \left(\frac{r_{b}}{l_{-}} \right) \mu
\ena
and taking the limit, we will see that, in Fig. \ref{11} (Right), both the $\mu=1$ boundary and the phase boundary between the BG and BH phases go to infinity,
\footnote{
For the latter, from the equation (\ref{negativeend}), $T_{end,AdS}$ becomes 
\beann
T_{end,AdS} \simeq \frac{1}{4\pi \mu_{AdS}} \sqrt{\frac{3}{G}} \left( \frac{r_{b}}{l} \right)^2 \to \infty .
\enann
}
 and the phase boundary between the empty phase and the BH phase will be $T_{AdS} = \frac{1}{\pi l_{-}} \sqrt{1-G\mu_{AdS}^2}$. 
\footnote{
For large $r_{b}/l_{-}$, the free energy of the BH can be written as
\beann
\left( \frac{r_{b}}{l_{-}}\right)F = \frac{r_{H}}{l_{-}^2} \big( l_{-}^2(1-G\mu_{AdS}^2) - r_{H}^2 \big) + \MC{O}\left( \left(l_{-}/r_{b}\right)^2 \right)
\enann
Therefore, in the limit, the horizon radius at the transition is $r_{H}= l_{-} \sqrt{1-G\mu^2_{AdS}}$. Substituting this into $T_{AdS}$, we get,
\beann
T_{AdS} = \frac{r_{b}}{l_{-}} T= \frac{1}{\pi l_{-}} \sqrt{1-G \mu_{AdS}^2} + \MC{O}\left( \left(l_{-}/r_{b}\right) \right) .
\enann
}

\subsection{positive $\Lambda$ and {\it bad} BG saddle}
In the case of pure gravity, when $\Lambda >0$, it has been shown that there exist {\it bad} BG saddles and thermodynamics becomes ill-defined due to their contribution \cite{Miyashita, DraperFarkas, BanihashemiJacobson}. Before proceeding to the analysis of the case of the Einstein-Maxwell system, let me briefly review the case of pure gravity.

In the case of pure gravity, the existence of BH saddles and {\it bad} BG saddles depends on $\Lambda$. Let the radius of the bolt be $R$. For a given $R$, we could know the value of $f(r_{b})$. If it is not positive, it means that there is no corresponding solution. The value is
\bea
f(r_{b})= \frac{(r_{b}-R)}{3 r_{b}} (3- \Lambda r_{b}^2 -\Lambda r_{b}R - \Lambda R^2)
\ena
Therefore the conditions for the existence of the solution are\\
~\\
 \hspace{1cm} $(3- \Lambda r_{b}^2 -\Lambda r_{b}R - \Lambda R^2):$ positive \& $(r_{b} - R)$: positive \\
\hspace{5.5cm} or \\
  \hspace{1cm}  $(3- \Lambda r_{b}^2 -\Lambda r_{b}R - \Lambda R^2):$ negative \& $(r_{b} - R )$: negative\\
 ~\\
The former corresponds to the BH saddle and the latter to the {\it bad} BG saddle. Depending on the value of $\Lambda r_{b}^2$, the possible ranges of $R$ for BH and {\it bad} BG saddles are different. I classify them into three classes and name them as follows:
\begin{itemize}
\item \{{\it good}, {\it bad}\}$_{I}$ : $0<\Lambda r_{b}^2 <1$ \\
Both BH and {\it bad} BG saddles exist. \\
$ R\in (0, r_{b}) $ for BH and $R \in (\frac{-\Lambda r_{b}+ \sqrt{\Lambda^2 r_{b}^2 + 4\Lambda (3-\Lambda r_{b}^2) }}{2\Lambda}, \infty) $ for {\it bad} BG.
\item \{{\it good}, {\it bad}\}$_{II}$ : $1<\Lambda r_{b}^2<3$ \\
Both BH and {\it bad} BG saddles exist. \\
$R\in (0, \frac{-\Lambda r_{b}+ \sqrt{\Lambda^2 r_{b}^2 + 4\Lambda (3-\Lambda r_{b}^2) }}{2\Lambda})$ for BH and $R\in (r_{b}, \infty)$ for {\it bad} BG.
\item \{{\it bad}\} : $3 \leq \Lambda r_{b}^2$ \\
Only {\it bad} BG saddles exist.\\
$R\in (r_{b}, \infty)$ for {\it bad} BG.
\end{itemize}
Here {\it good} in parentheses stands for BH and {\it bad} stands for {\it bad} BG. The special case is $\Lambda r_{b}^2= 1$, where the maximum bolt radius of BH and the minimum bolt radius of {\it bad} BG coincide $r_{b}=\frac{-\Lambda r_{b}+ \sqrt{\Lambda^2 r_{b}^2 + 4\Lambda (3-\Lambda r_{b}^2) }}{2\Lambda}$ and $R=r_{b}$ corresponds to Euclidean Nariai space. The reason why I use the term ``{\it bad} BG'' is that their heat capacities are always negative and dominantly contribute to the partition function, making the system thermodynamically unstable \cite{Miyashita, DraperFarkas, BanihashemiJacobson}.

Now let's extend the analysis of pure gravity to Einstein-Maxwell, i.e. to the case of $\mu \neq 0$. As before, we may be able to judge the existence of the solution for a given $R$ and $\mu$ by the positivity of $f(r_{b})$. This is given by
\bea
f(r_{b})= \frac{(r_{b}-r_{H})^2}{3 r_{b}} \frac{(3- \Lambda r_{b}^2 -\Lambda r_{b}r_{H} - \Lambda r_{H}^2)}{(r_{b} - r_{H} (1-G\mu^2))}
\ena
Therefore the conditions for the existence of a solution are\\
~\\
 \hspace{1cm} $(3- \Lambda r_{b}^2 -\Lambda r_{b}r_{H} - \Lambda r_{H}^2):$ positive \& $(r_{b} - r_{H} (1-G\mu^2))$: positive  \\
\hspace{5.5cm} or \\
  \hspace{1cm}  $(3- \Lambda r_{b}^2 -\Lambda r_{b}r_{H} - \Lambda r_{H}^2):$ negative \& $(r_{b} - r_{H} (1-G\mu^2))$: negative \\
 ~\\
The former corresponds to the BH saddle and the {\it good} BG saddle, the latter corresponds to the {\it bad} BG saddle. This time, depending on the value of $\Lambda r_{b}^2$ and $\sqrt{G} \mu$, the possible ranges of $R$ for BH/{\it good} BG and {\it bad} BG saddles are different and I classify them into five classes and name them as follows:
\begin{itemize}
\item \{{\it good}, {\it bad}\}$_{I}$ : $ 0<\Lambda r_{b}^2<\frac{3(1-G\mu^2)^2}{3-3G\mu^2+ G^2 \mu^4} , ~ \sqrt{G}\mu<1$ \\
Both BH/{\it good} BG and {\it bad} BG saddles exist. \\
$ R\in \left(0, \frac{r_{b}}{1-G\mu^2} \right) $ for BH/{\it good} BG and $R \in (\frac{-\Lambda r_{b}+ \sqrt{\Lambda^2 r_{b}^2 + 4\Lambda (3-\Lambda r_{b}^2) }}{2\Lambda}, \infty) $ for {\it bad} BG.
\item \{{\it good}, {\it bad}\}$_{II}$ : $\frac{3(1-G\mu^2)^2}{3-3G\mu^2+ G^2 \mu^4} <\Lambda r_{b}^2<3, ~ \sqrt{G}\mu<1$ \\
Both BH/{\it good} BG and {\it bad} BG saddles exist. \\
$R\in (0, \frac{-\Lambda r_{b}+ \sqrt{\Lambda^2 r_{b}^2 + 4\Lambda (3-\Lambda r_{b}^2) }}{2\Lambda})$ for BH/{\it good} BG and $R\in \left(\frac{r_{b}}{1-G\mu^2}, \infty\right)$ for {\it bad} BG.
\item \{{\it bad}\} : $3\leq \Lambda r_{b}^2, ~ \sqrt{G}\mu<1$ \\
Only {\it bad} BG saddles exist.\\
$R\in \left(\frac{r_{b}}{1-G\mu^2}, \infty\right)$ for {\it bad} BG.
\item \{{\it good}\} : $0<\Lambda r_{b}^2 <3, ~ \sqrt{G}\mu\geq1$ \\
Only BH/{\it good} BG saddles exist. \\
$R\in (0, \frac{-\Lambda r_{b}+ \sqrt{\Lambda^2 r_{b}^2 + 4\Lambda (3-\Lambda r_{b}^2) }}{2\Lambda})$ for BH/{\it good} BG.
\item $\emptyset$ : $3 \leq \Lambda r_{b}^2, ~ \sqrt{G}\mu\geq1$ \\
No saddles exist.\\
\end{itemize}
This time, {\it good} in parentheses stands for both BH and {\it good} BG. Here, the term ``{\it good} BG'' means that their heat capacities are positive (when they dominantly contribute), as are the BGs that appear when $\Lambda \leq 0$. The above classification is summarized in Fig. \ref{13}.
                                                 %
\iffigure
\begin{figure}[h]
\begin{center}
	\includegraphics[width=8cm]{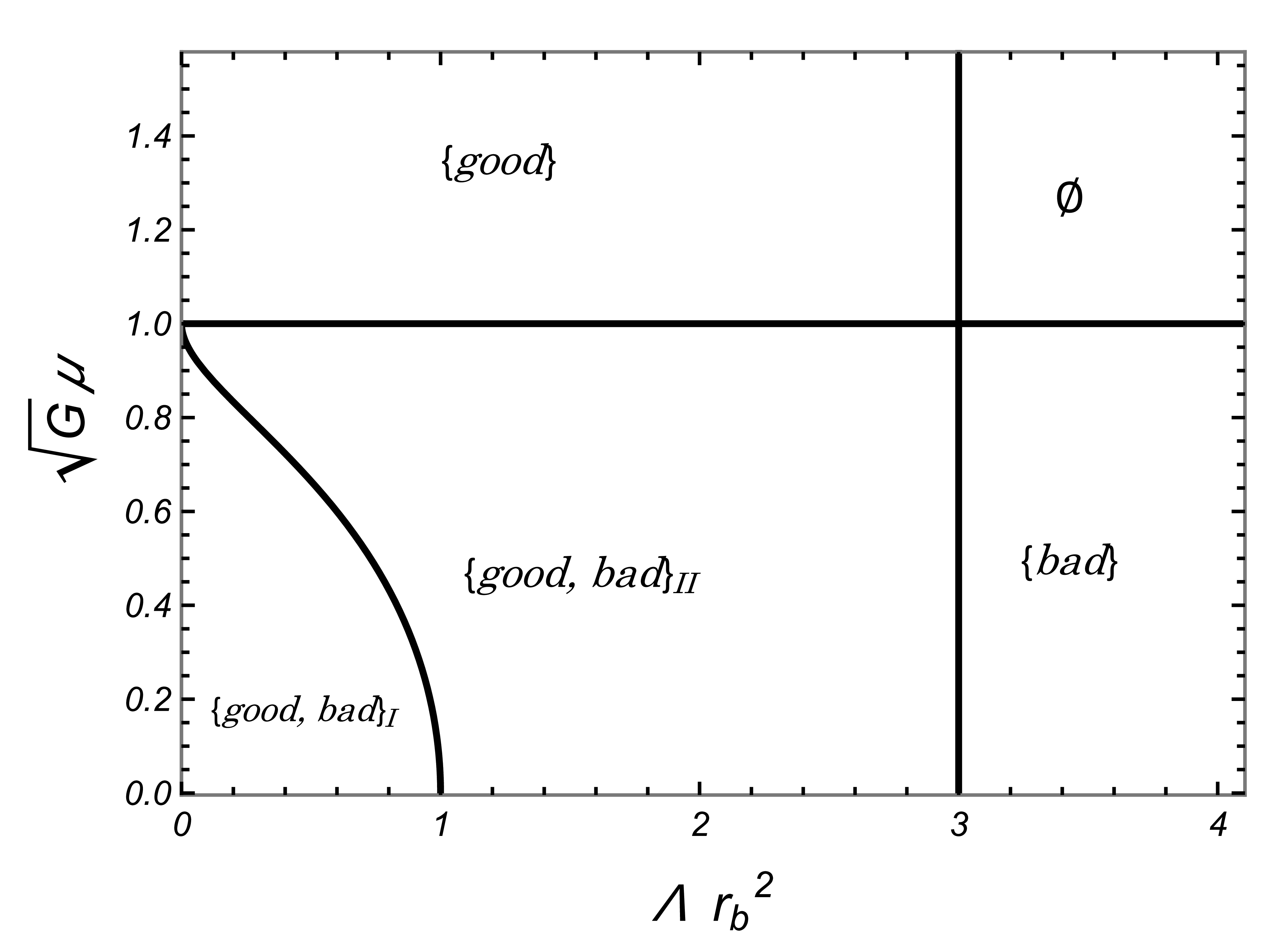} 
	\caption{Classification of the existence and the possible range of the bolt radius $R$. It depends on $\Lambda r_{b}^2$ and $\sqrt{G} \mu$. Below $\sqrt{G} \mu<1$, it is qualitatively the same as the pure gravity case $\mu=0$, i.e. there always exist {\it bad} BG saddles and they make the system thermodynamically unstable. }
\label{13}
\end{center}
\end{figure}
\fi
                                                 %
Below $\sqrt{G} \mu<1$, it is qualitatively same as the pure gravity case $\mu=0$, i.e. there always exist {\it bad} BG saddles and they make the system thermodynamically unstable. On the other hand, something strange happens when $\sqrt{G}\mu \geq 1, 0<\Lambda r_{b}^2 <3$. Recall that there do not exist any BH/BG saddles in this range of $\mu$ when $\Lambda \leq0$. This time, however, BH/{\it good} BG saddles appear and 
{\it bad} BG saddles that exist when $\sqrt{G} \mu<1$ disappear.

Let's focus on the case of $1 \leq \sqrt{G}\mu , 0<\Lambda r_{b}^2 <3$. The behavior of the free energy versus temperature is almost the same as in the case of $\Lambda<0$ (Fig. \ref{9}), except for the behavior between the one with cusp (Fig. \ref{9} (Left)) and the one without cusp (Fig. \ref{9} (Right)). 
                                                 %
\iffigure
\begin{figure}[h]
\begin{center}
	\includegraphics[width=5.2cm]{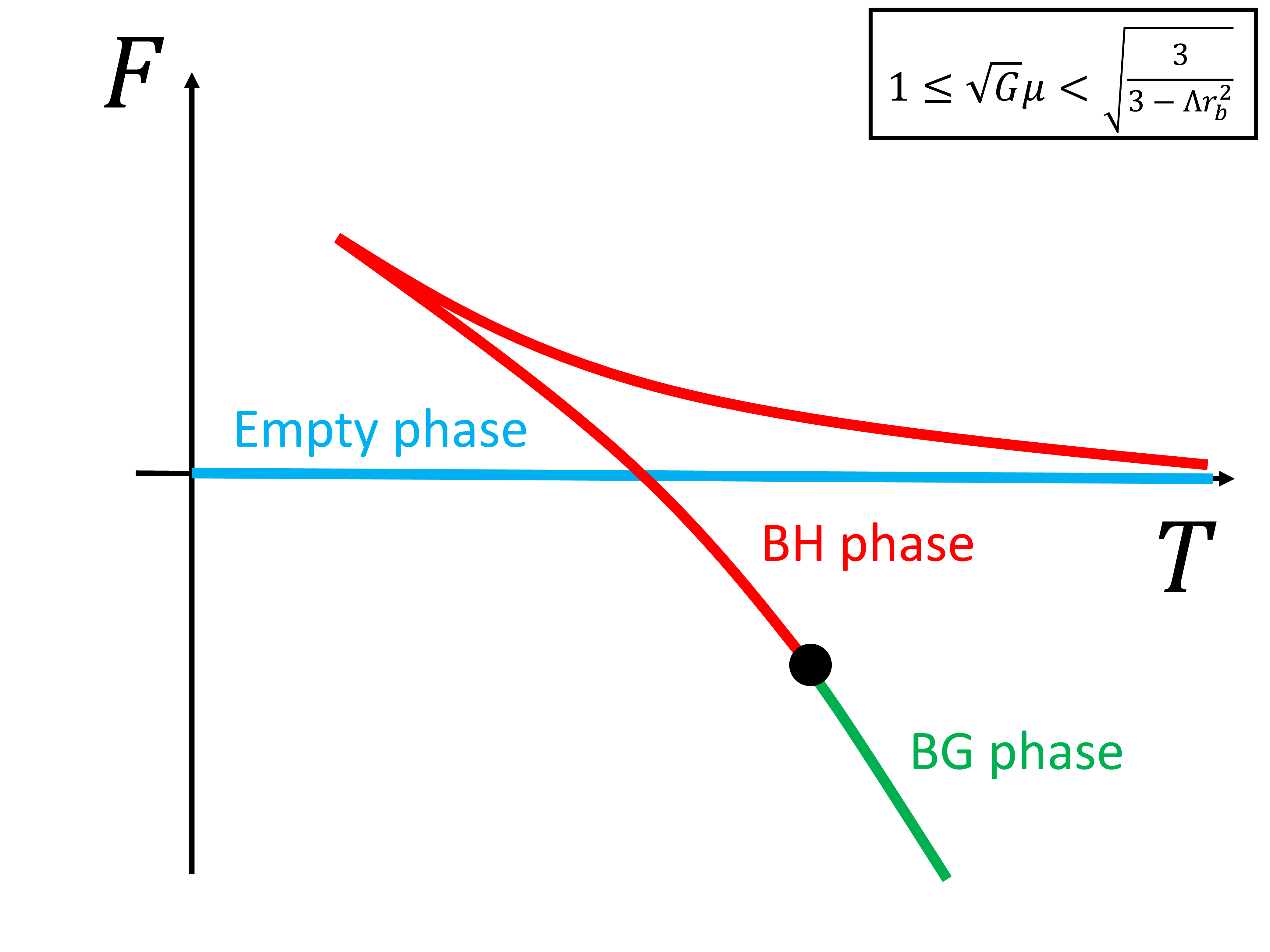} ~
	\includegraphics[width=5.2cm]{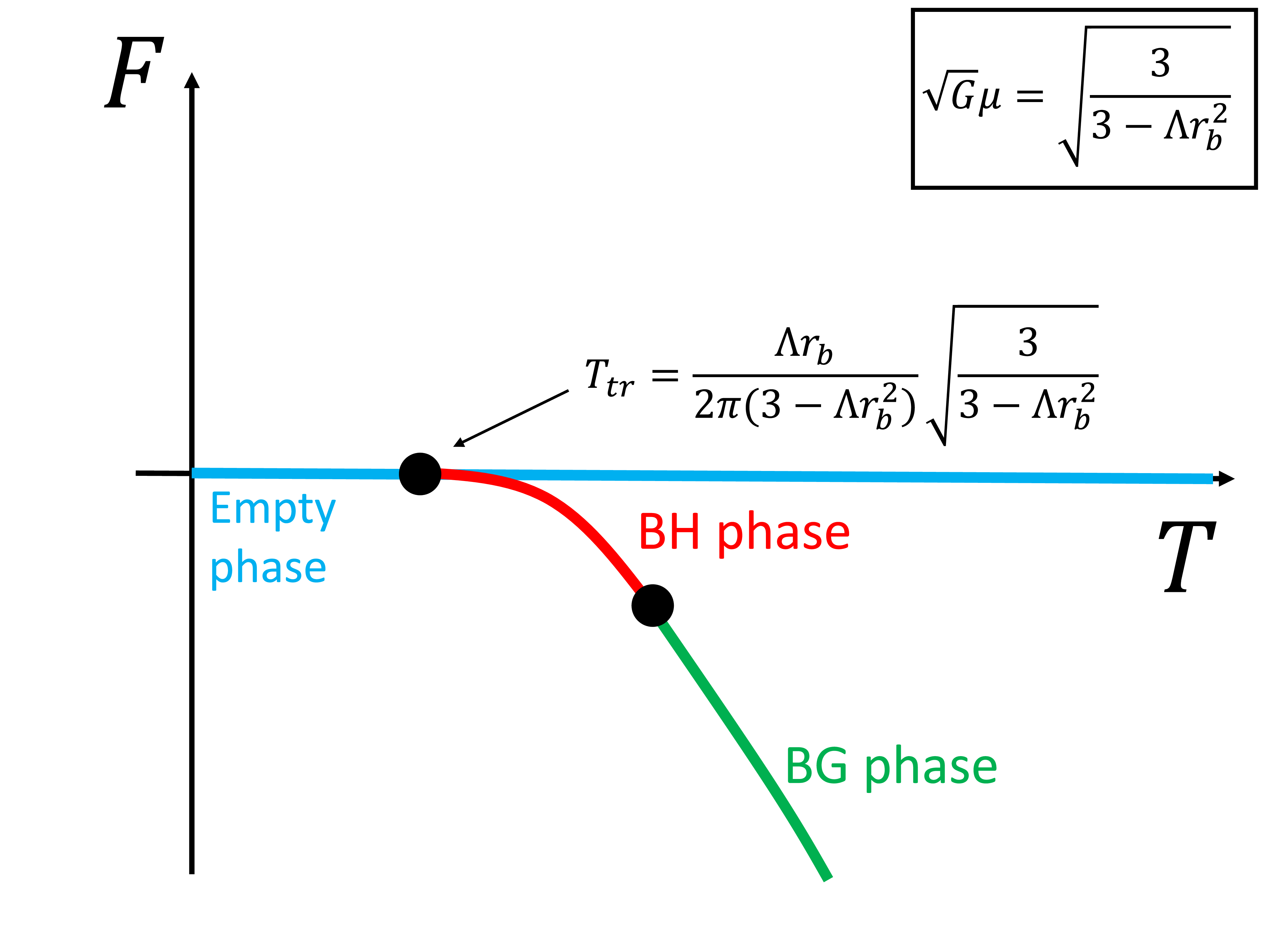} ~
	\includegraphics[width=5.2cm]{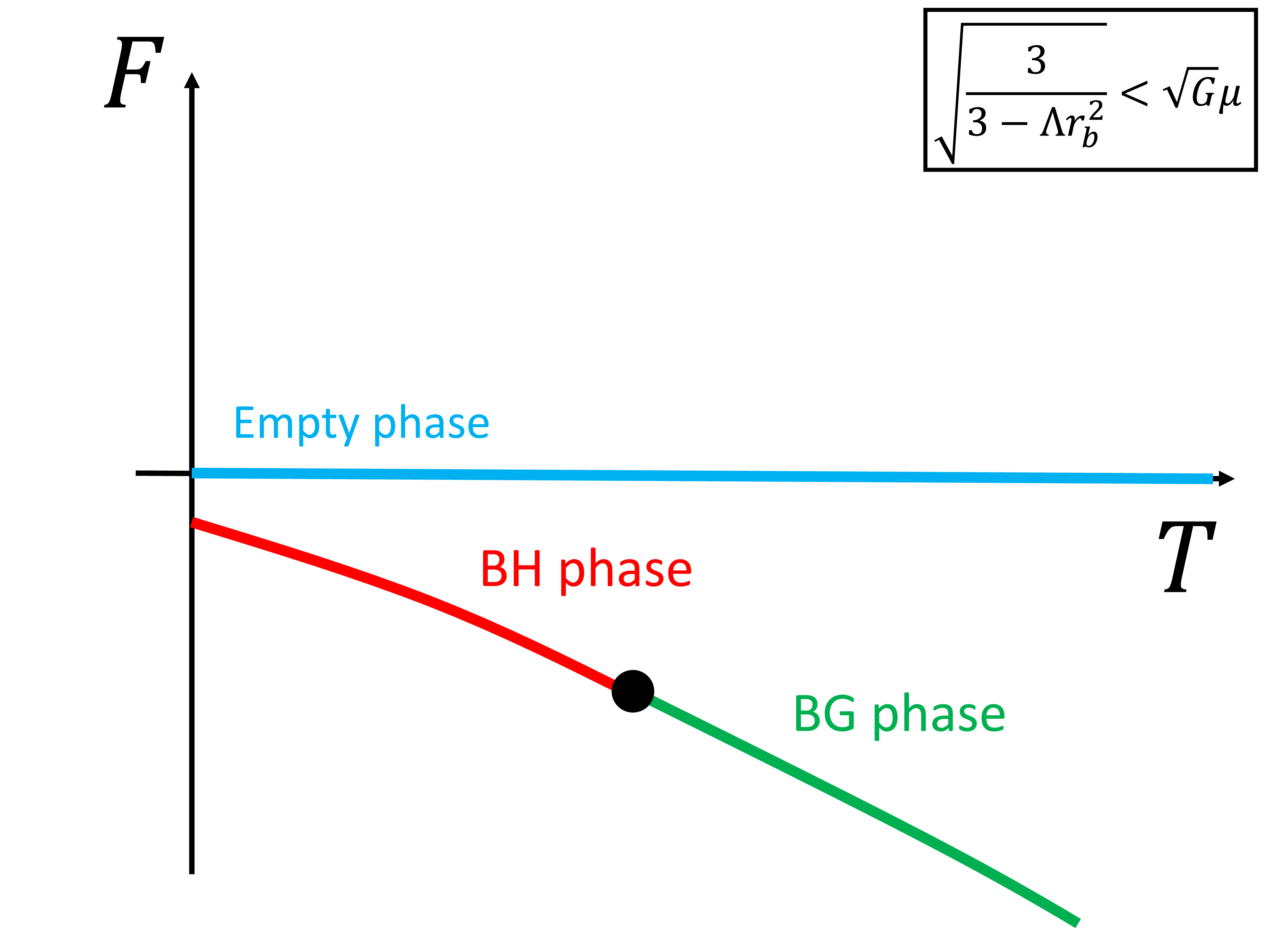}  
	\caption{Qualitative behaviors of the free energy when $0<\Lambda r_{b}^2 < 3 $. The position of the boundary between BH and BG depends on $\mu$ and $\Lambda$. Because of this dependence, the phase diagram may be classified as Fig. \ref{15}.}
\label{14}
\end{center}
\end{figure}
\fi
                                                 %
As shown in Fig. \ref{14}, the cusp structure appears when $1\leq \mu< \sqrt{\frac{3}{3-\Lambda r_{b}^2}}$ and does not appear when $ \sqrt{\frac{3}{3-\Lambda r_{b}^2}}< \mu $. When $\mu= \sqrt{\frac{3}{3-\Lambda r_{b}^2}}$, the BH/{\it good} BG branch is smoothly connected to the empty branch at $T=\frac{\Lambda r_{b}}{2\pi (3-\Lambda r_{b}^2)} \sqrt{\frac{3}{3-\Lambda r_{b}^2}} $. As far as I know, this kind of second order phase transition between empty phase and BH(/{\it good} BG) phase (in (grand) canonical ensembles) has not been observed before in gravitational thermodynamics.
\footnote{
For second order phase transitions between BH phases, for example, it was observed in \cite{LugoMorenoSchaposnik}, in the Einstein-Yang-Mills-Higgs system with AdS boundary condition.   
}
Additionally, there is a difference in the ``transition'' temperature $T_{end}$ between BH and {\it good} BG. Recall that it is given by (\ref{negativeend}). When we regard it as a function of $\mu$, the zero is given by $\sqrt{G} \mu_{zero} = \sqrt{\frac{1-\Lambda r_{b}^2}{1-2\Lambda r_{b}^2}}$. When $\Lambda < 0$, it is always real and less than $1$, and there always exist some critical value of the chemical potential above which the phase becomes a BG phase for all temperatures as in Fig. \ref{11}. However, when $\Lambda r_{b}^2 \geq \frac{1}{2}$, $\mu_{zero}$ is no longer real and the low temperature region is always BH phase. In addition, when $\Lambda r_{b}^2 \geq 1$, $T_{end}$ itself is no longer real. This means that there is no BG phase. Therefore, phase diagrams are classified as Fig. \ref{15}. 
                                                 %
\iffigure
\begin{figure}[h]
\begin{center}
	\includegraphics[width=5.3cm]{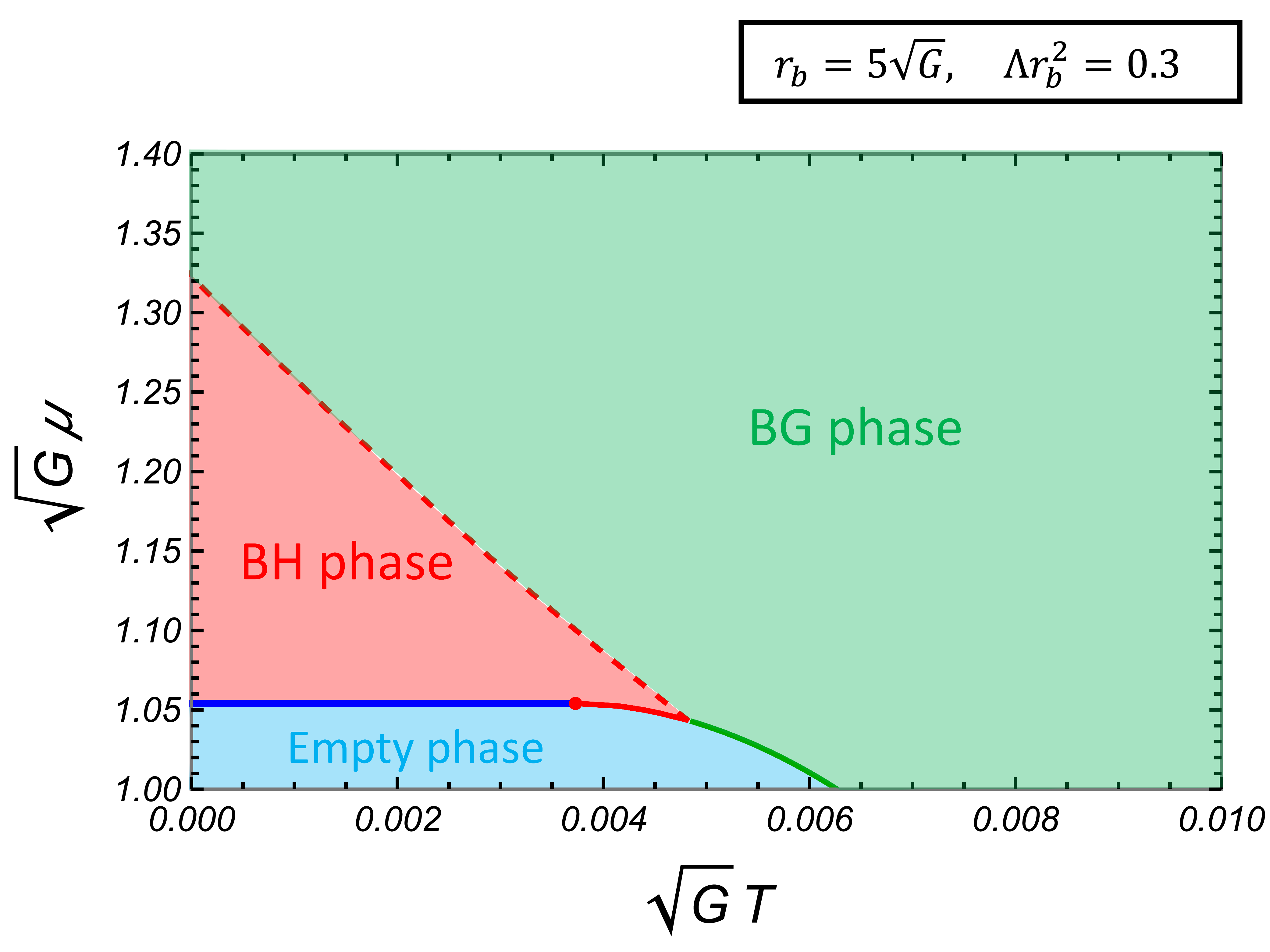}
	\includegraphics[width=5.3cm]{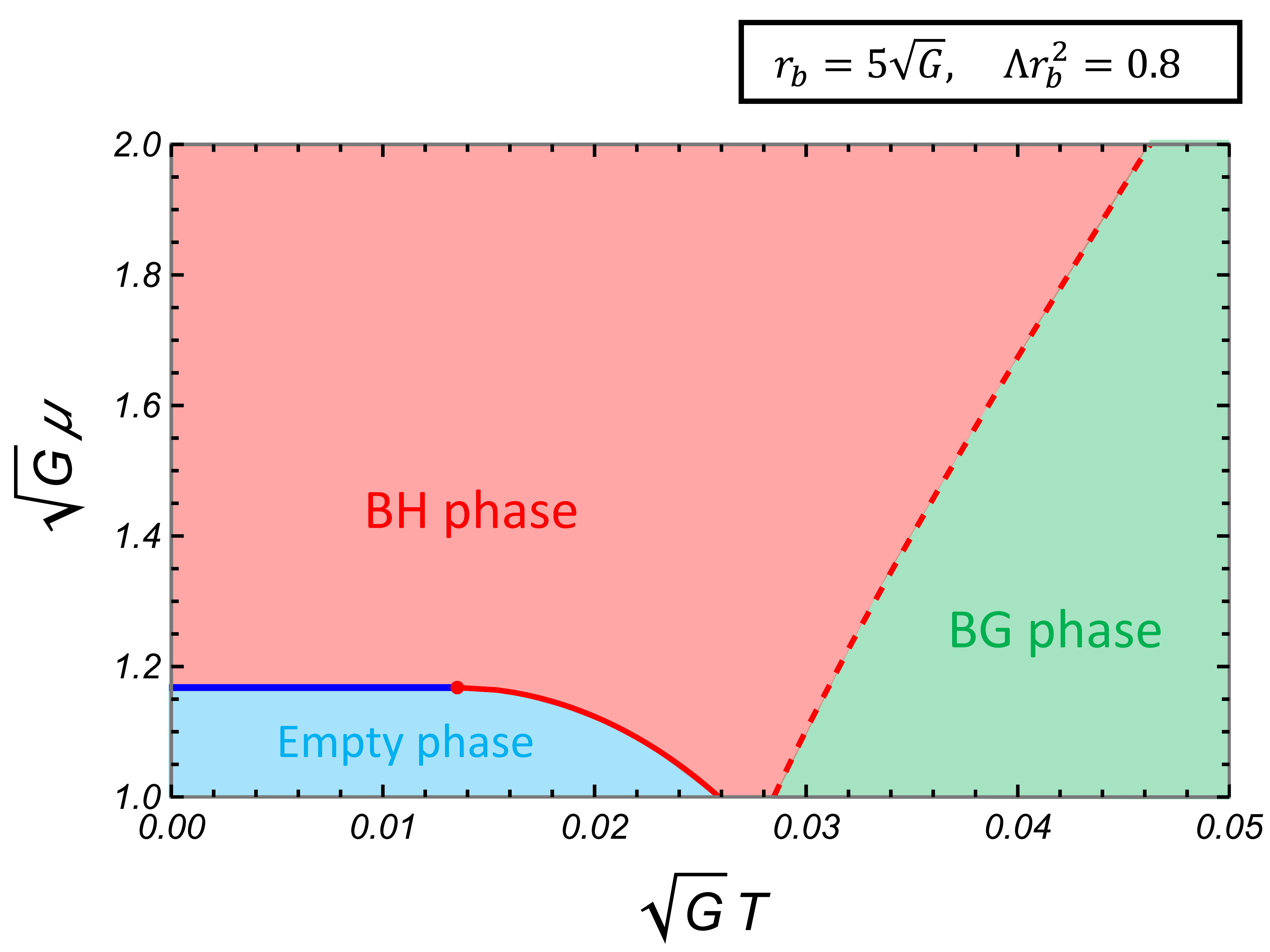}
	\includegraphics[width=5.3cm]{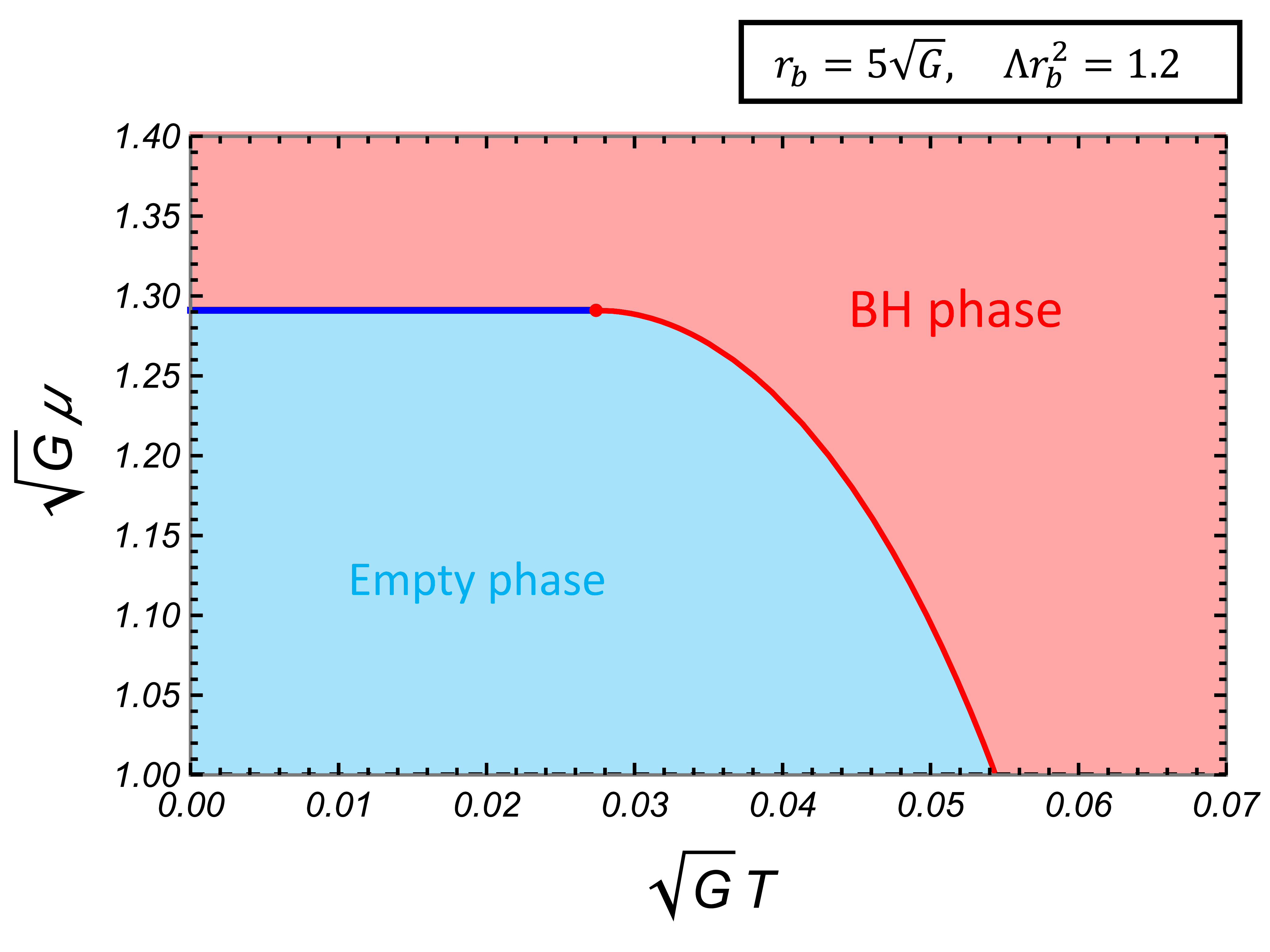} 
	\caption{Examples of the phase diagram above $\mu=1$. Note that below $\mu=1$, there exist {\it bad} BG saddles and the system is thermodynamically unstable if their contribution is included in the partition function. (Left) An example for the case $0<\Lambda r_{b}^2 < \frac{1}{2}$. In this case, the boundary curve between BH and BG intersects the $\mu$ axis at $\sqrt{G} \mu = \sqrt{G} \mu_{zero} = \sqrt{\frac{1-\Lambda r_{b}^2}{1-2\Lambda r_{b}^2}}$.  The lower end of the curve will always be on the boundary curve of the empty phase. (Middle) An example of the case $\frac{1}{2} \leq \Lambda r_{b}^2<1 $. In this case, the boudary curve does not intersect to the axis. Note that the lower end of the curve will be on the boundary curve of empty phase when $0< \Lambda r_{b}^2 \leq \frac{8}{11} \simeq 0.727$ and on the $\mu=1$ axis when $\frac{8}{11} \leq \Lambda r_{b}^2$. This right panel is a example of the latter. (Right) An example of the case $1 \leq \Lambda r_{b}^2<3 $. In this case, the BG phase does not exist.}
\label{15}
\end{center}
\end{figure}
\fi
                                                 %

In this paper, I assume that all Euclidean saddles contribute to the partition function, in particular both {\it good} BGs and {\it bad} BGs. Therefore, under this assumption, thermodynamics is well-defined only when $0<\Lambda r_{b}^2 < 3, \mu\geq1$ in the case of positive $\Lambda$. As I have pointed out several times, this assumption is not trivial and could be incorrect. In particular, if we make an alternative assumption that {\it all Euclidean saddles except bad BGs contribute to the partition function}, the thermodynamics is well-defined even for $0<\Lambda r_{b}^2 < 3, \mu<1$ since there exist BH and {\it good} BG saddles. Thermodynamics under the alternative assumption is discussed in Appendix B.

%

\section{Summary}
The main objective of this paper was to answer the following two questions;
\begin{itemize}
\item Recently, in \cite{Miyashita, DraperFarkas, BanihashemiJacobson}, it was found that there exists a new type of saddle point geometries and that they cause the thermodynamical instability of pure gravity with a positive $\Lambda$. In \cite{Miyashita} and here, I call this type of geometry "bag of gold(BG)". BG is similar to BH, but one difference is that the area of the bolt is larger than that of the boundary. So one question is:

Are there any BGs that do not lead to thermodynamic instability?
\item In \cite{BasuKrishnanSubramanian}, it has been reported that the free energy in the Einstein-Maxwell system with box boundary condition shows some peculiar behavior. The BH branch terminates at some finite temperature, and the free energies of the empty saddle and the BH saddle do not show the Hawking-Page phase structure even for sufficiently small chemical potential $\mu$. However, for other cases, such as pure gravity with box boundary condition($\Lambda \leq 0$), that with AdS boundary condition, and the Einstein-Maxwell system with AdS boundary condition (with sufficiently small $\mu$), they all show the Hawking-Page phase structure. 

Can we somehow resolve this discrepancy?  
\end{itemize}
In this paper, I gave a positive answer to these questions at the same time by showing the existence of BGs in the Einstein-Maxwell system and investigating these properties. When $\Lambda\leq 0$, BGs in the Einstein-Maxwell system are {\it good}, in the sense that they are thermodynamically stable and dominant in the path integral if they are included. (In contrast to the {\it bad} BGs, which cause thermodynamical instability of the system in the case of pure gravity with a positive $\Lambda$.) I also showed that the BG branch smoothly connects to the BH branch and, as a result, the system exhibits the Hawking-Page phase structure at small $\mu$.

Another objective is to investigate the thermodynamical properties of the Einstein-Maxwell system under the assumption that {\it all Euclidean saddles contribute to the partition function}. I have studied  the grand canonical ensembles in the main text and the canonical and microcanonical ensembles in Appendix A. Some of the properties related to the existence of BGs are summarized in Table \ref{T1} for grand canonical ensembles and Table \ref{T2} for (micro)canonical ensembles.
\begin{table}
\begin{center}
\begin{tabular}{| l  l || l | l | l |}
\hline 
 & & stability & BG & entropy bound  \\
\hline \hline
$ \Lambda \leq 0 $ & ($\sqrt{G}\mu<1$) & stable & $ {\it good}  $ & $\frac{\pi r_{b}^2}{G}\frac{1}{(1-G\mu^2)^2} $\\
 & ($\sqrt{G}\mu>1$)  & N/A & N/A & N/A \\
\hline 
$ 0\leq \Lambda r_{b}^2 \leq 1  $ &  ($\sqrt{G}\mu<1$) & unstable & {\it good} \& {\it bad}  & $\infty$  \\ 
 & ($\sqrt{G}\mu>1$) & stable & {\it good} & $\frac{\pi r_{b}^2}{G}\frac{1}{(1-G\mu^2)^2} $  \\
\hline
$ 1\leq \Lambda r_{b}^2 \leq 3  $ & ($\sqrt{G}\mu<1$) & unstable & {\it good} \& {\it bad}  & $\infty$  \\
 & ($\sqrt{G}\mu>1$) & stable & {\it BH} & $\frac{\pi r_{b}^2}{G} $  \\
\hline  
$ 3<\Lambda r_{b}^2 $ & ($\sqrt{G}\mu<1$) & unstable & {\it bad} & $\infty$  \\
 & ($\sqrt{G}\mu>1$) & N/A & N/A & N/A  \\
\hline 
\end{tabular}
\caption{Some thermodynamical properties of the Einstein-Maxwell system in grand canonical ensembles. In the column of BG, {\it good} means that both {\it good} BGs and BHs exist, {\it BH} means that only BHs exist. N/A means that there are no Euclidean saddles in this parameter region. }
\label{T1}
\end{center}
\end{table}
\begin{table}
\begin{center}
\begin{tabular}{| l  l || l | l | l |}
\hline 
 & & stability & BG & entropy bound  \\
\hline \hline
$ \Lambda \leq 0 $ & ($\sqrt{G}Q< r_{b} \sqrt{1-\Lambda r_{b}^2}$) & stable & $ {\it good}  $ & $\frac{\pi r_{b}^2}{G}\frac{1}{(1-G\mu^2)^2} $\\
 & ($\sqrt{G}Q> r_{b} \sqrt{1-\Lambda r_{b}^2}$)  & unstable & {\it bad} & $\frac{\pi (\sqrt{1-4\Lambda G Q^2}-1)}{-2G\Lambda}$ \\
\hline 
$ 0\leq \Lambda r_{b}^2 \leq \frac{1}{2}  $ &  ($\sqrt{G}Q< r_{b} \sqrt{1-\Lambda r_{b}^2}$) & unstable & {\it BH} \& {\it bad}  & $\infty$  \\ 
 &  ($\sqrt{G}Q< r_{b} \sqrt{1-\Lambda r_{b}^2}$)& unstable & {\it bad} & $\infty$   \\
\hline
$ \frac{1}{2} \leq \Lambda r_{b}^2 \leq 3  $ &  ($Q<Q_{cr}(\Lambda, r_{b})$) & unstable & {\it BH} \& {\it bad}  & $\infty$  \\ 
 & ($Q>Q_{cr}(\Lambda, r_{b})$) & unstable & {\it bad} & $\infty$   \\
\hline
$ 3<\Lambda r_{b}^2 $ &  & unstable & {\it bad} & $\infty$ \\
\hline 
\end{tabular}
\caption{Some thermodynamical properties of the Einstein-Maxwell system in canonical ensembles (and microcanonical ensembles). In the column of BG, {\it good} means that both {\it good} BGs and BHs exist, {\it BH} means that only BHs exist, as in Table 1. Stability means thermodynamical stability in canonical ensembles. $Q_{cr}(\Lambda, r_{b})$ is given by (\ref{eqQcr}) in Appendix A. }
\label{T2}
\end{center}
\end{table}
In general, a positive $\Lambda$ causes the existence of {\it bad} BGs and the appearance of {\it bad} BGs causes the entropy bound to be $\infty$, except for the two cases in the tables. One is the case of $0<\Lambda r_{b}^2 < 3, \sqrt{G}\mu >1$ in grand canonical ensembles. In this case, although $\Lambda$ is positive, there are no BGs and the system is thermodynamically stable. The other is the case of $\Lambda \leq 0, \sqrt{G}Q < r_{b} \sqrt{1-\Lambda r_{b}^2}$ in (micro)canonical ensembles. In this case, although $\Lambda$ is not positive, there exist {\it bad} BGs. However, the entropy bound is still finite and, remarkably,  does not depend on the boundary size $r_{b}$. At the moment, I have no idea about the implications of these cases.

Until recently, the importance (or sometime existence) of BG saddles in gravitational thermodynamics has been overlooked in the literature, perhaps because of their peculiarity.
\footnote{
I suspect there are, at least, two reasons. One is that the horizon is larger than the boundary sphere, so one might feel that it is not the inside but the outside. The other is that if we extend the geometry beyond the boundary, some BGs have a singularity. Of course, these are not real problems.  
}
Also, in the recent study \cite{Miyashita, DraperFarkas, BanihashemiJacobson}, the role of BGs is examined for pure gravity case but it turned out that they have a bad property, i.e. they cause thermodynamical instability. So the importance of BG saddles was unclear and one might argue that they should always be excluded from the path integral since they all seem to have {\it bad} properties and we do not know the true integration contour of the Euclidean gravitational path integral \cite{GibbonsHawkingPerry}. In this paper, I show the existence of {\it good} BGs and their necessity for well-defined thermodynamics in the Einstein-Maxwell system. Presumably, from the results in this paper, there is little doubt that some BGs play important roles in gravitational thermodynamics. However, it is still unclear whether all BGs must be included in the path integral or not.

\section*{Acknowledgement}
This work is supported in part by the National Science and
Technology Council (No. 111-2112-M-259-016-MY3).

\appendix
\section{Canonical Ensembles and Microcanonical Ensembles}
In this Appendix, I consider canonical ensembles and microcanonical ensembles. For each type of ensemble, I consider the $\Lambda \leq 0$ case, and the $\Lambda>0$ case in turn. 

In these types of ensembles, we fix $Q$ instead of $\mu$. In the case of grand canonical ensembles which was discussed in the main text, for a given $\mu$, there could exist both BH and BG phases as shown in Fig. \ref{7}. However, we will see that, in canonical and microcanonical ensembles, there is a critical value of $Q$ below which only BH phase exists and above which only BG phase exists. 
\subsection{canonical ensembles}
\subsubsection{$\Lambda \leq 0$}
In order to check the existence of solutions for a given $R$ (radius of bolt) and $Q$, the positivity of $(\theta(r_{b}-R)-\theta(R-r_{b}))f'(R)$ may be useful. For the case of (micro)canonical ensemble, it is given by
\bea
f'(R)= -\frac{GQ^2}{R^3} + \frac{1}{R} -\Lambda R 
\ena 

It is easy to check that when $\Lambda=0$, $f'(R) = \frac{1}{R^3}(R^2 - G Q^2)$ and the zeros are $R= \pm \sqrt{G} Q$. Therefore when $\Lambda=0$,
\begin{itemize}
\item $0 \leq \sqrt{G} Q < r_{b}$: ~~ $R\in \left[ \sqrt{G} Q, r_{b} \right)$
\item $r_{b} < \sqrt{G} Q $: ~~ $R\in \left( r_{b}, \sqrt{G} Q \right)$
\end{itemize}
Only BH saddles exist in the first case and only BG saddles in the second case. 

A similar result holds for $\Lambda \leq 0$. First, $f'(R)$ can be rewritten as
\bea
f'(R)=\frac{1}{R^3}(-\Lambda R^4 + R^2 - GQ^2)
\ena
The equation $g(R)\equiv R^3 f'(R)=(-\Lambda R^4 + R^2 - GQ^2)=0$ has one positive root $R=\sqrt{ \frac{\sqrt{1-4 \Lambda G Q^2} -1  }{-2\Lambda}  }$. This root is less than $r_{b}$ when $\DS 0 \leq  \sqrt{G} Q < r_{b} \sqrt{1-\Lambda r_{b}^2}$ and greater when  $\DS  r_{b} \sqrt{1-\Lambda r_{b}^2}< \sqrt{G}Q $.
Then, when $\Lambda \leq 0$,
\begin{itemize}
\item $\DS 0 \leq  \sqrt{G} Q < rb \sqrt{1-\Lambda r_{b}^2}$: ~~ $R\in \left[ \sqrt{ \frac{\sqrt{1-4 \Lambda G Q^2} -1  }{-2\Lambda}  }, r_{b} \right)$
\item $\DS  rb \sqrt{1-\Lambda r_{b}^2}< \sqrt{G}Q $: ~~ $R\in \left( r_{b}, \sqrt{ \frac{\sqrt{1-4 \Lambda G Q^2} -1  }{-2\Lambda}  } \right)$
\end{itemize}
Again, only BH saddles exist in the first case and only BG saddles in the second case. When $\sqrt{G}Q=r_{b} \sqrt{1-\Lambda r_{b}^2}$, there are no BHs and BGs, but BRs.

The behavior of the free energy $F_{c}$ versus temperature depends on $Q$ for a fixed $\Lambda, r_{b}$. I show the qualitative behaviors of the free energies in Fig. \ref{A1}. 
                                                 %
\iffigure
\begin{figure}[h]
\begin{center}
	\includegraphics[width=6.5cm]{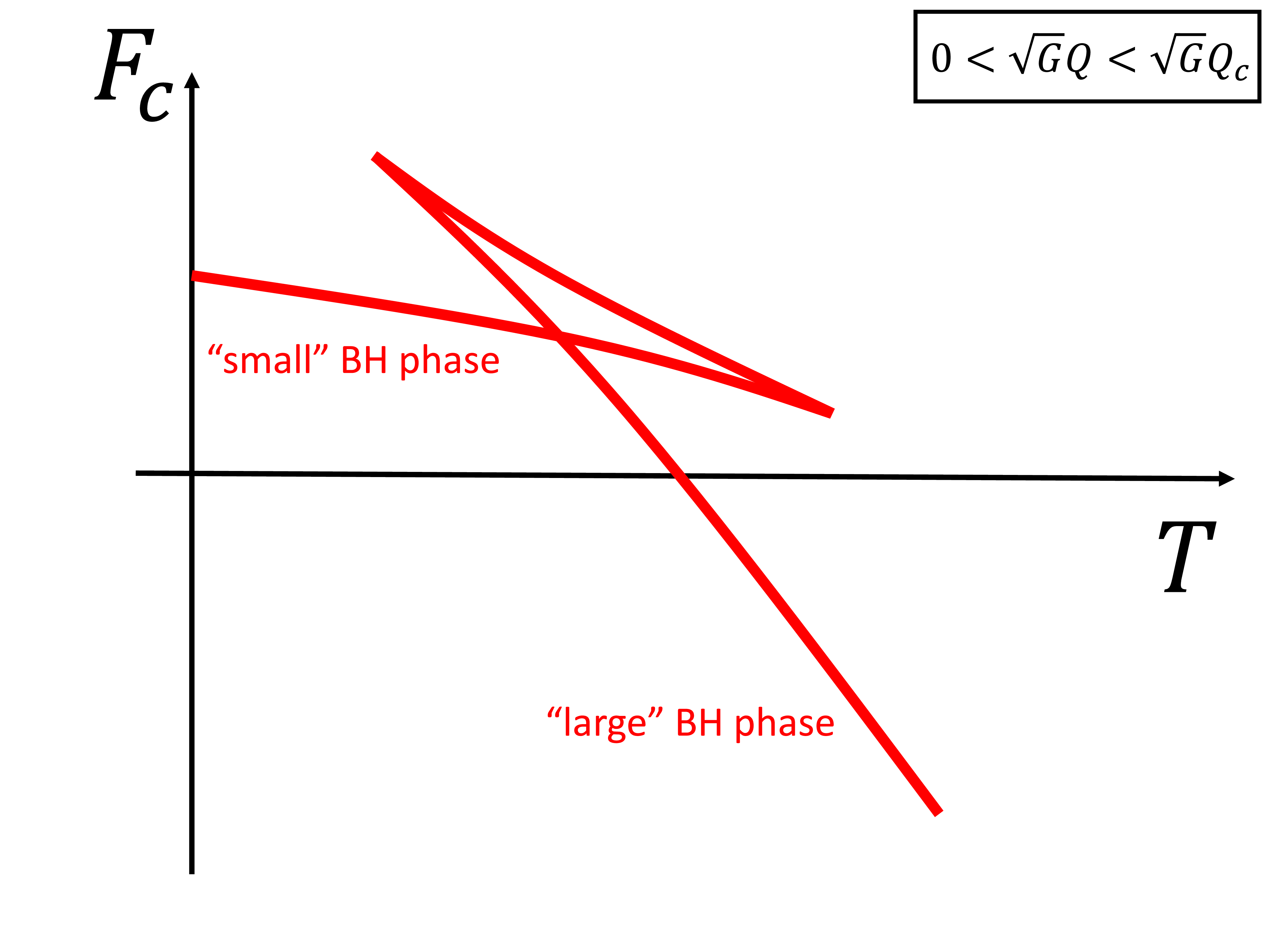} ~
	\includegraphics[width=6.5cm]{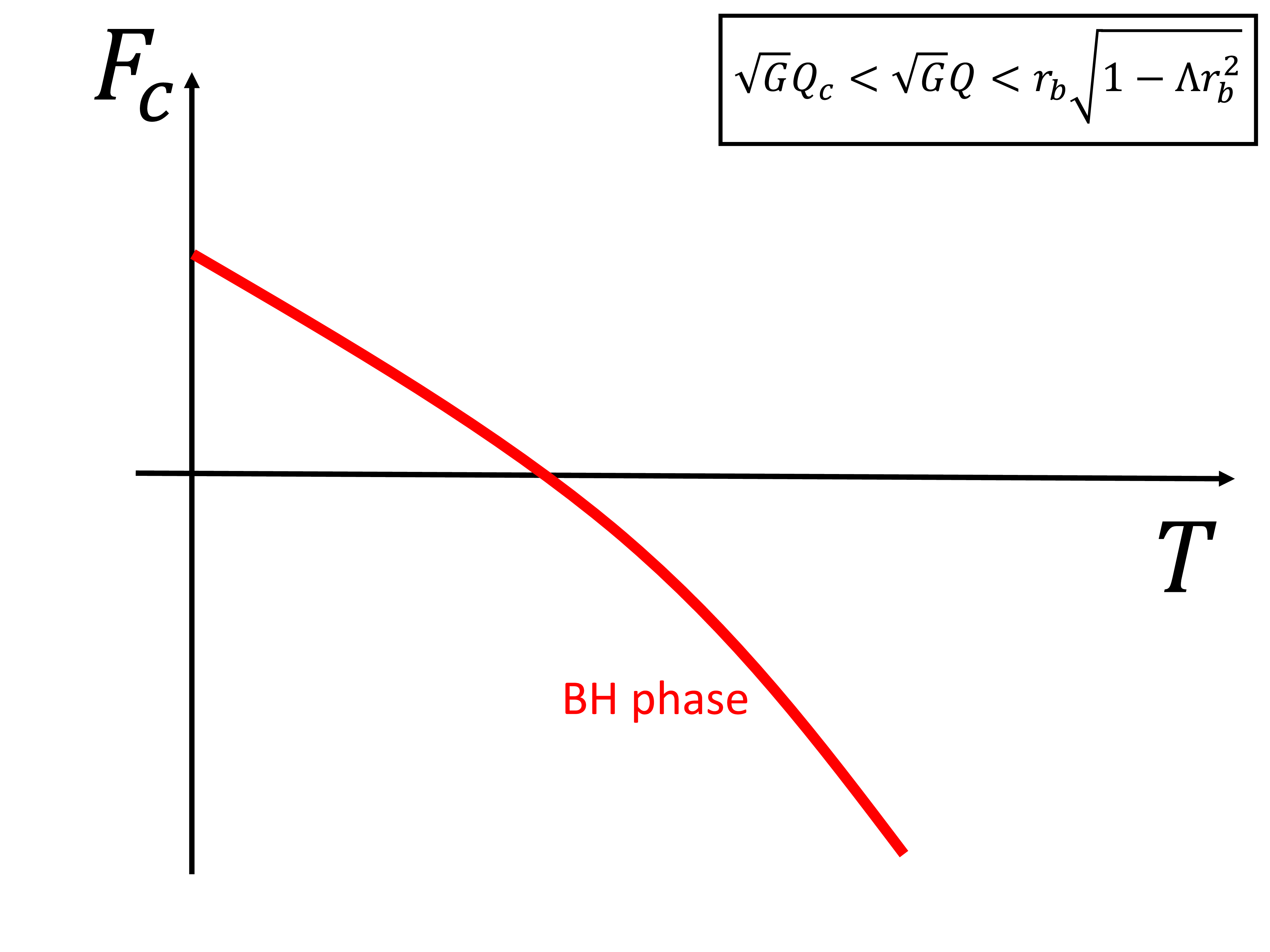} ~\\
	\includegraphics[width=6.5cm]{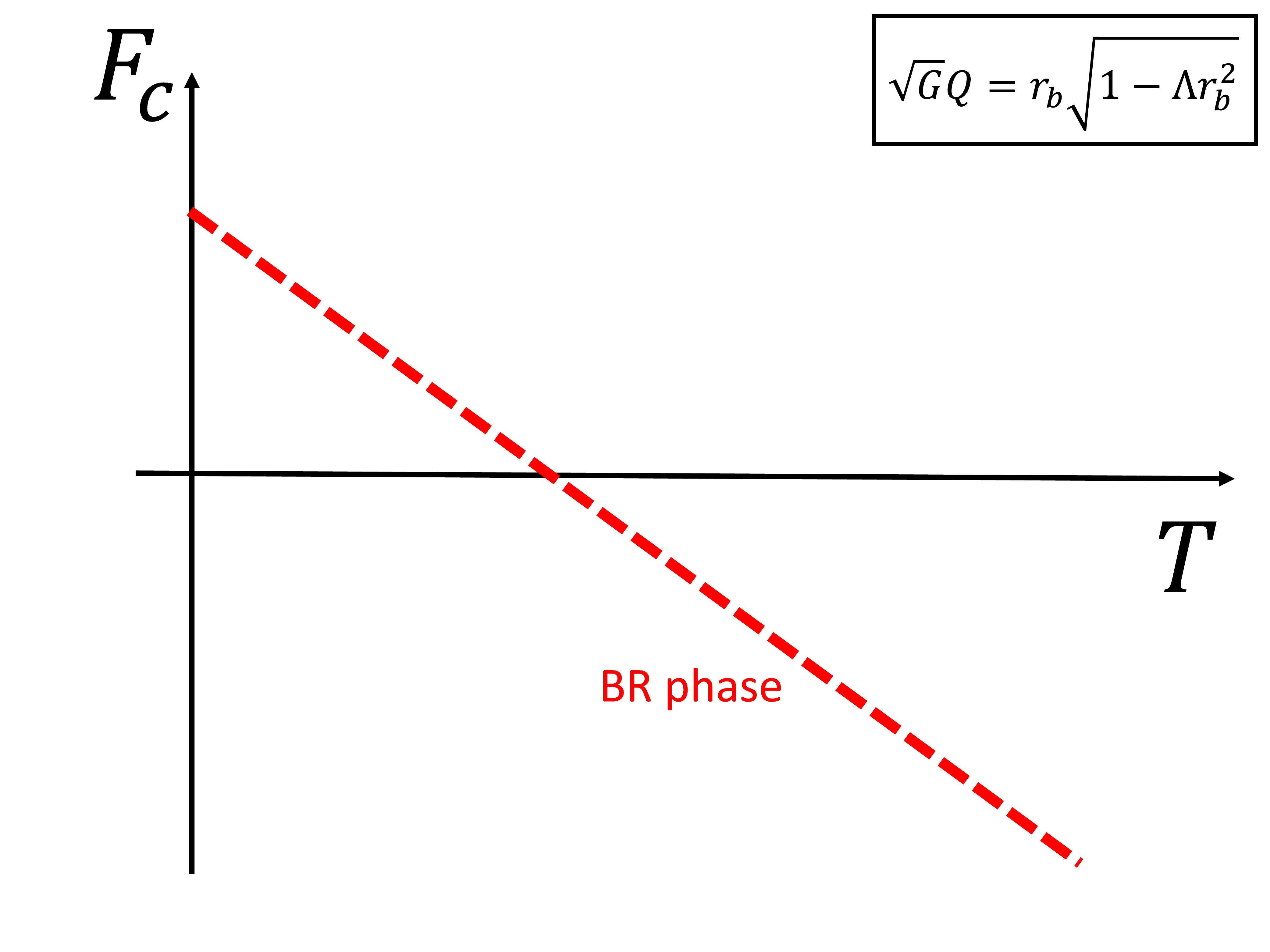} ~
	\includegraphics[width=6.5cm]{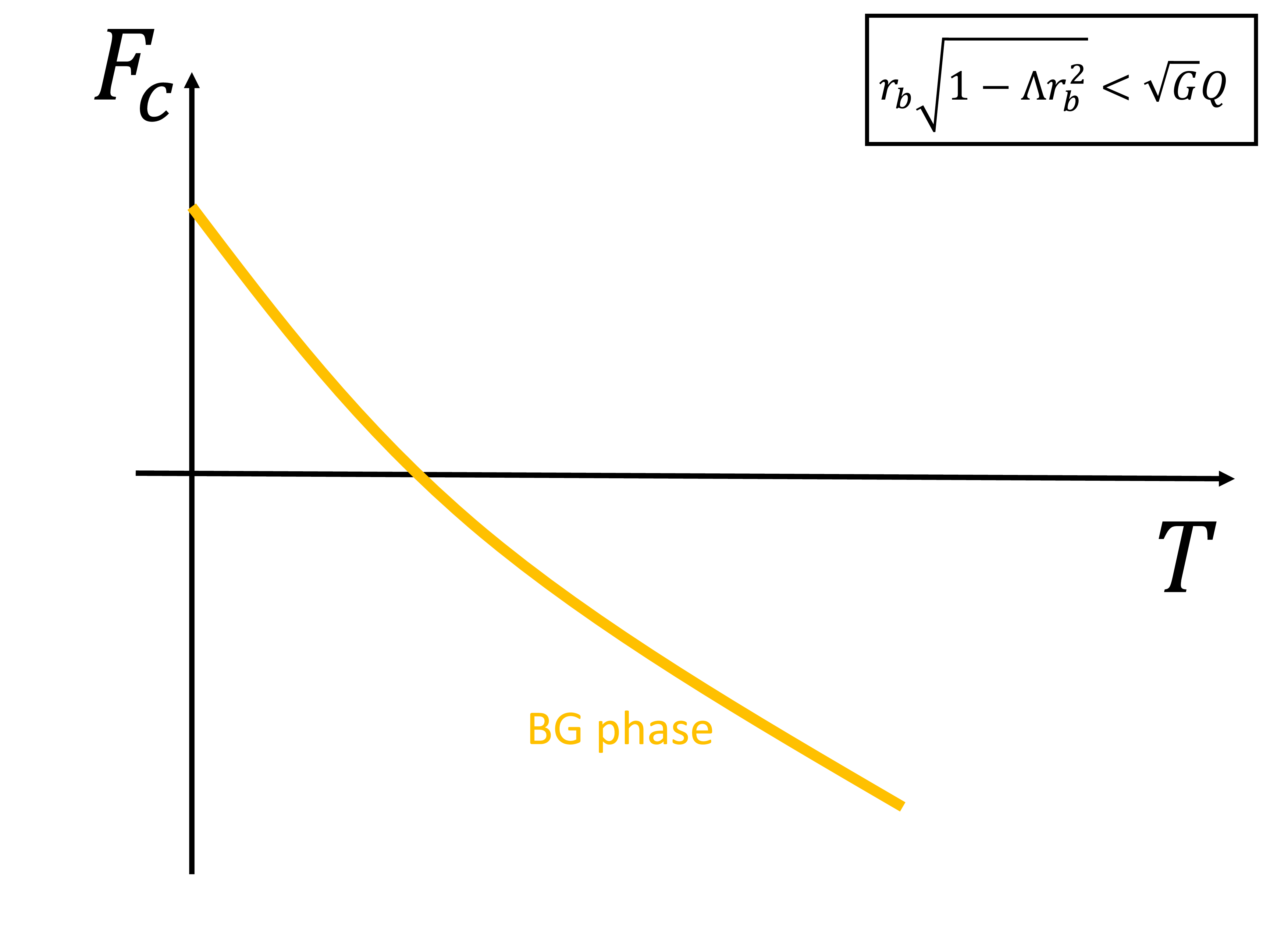} 
	
	\caption{Qualitative behaviors of free energy. }
\label{A1}
\end{center}
\end{figure}
\fi
                                                 %
For $0< \sqrt{G}Q < r_{b}\sqrt{1-\Lambda r_{b}^2}$, they are qualitatively the same as the AdS boundary case \cite{ChamblinEmparanJohnsonMyers}; below some critical value $Q_{c}$, there are two phases, ``small'' BH phase and ``large'' BH phase. (The parameter dependence of $Q_{c}$ and that of the corresponding temperature $T_{c}$ are shown in Fig. \ref{A2}.) 
                                                 %
\iffigure
\begin{figure}[h]
\begin{center}
	\includegraphics[width=7cm]{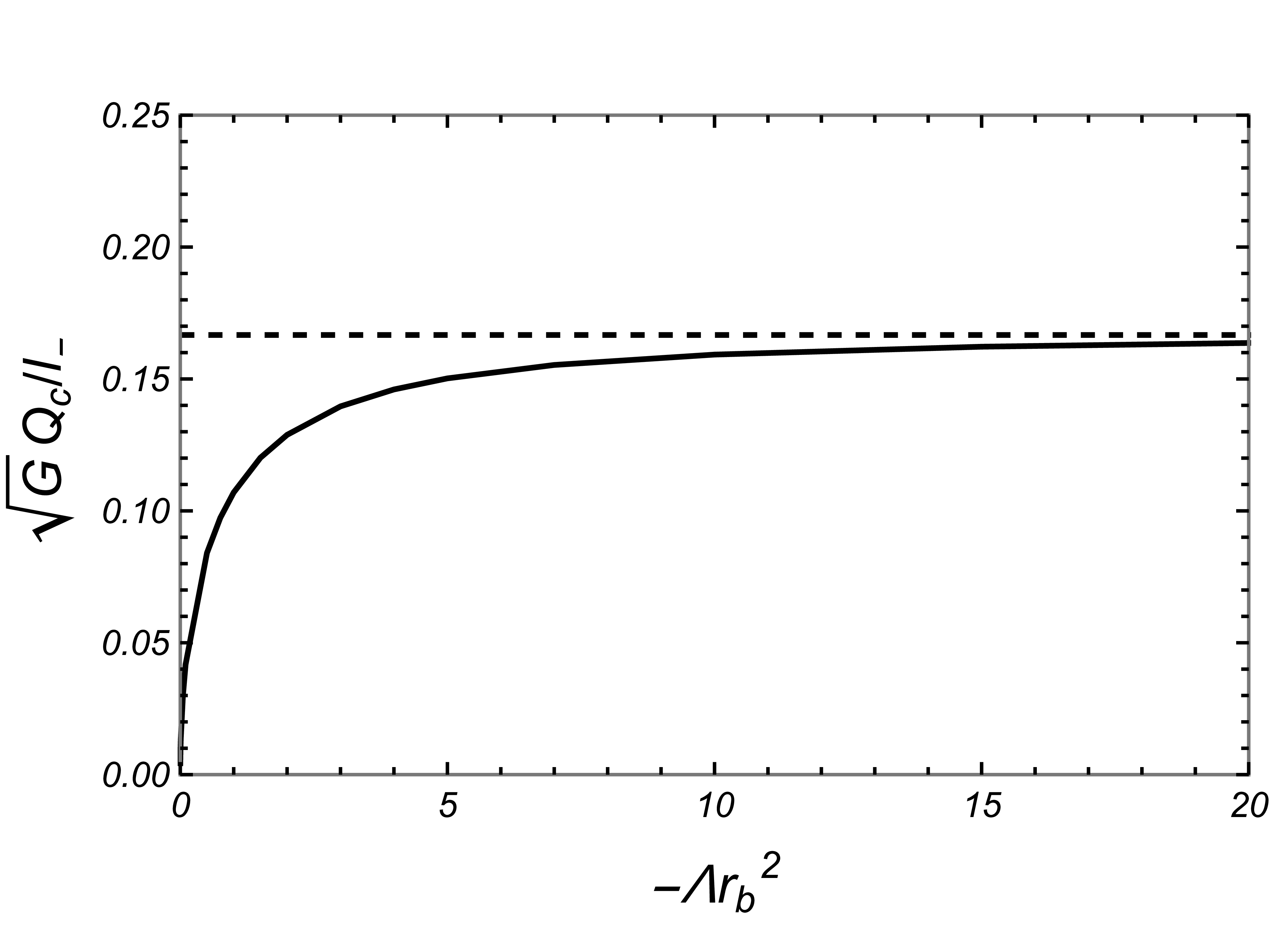} ~~~
	\includegraphics[width=7cm]{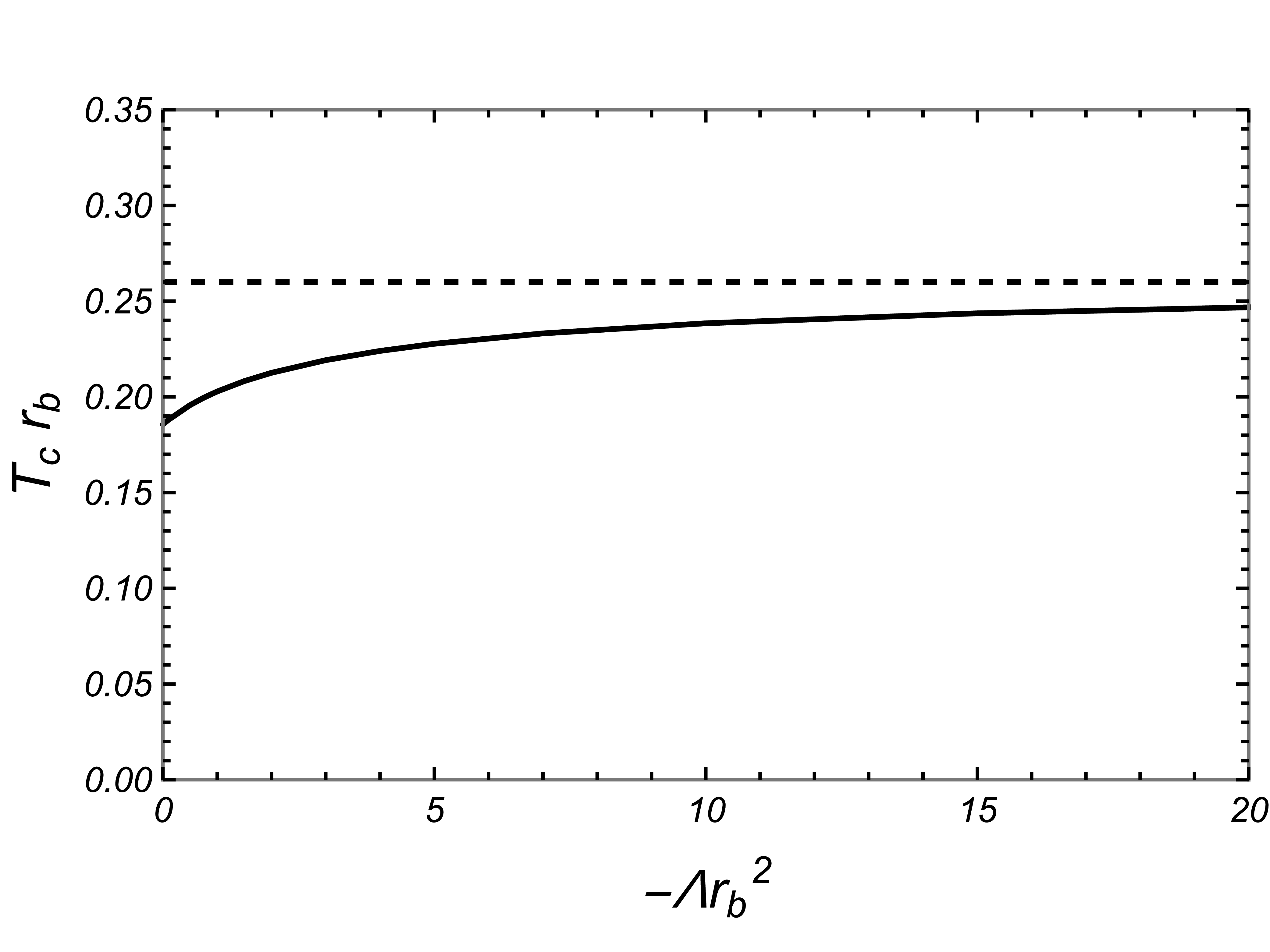} ~
	
	\caption{The parameter dependence of $Q_{c}$ and the corresponding temperature $T_{c}$. The combinations $\frac{\sqrt{G}Q_{c}}{l_{-}}$ and $T_{c} r_{b}$ depend only on $-\Lambda r_{b}^2$. (Left) $\frac{\sqrt{G}Q_{c}}{l_{-}}$. The dashed line represents that of AdS boundary condition $\frac{\sqrt{G}Q_{c,AdS}}{l_{-}}= \frac{1}{6} \simeq 0.1667$ \cite{ChamblinEmparanJohnsonMyers}.  (Right) $T_{c} r_{b}$. The dashed line represents $T_{c, AdS} l_{-} = \sqrt{\frac{2}{3}}\frac{1}{\pi} \simeq 0.2599 $ \cite{ChamblinEmparanJohnsonMyers}. (Recall that in the limit $r_{b}/l_{-} \to \infty$, the temperature of the box boundary condition and that of the AdS boundary condition are related by $T_{AdS} = \left( \frac{r_{b}}{l_{-}} \right) T$.)   }
\label{A2}
\end{center}
\end{figure}
\fi
                                                 %
For $\sqrt{G}Q_{c}<\sqrt{G}Q< r_{b}\sqrt{1-\Lambda r_{b}^2}$, there is only one phase. When $ \sqrt{G}Q = r_{b}\sqrt{1-\Lambda r_{b}^2}$, there are no BHs and BGs, but BRs. Since the energy and the entropy of BR spacetime do not depend on the temperature, the free energy $F_{c}=E-TS$ of BR spacetime becomes a linear function as shown in Fig. \ref{A1}. (See Appendix C.) Above $r_{b}\sqrt{1-\Lambda r_{b}^2}$, the phase becomes the BG phase. This BG is ${\it bad}$ in the sense that the second derivative of $F_{c}$ with respect to temperature is positive. Therefore, when $r_{b}\sqrt{1-\Lambda r_{b}^2} < \sqrt{G}Q $, the system would be thermodynamically unstable. The phase diagram is shown in Fig. \ref{A3}. 
                                                 %
\iffigure
\begin{figure}[h]
\begin{center}
	\includegraphics[width=8cm]{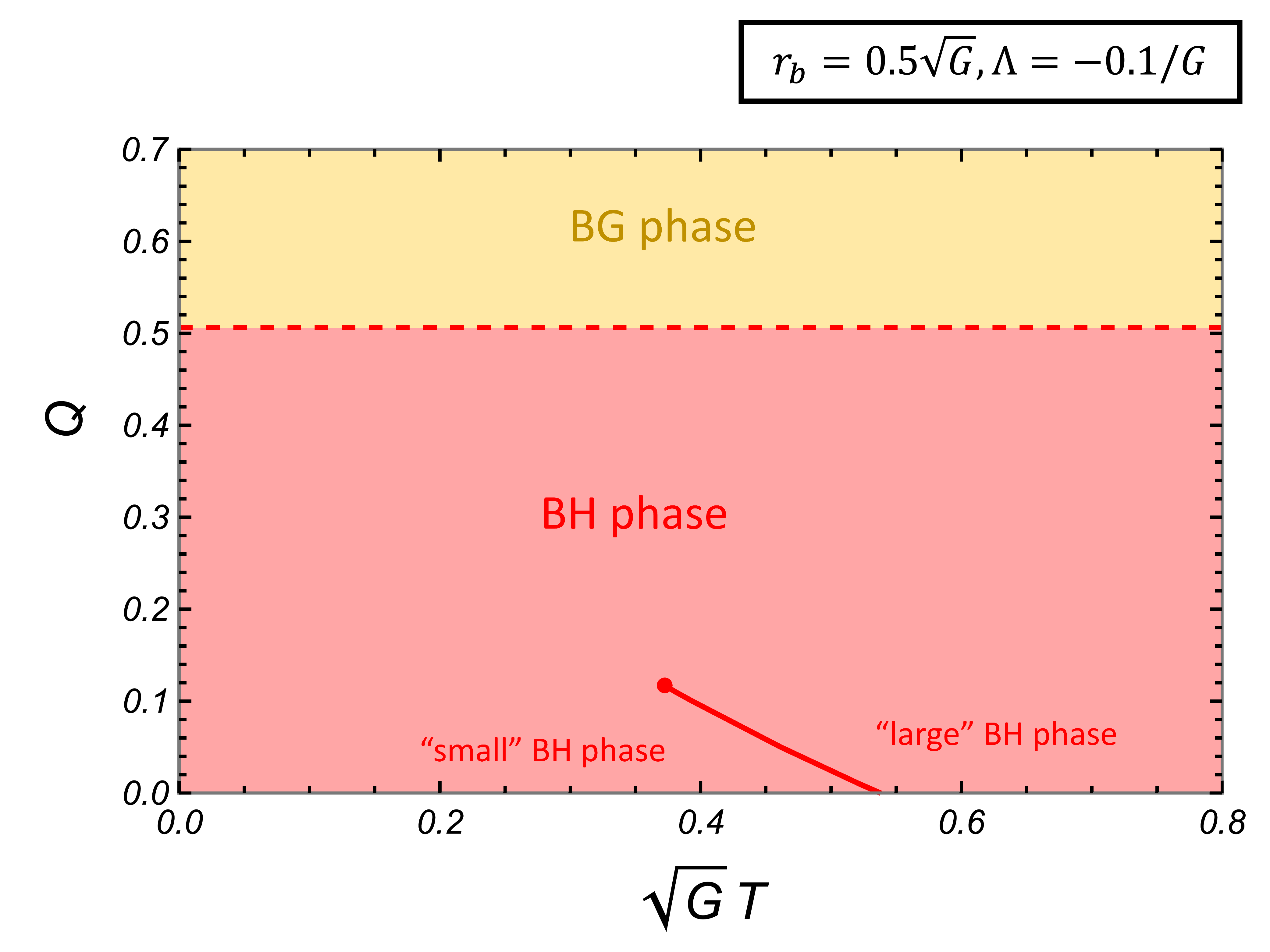} ~

	\caption{An example of phase diagram of canonical ensembles when $\Lambda \leq 0$ ($r_{b}=0.5\sqrt{G}, \Lambda = -0.1/G$). For sufficiently low $Q$, there are two phases, "small" BH and "large" BH. The red dot represents the critical point $(\sqrt{G}T_{c}, Q_{c})$ whose parameter dependence is shown in Fig. \ref{A2}. The dashed line $\sqrt{G}Q=r_{b}\sqrt{1-\Lambda r_{b}^2}$ represents the BR phase. Above the line, the phase becomes the BG "phase," which is thermodynamically unstable.  }
\label{A3}
\end{center}
\end{figure}
\fi
                                                 %
\subsubsection{$\Lambda > 0$}
In the case of pure gravity or grand canonical ensemble in the Einstein-Maxwell system, the existence of BH/BG saddles depends on the parameters $\Lambda r_{b}^2, \sqrt{G}\mu$ and can be judged by the positivity of $f(r_{b})$. Similarly, the existence of BH/BG saddles in the case of the (micro)canonical ensemble in the Einstein-Maxwell system  depends on the parameters $\Lambda r_{b}^2, \sqrt{G\Lambda}Q $ and may be judged by the positivity of
\footnote{
However, as we will see shortly, this is not a sufficient condition.
}
\bea
f(r_{b})= \frac{r_{b}-R}{3r_{b}^2 R} \left( -3GQ^2 + r_{b}R \left( 3- \Lambda r_{b}^2 - \Lambda r_{b}R - \Lambda R^2 \right) \right)  .
\ena 
Therefore, as in the grand canonical case, the necessary conditions for the existence of solutions are\\
~\\
 \hspace{1cm} $\left( -3GQ^2 + r_{b}R \left( 3- \Lambda r_{b}^2 - \Lambda r_{b}R - \Lambda R^2 \right) \right) :$ positive \& $(r_{b} - R)$: positive \\
\hspace{7.5cm} or \\
  \hspace{1cm}  $\left( -3GQ^2 + r_{b}R \left( 3- \Lambda r_{b}^2 - \Lambda r_{b}R - \Lambda R^2 \right) \right) :$ negative \& $(r_{b} - R )$: negative\\
 ~\\
Let's define the function $h(R) \equiv -3GQ^2 + r_{b}R \left( 3- \Lambda r_{b}^2 - \Lambda r_{b}R - \Lambda R^2 \right) $ and examine its parameter dependence on positivity. We can easily check that $h(0)=-3GQ^2, h'(0)=r_{b}(3- \Lambda r_{b}^2) $ and the larger root of $h'(R)$ is given by
\bea
R_{max}= \frac{- \Lambda r_{b}+\sqrt{9\Lambda - 2\Lambda^2 r_{b}^2} }{3\Lambda} .
\ena 
And the equation $h(R_{max})=0$ gives the critical value of $Q$;
\bea
\sqrt{G}Q_{cr}= \frac{1}{9}\sqrt{ \frac{ r_{b} \left( 18\sqrt{ \Lambda(9-2\Lambda r_{b}^2)  } + \Lambda r_{b} \left( -27+ \Lambda r_{b}^2 -4r_{b}\sqrt{\Lambda(9-2 \Lambda r_{b}^2)  } \right)   \right) }{ \Lambda } } \label{eqQcr}
\ena
As an example, the qualitative behavior of $h(R)$ in the case of $0<\Lambda r_{b}^2<3, 0<Q<Q_{cr}$ is shown in Fig. \ref{A4}. 
                                                 %
\iffigure
\begin{figure}[h]
\begin{center}
	\includegraphics[width=8cm]{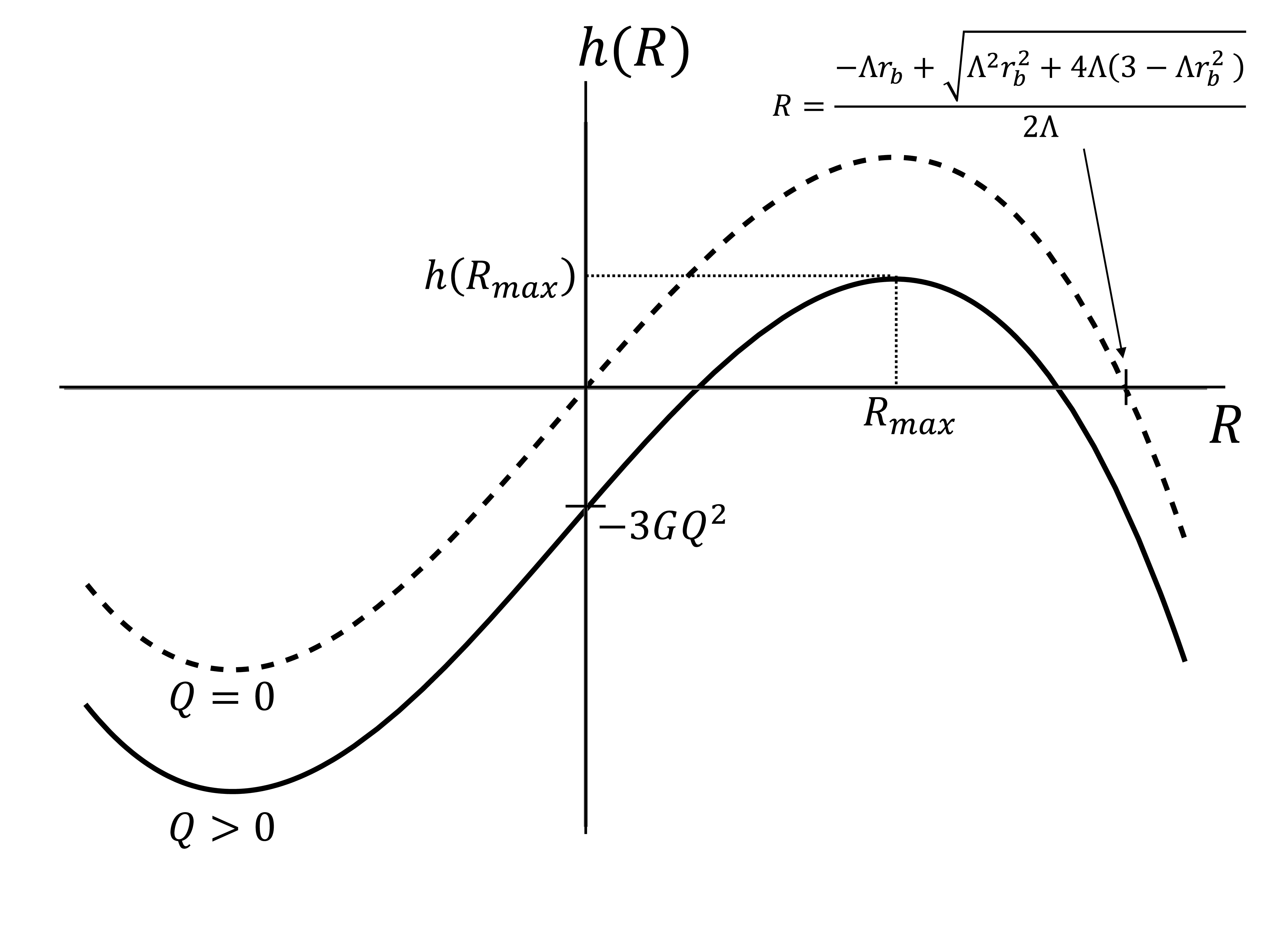} ~

	\caption{Qualitative behavior of $h(R)$ in the case of $0<\Lambda r_{b}^2<3, 0<Q<Q_{cr}$ (The solid curve). The dashed curve represents $h(R)$ in the case of $0<\Lambda r_{b}^2<3, Q=0$.    }
\label{A4}
\end{center}
\end{figure}
\fi
                                                 %
The positive roots are always exist in the range $R\in\left[0, \frac{ -\Lambda r_{b} + \sqrt{\Lambda^2 r_{b}^2 + 4\Lambda(3-\Lambda r_{b}^2) } }{2\Lambda} \right]$. For sufficiently small $Q$, the larger root is greater than $r_{b}$ and the maximum horizon radius of BH saddle is $r_{b}$.  When $r_{b}\sqrt{1-\Lambda r_{b}^2}<\sqrt{G}Q < \sqrt{G}Q_{cr}$, the maximum horizon radius is less than $r_{b}$ and is given by the larger root. When either $3<\Lambda r_{b}^2$ or $Q_{cr}<Q$, $h(R)<0$ for $R>0$, that is, there are no BH saddles.

However, we need an additional condition, namely the positivity of $(\theta(r_{b}-R)-\theta(R-r_{b}))f'(R)$. The equation $g(R)\equiv R^3 f'(R) = (-\Lambda R^4 + R^2 -GQ^2)=0$ has two positive roots $R=\sqrt{\frac{1 \pm \sqrt{1-4\Lambda GQ^2}}{2\Lambda}} $ when $GQ^2 < \frac{1}{4\Lambda}$. Therefore, for $0<\Lambda r_{b}^2 < \frac{1}{2}, r_{b}\sqrt{1-\Lambda r_{b}^2}<\sqrt{G}Q <\sqrt{G}Q_{cr}$, there are no BH saddles even though $h(R)$ is positive in the range $\big({\rm (the ~ smaller ~ positive ~ root ~ of ~ }h(R)=0), r_{b} \big)$.

With a little more algebra, we can classify the parameter space as follows and summarized in Fig. \ref{A5}:
\begin{itemize}
\item \{{\it BH}, {\it bad}\}$_{I}$ : $ 0<\Lambda r_{b}^2<1 , ~ \sqrt{G}Q< r_{b}\sqrt{1- \Lambda r_{b}^2 }$ \\
Both BH and {\it bad} BG saddles exist. \\
$ R\in \left(\sqrt{\frac{1 - \sqrt{1-4\Lambda GQ^2}}{2\Lambda}}, r_{b} \right) $ for BH and $R \in \big({\rm (the ~ larger ~ positive ~ root ~ of ~ }h(R)=0), \infty\big) $ for {\it bad} BG.
\item \{{\it BH}, {\it bad}\}$_{II}$ : $\frac{1}{2} <\Lambda r_{b}^2<1, ~ r_{b}\sqrt{1-\Lambda r_{b}^2}<\sqrt{G}Q < \sqrt{G}Q_{cr}$ \\
\hspace{1.6cm} and $1 <\Lambda r_{b}^2<3, ~ 0<\sqrt{G}Q < \sqrt{G}Q_{cr}$ \\
Both BH and {\it bad} BG saddles exist. \\
$R\in \left(\sqrt{\frac{1 - \sqrt{1-4\Lambda GQ^2}}{2\Lambda}},  {\rm (the ~ larger ~ positive ~ root ~ of ~ }h(R)=0) \right)$ for BH and $R\in (r_{b}, \infty)$ for {\it bad} BG.
\item \{{\it bad}, {\it bad}\}$_{I}$ : $0< \Lambda r_{b}^2<\frac{1}{2}, ~ r_{b}\sqrt{1-\Lambda r_{b}^2}<\sqrt{G}Q<\sqrt{G}Q_{cr}$ \\
Only {\it bad} BG saddles exist. \\
$R\in \left(r_{b}, \sqrt{\frac{1 - \sqrt{1-4\Lambda GQ^2}}{2\Lambda}}\right) \cup \big({\rm (the ~ larger ~ positive ~ root ~ of ~ }h(R)=0), \infty \big)$ for {\it bad} BG.
\item \{{\it bad}, {\it bad}\}$_{II}$ : $0< \Lambda r_{b}^2<\frac{1}{2}, ~ \sqrt{G}Q_{cr}<\sqrt{G}Q<\frac{1}{2}$ \\
Only {\it bad} BG saddles exist. \\
$R\in \left(r_{b}, \sqrt{\frac{1 - \sqrt{1-4\Lambda GQ^2}}{2\Lambda}}\right) \cup \left(\sqrt{\frac{1 + \sqrt{1-4\Lambda GQ^2}}{2\Lambda}}, \infty\right)$ for {\it bad} BG.
\item \{{\it bad}\} : The other region \\
Only {\it bad} BG saddles exist.\\
$R\in (r_{b}, \infty)$ for {\it bad} BG.
\end{itemize}
                                                 %
\iffigure
\begin{figure}[h]
\begin{center}
	\includegraphics[width=8cm]{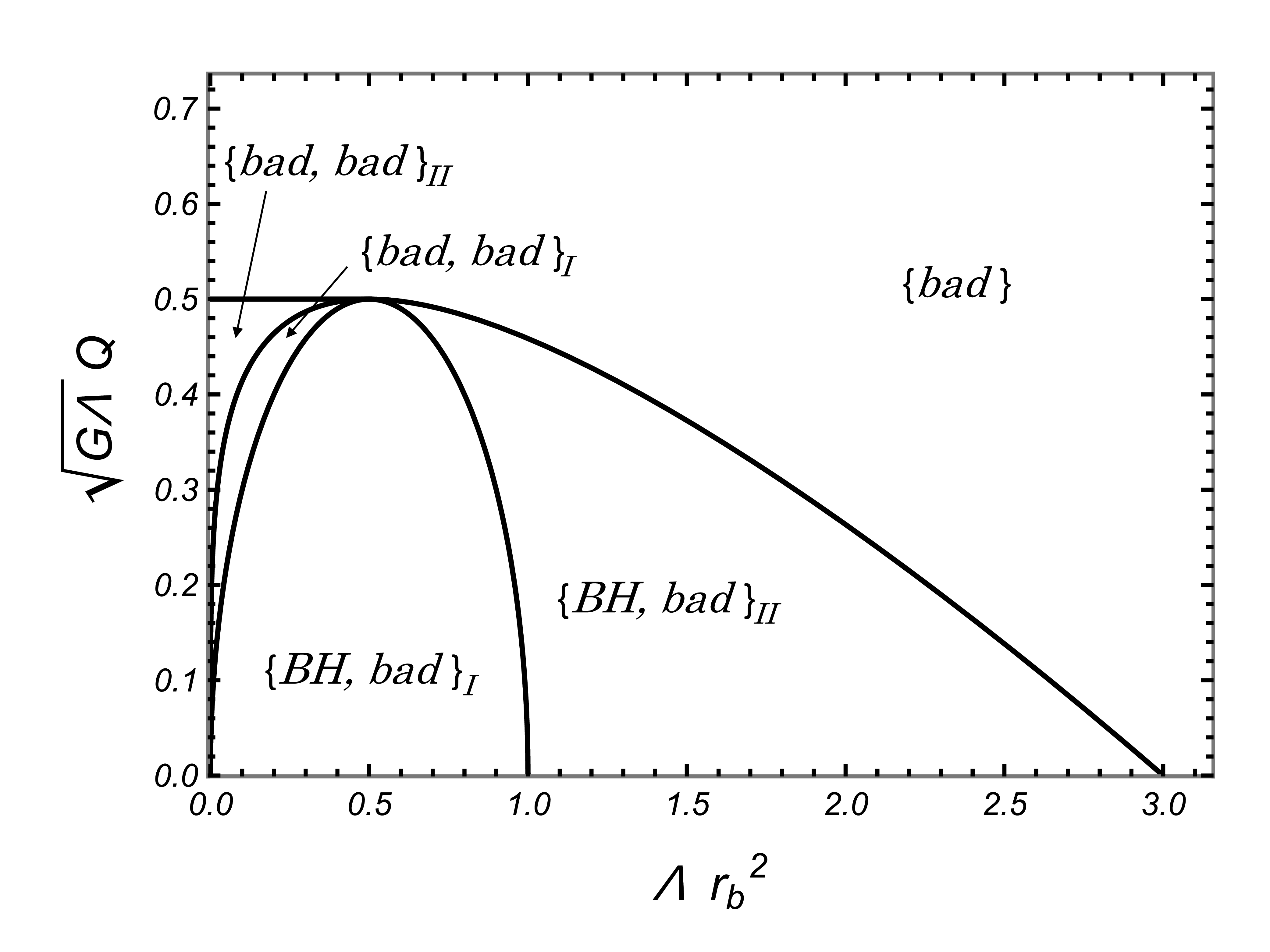} ~

	\caption{Classification of the existence and the possible range of the bolt radius $R$ in the case of the (micro)canonical ensemble. The boundary curve of \{{\it BH}, {\it bad}\}$_{I}$ is given by $\sqrt{G\Lambda}Q = \sqrt{\Lambda r_{b}^2 (1-\Lambda r_{b}^2)}$. The outer boundaries of \{{\it bad}, {\it bad}\}$_{I}$ and \{{\it BH}, {\it bad}\}$_{II}$ form a single smooth curve $\sqrt{G\Lambda} Q = \sqrt{G\Lambda}Q_{cr}(\Lambda, r_{b})$.  }
\label{A5}
\end{center}
\end{figure}
\fi
                                                 %
The notation is similar to that of the grand canonical case. However, since there are no good BGs, I used {\it BH} instead of {\it good}.
\footnote{
Note that I only used {\it good} for BH at the beginning of subsection 4.2, when I briefly discussed the case of pure gravity with a positive $\Lambda$. Here, I use {\it BH} because I want to emphasize that there are no good BGs in the case of the (micro)canonical ensemble.  
}

In the (micro)canonical case with a positive $\Lambda$, there always exist {\it bad} BGs. So the system may be thermodynamically unstable.

\subsection{microcanonical ensembles}
\subsubsection{$\Lambda \leq 0$}
The qualitative behavior of the entropy depends on $Q$ for fixed $\Lambda$ and $r_{b}$, similar to the canonical ensemble free energy (Fig. \ref{A1}). It is shown in Fig. \ref{A6}. 
                                                 %
\iffigure
\begin{figure}[h]
\begin{center}
	\includegraphics[width=7cm]{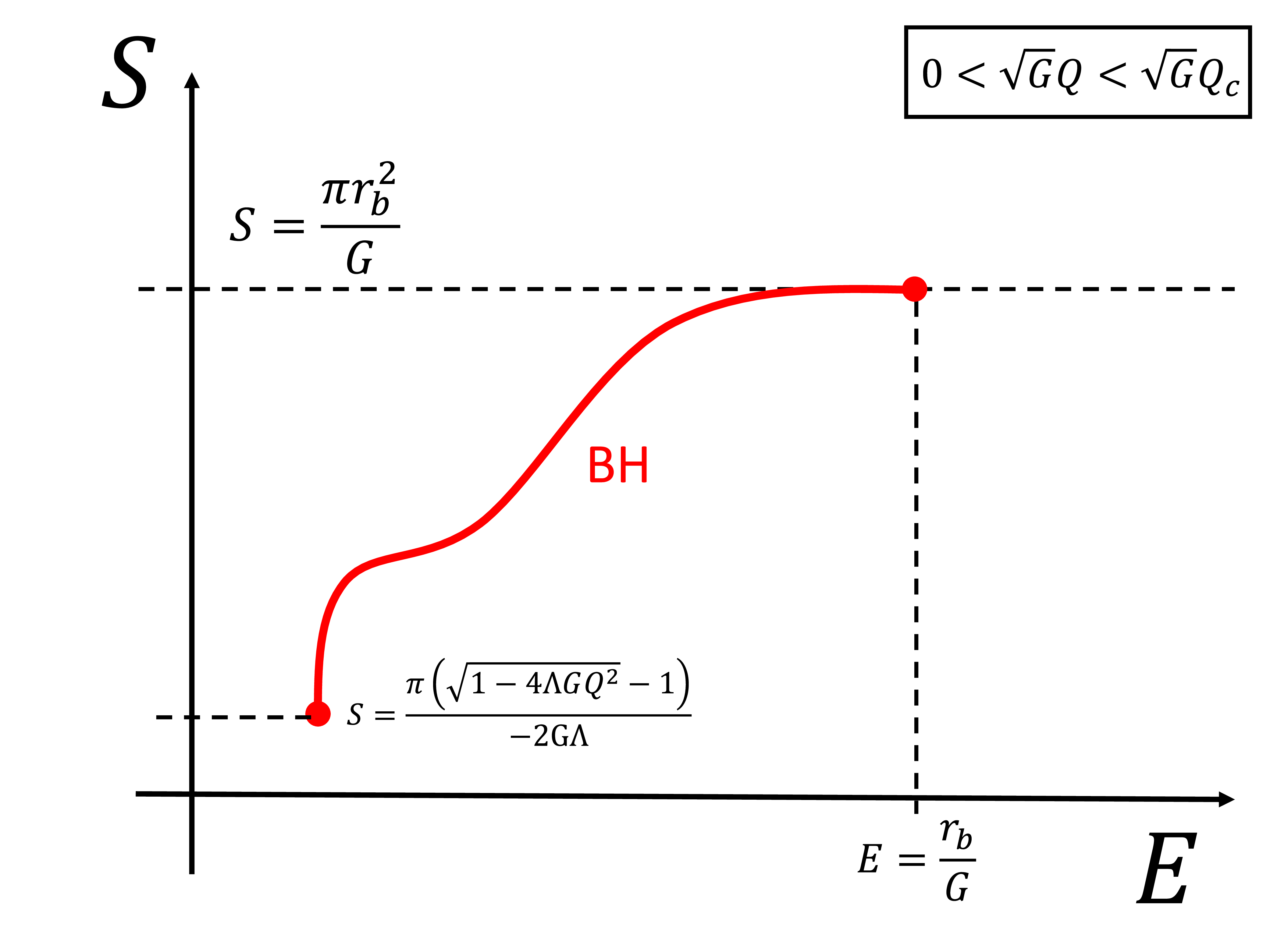} 
	\includegraphics[width=7cm]{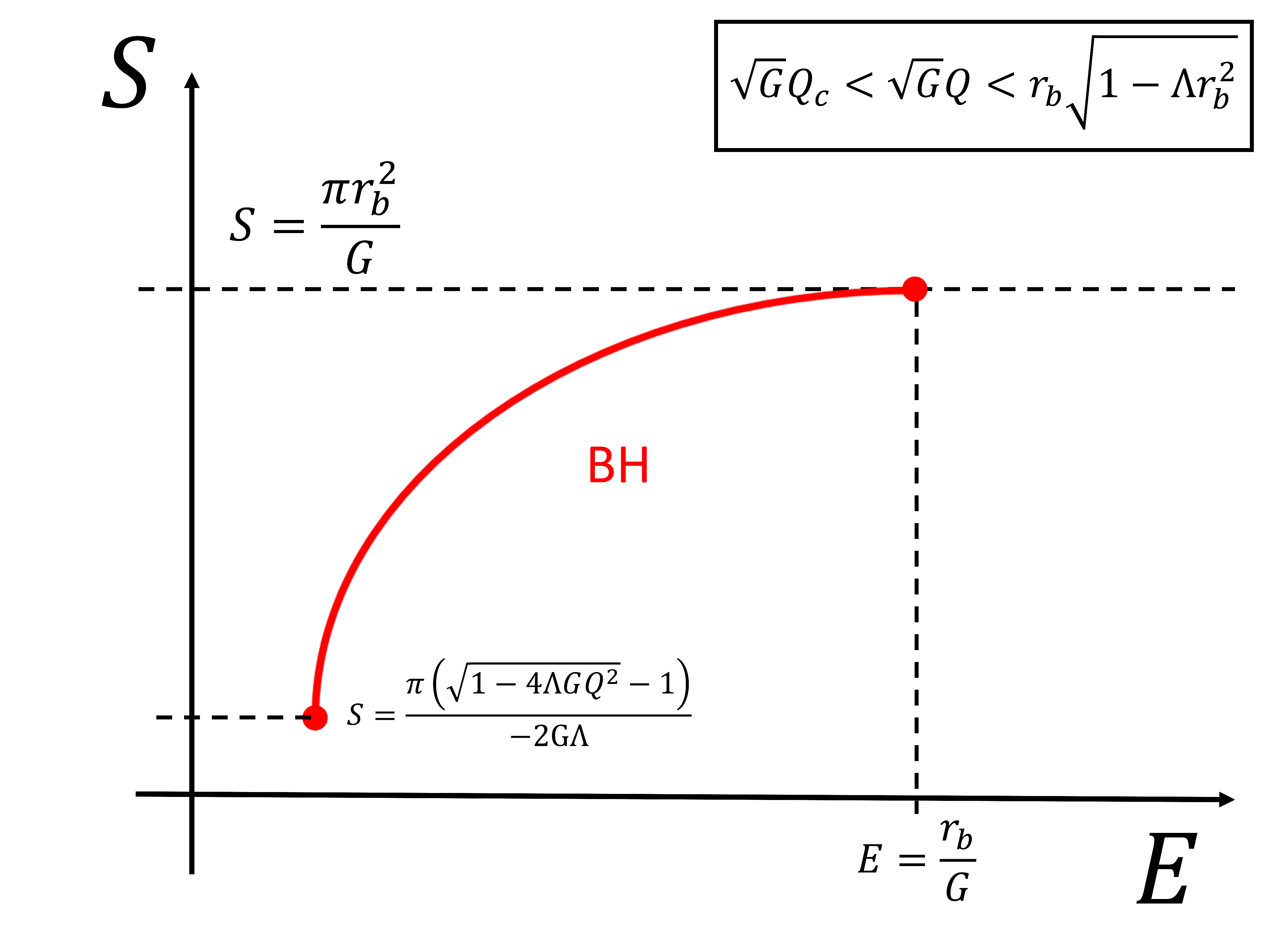}
	\includegraphics[width=7cm]{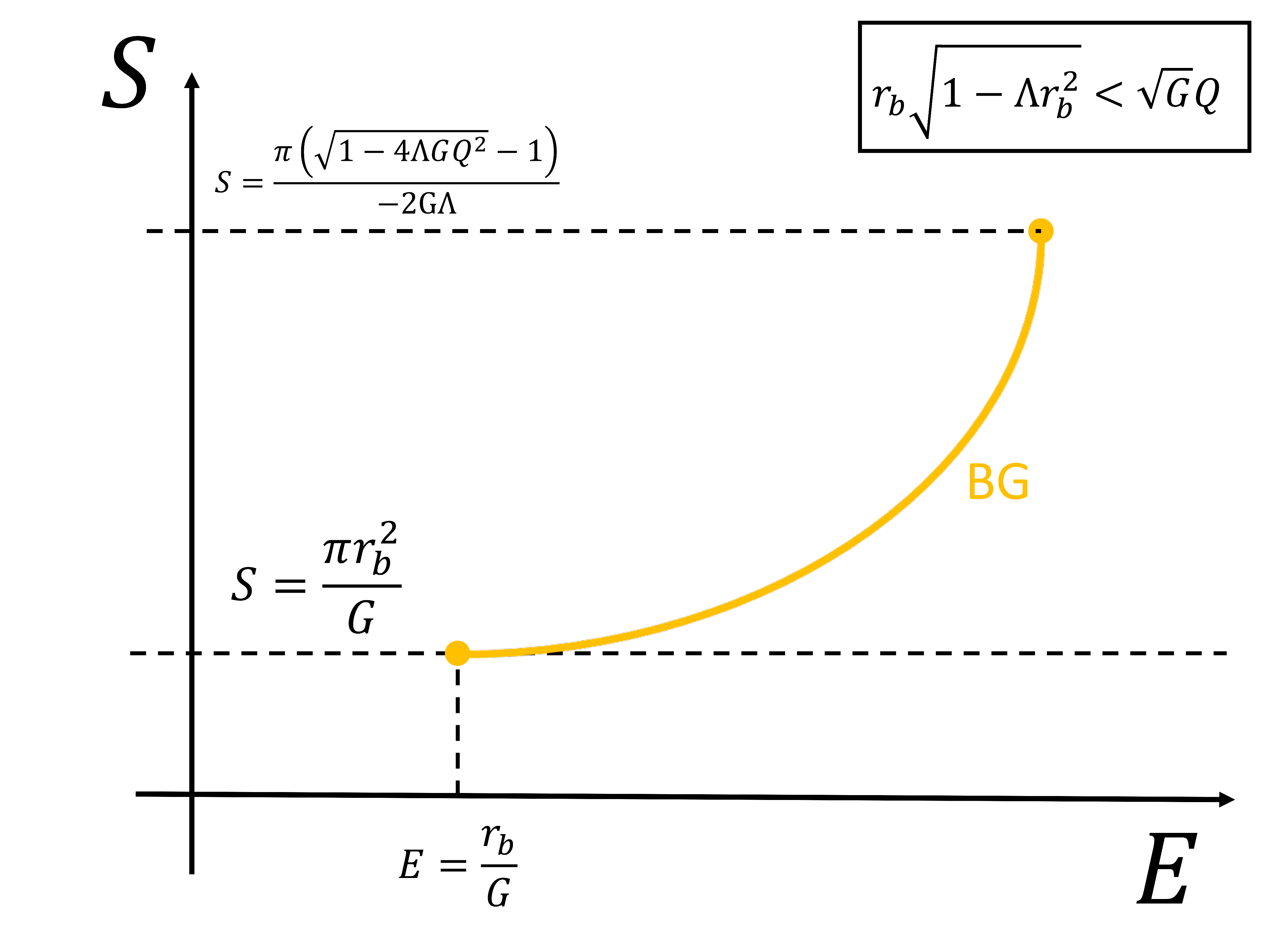} 

	\caption{Qualitative behaviors of entropy ($\Lambda\leq 0$). }
\label{A6}
\end{center}
\end{figure}
\fi
                                                 %
For $0<\sqrt{G}Q< r_{b}\sqrt{1-\Lambda r_{b}^2}$, only BH branch exists and the lowest entropy is given by $S= \frac{\pi(\sqrt{1-4\Lambda GQ^2}-1)}{-2G\Lambda}$. (When $\Lambda=0$, it is replaced by $S= \pi Q^2$.) The fact the highest entropy is given by $S=\frac{\pi r_{b}^2}{G}$ is same as the pure gravity case. On the other hand, for $r_{b}\sqrt{1-\Lambda r_{b}^2} < \sqrt{G}Q $, only BG branch exists and the lowest entropy is given by $S= \frac{\pi r_{b}^2}{G}$ and the highest one by $S= \frac{\pi(\sqrt{1-4\Lambda GQ^2}-1)}{-2G\Lambda}$. A notable point is that there exists a transition of the entropy bound. For sufficiently low $Q$, the bound is same as that of pure gravity, is given by the area of the boundary. And for high $Q$, it becomes irrelevant to the boundary geometry, depending only on $Q$ and $\Lambda$. It would be interesting to investigate whether this kind of property holds in gravity systems with other conserved charges, such as the SU(2) charge. Addressing this question is beyond the scope of the present paper. I will report on it in a future publication.

\subsubsection{$\Lambda > 0$}
Again, the qualitative behavior of the entropy depends on $Q$ for fixed $\Lambda$ and $r_{b}$, i.e. it depends on which parameter region the system is in. This is shown in Fig. \ref{A7}. (For the classification of the parameter space, see Fig. \ref{A5}.) 
                                                 %
\iffigure
\begin{figure}[h]
\begin{center}
	\includegraphics[width=7cm]{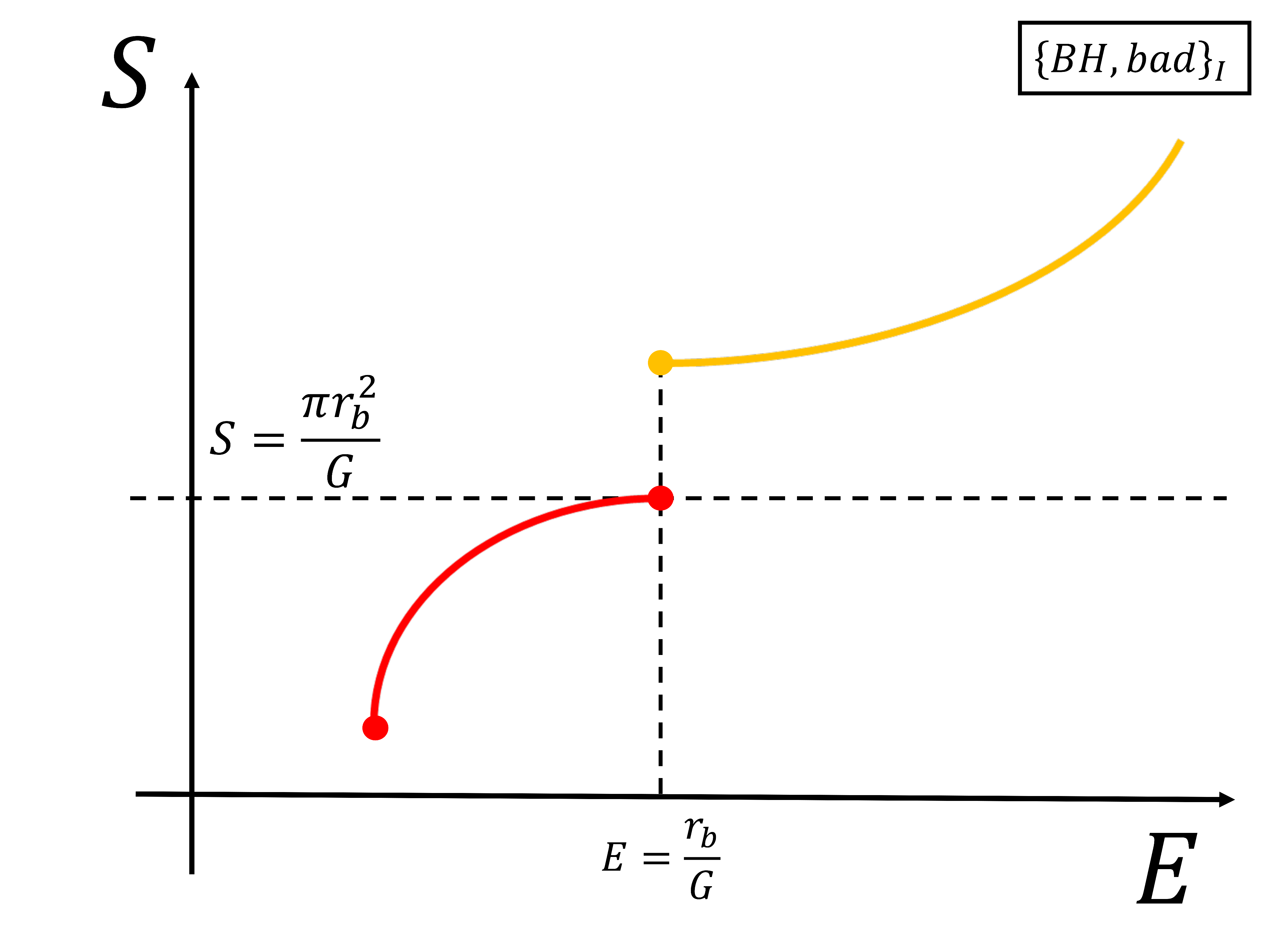} 
	\includegraphics[width=7cm]{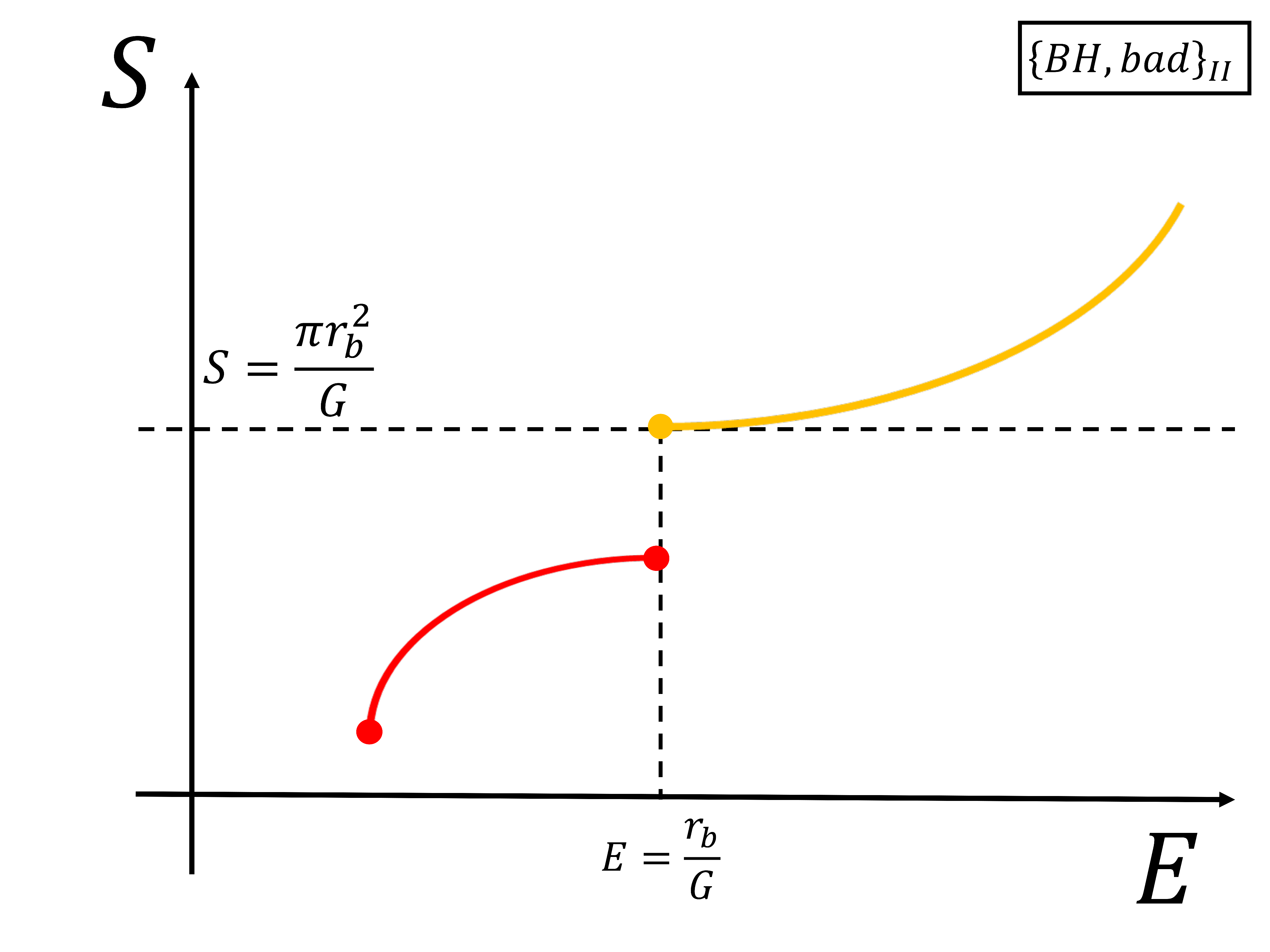}
	\includegraphics[width=7cm]{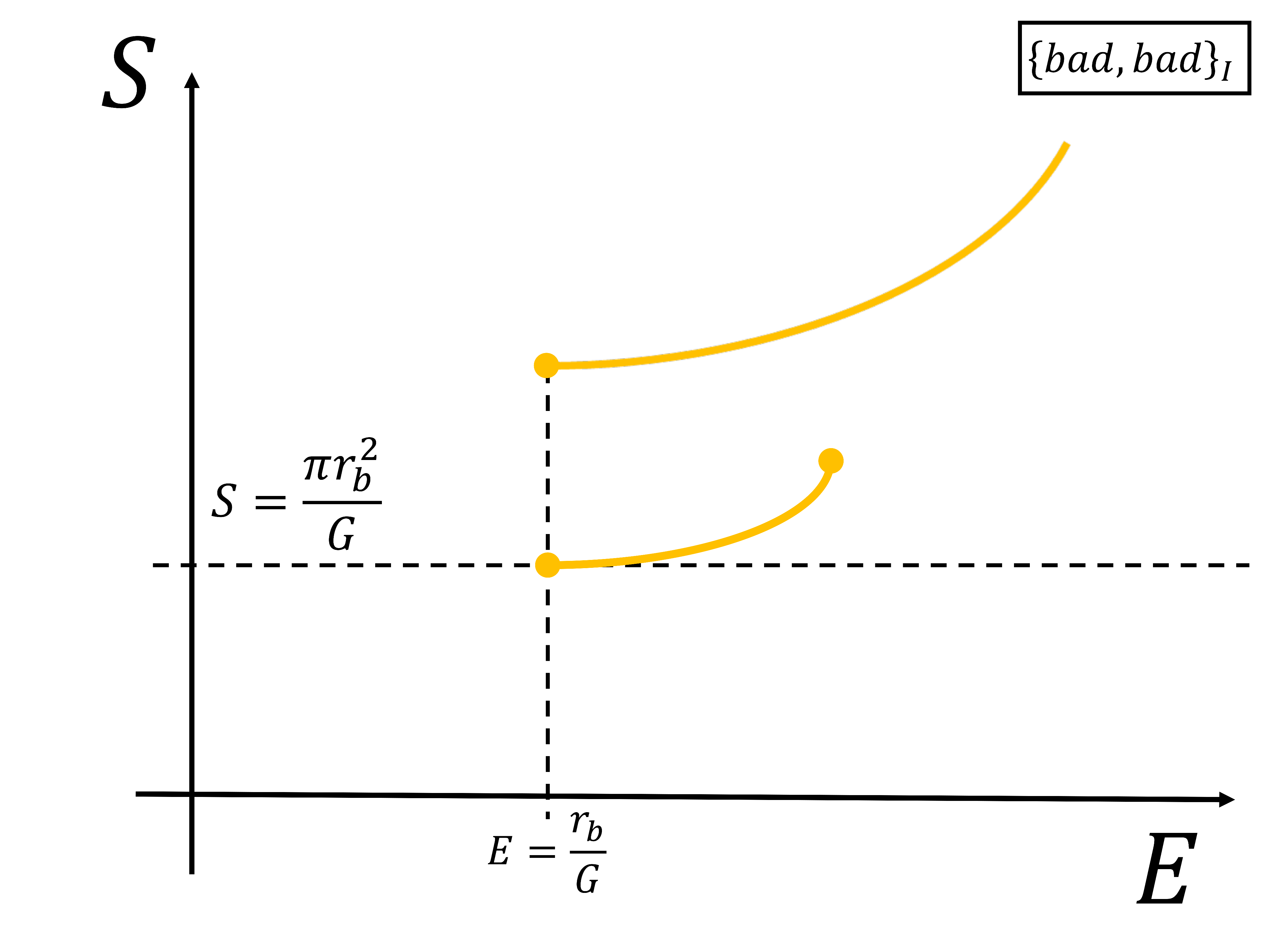} 
	\includegraphics[width=7cm]{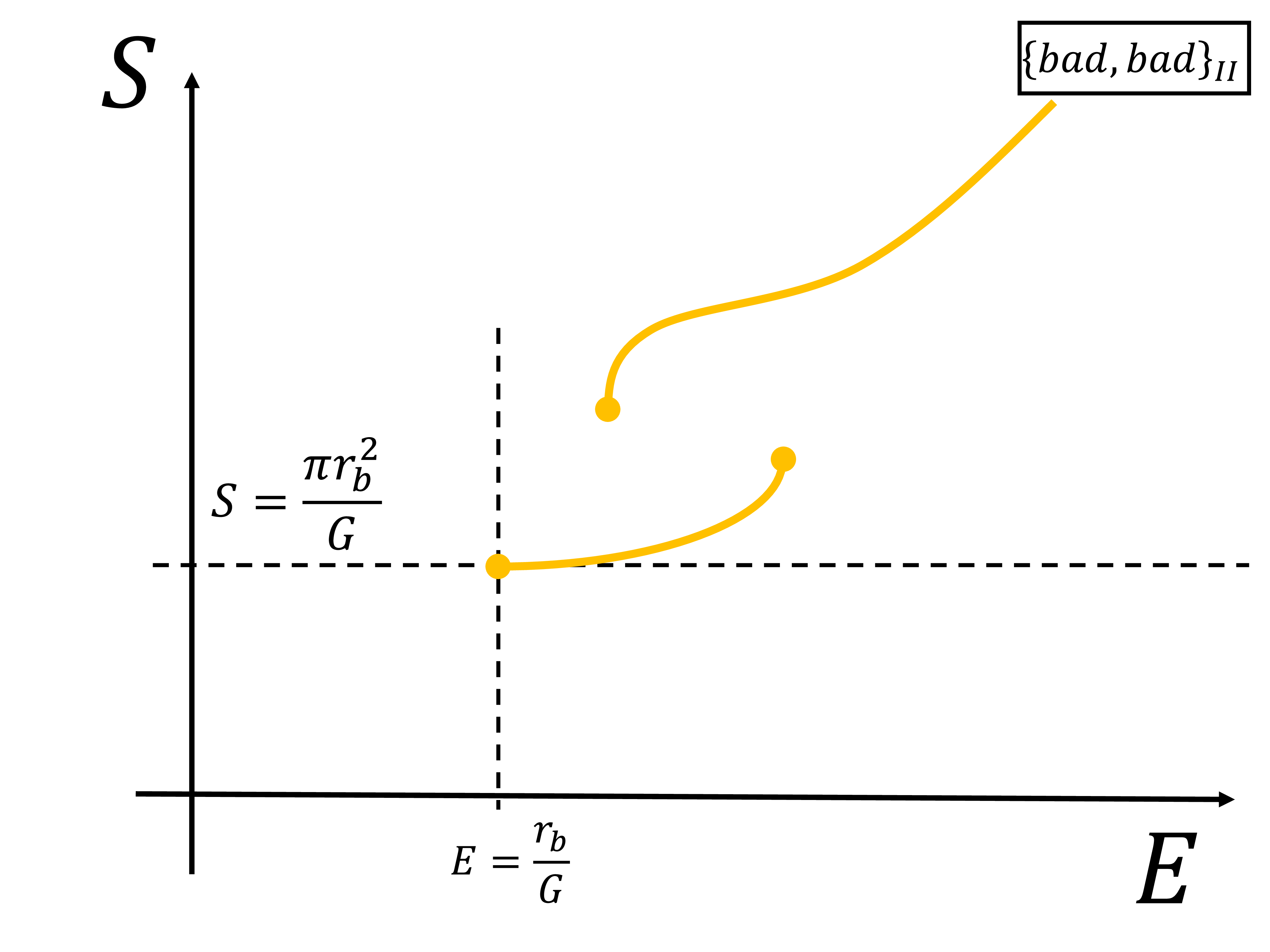}
	\includegraphics[width=7cm]{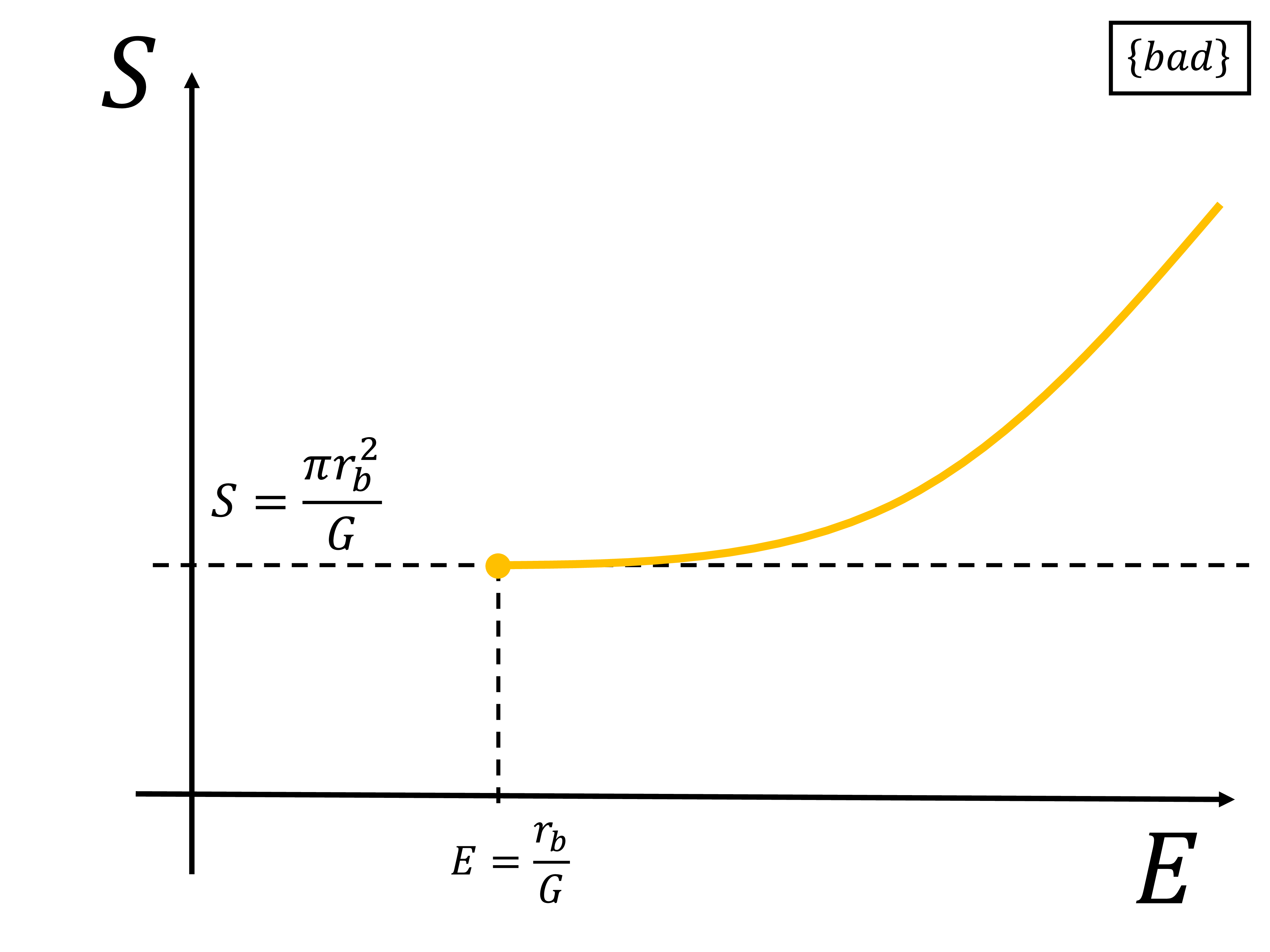} 

	\caption{Qualitative behaviors of entropy ($\Lambda> 0$). For the classification of the parameter space, see Fig. \ref{A5}. }
\label{A7}
\end{center}
\end{figure}
\fi
                                                 %
In any case, there are no entropy bounds. This property is the same as that of pure gravity. The qualitative behaviors of \{{\it BH, bad}\}$_{I}$, \{{\it BH, bad}\}$_{II}$ and \{{\it bad}\} are also same as those of pure gravity, except that the lowest entropy state is an extremal BH for the first two cases.

The behaviors of \{{\it bad, bad}\}$_{I}$ and \{{\it bad, bad}\}$_{II}$, which the system exhibits when $\sqrt{G}Q$ is relatively small compared to the dS length, are not seen in the pure gravity system. Probably the \{{\it bad, bad}\}$_{I}$ case is the more peculiar, since the lowest entropy can be arbitrarily large if we tune $\Lambda$ and $Q$. For the pure gravity case, the lowest entropy is $\frac{\pi r_{b}^2}{G}$ when the BH branch is absent. Actually, the lowest entropy of the lower branch for the \{{\it bad, bad}\}$_{I}$ case is $\frac{\pi r_{b}^2}{G}$. However, in this case, there always exists another branch that gives the dominant contribution. 

\clearpage
\section{Thermodynamics without {\it bad} BG}
In the main text and in Appendix A, I made the assumption that {\it all Euclidean saddles contribute to the partition function}. Since there exist {\it bad} BGs when $\Lambda>0$, this assumption makes the system thermodynamically unstable in that case. Here, instead of it, I make another assumption that {\it all Euclidean saddles \underline{except bad BGs} contribute to the partition function} and see thermodynamic properties in the case of $\Lambda>0$.

\subsection{grand canonical ensembles without {\it bad} BG}
Recall that when $0<\Lambda r_{b}^2 <3, \sqrt{G}\mu<1$, although there are BH/{\it good} BG saddles, {\it bad} BGs cause thermodynamical instability. Therefore, in the main text I focused only on the case of $0<\Lambda r_{b}^2 <3, \sqrt{G}\mu>1$. Examples of phase diagrams are shown in Fig. \ref{B1} under the alternative assumption. These are the extensions of the phase diagrams in Fig. \ref{15}.
                                                 %
\iffigure
\begin{figure}[h]
\begin{center}
	\includegraphics[width=5.3cm]{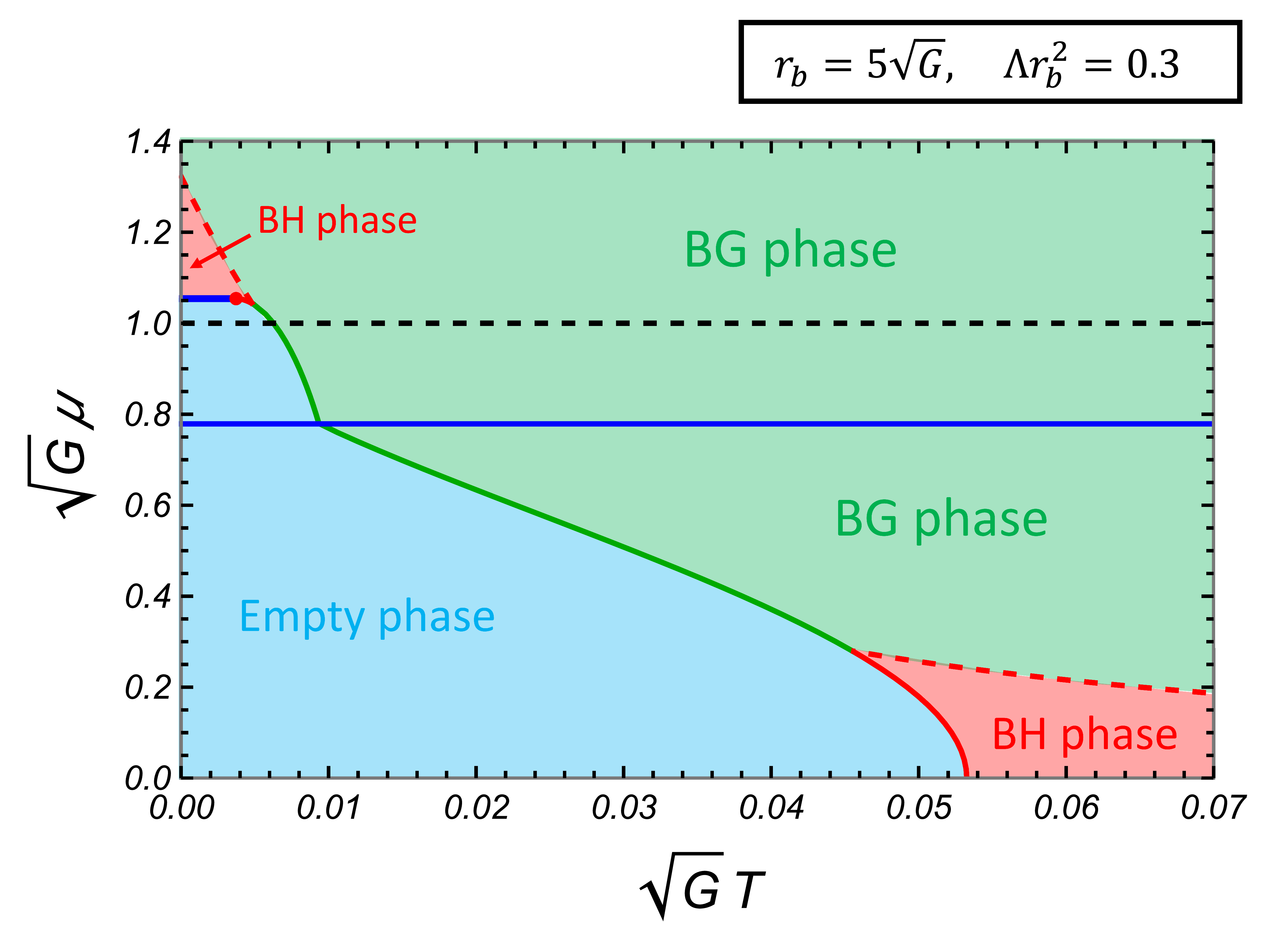} 
	\includegraphics[width=5.3cm]{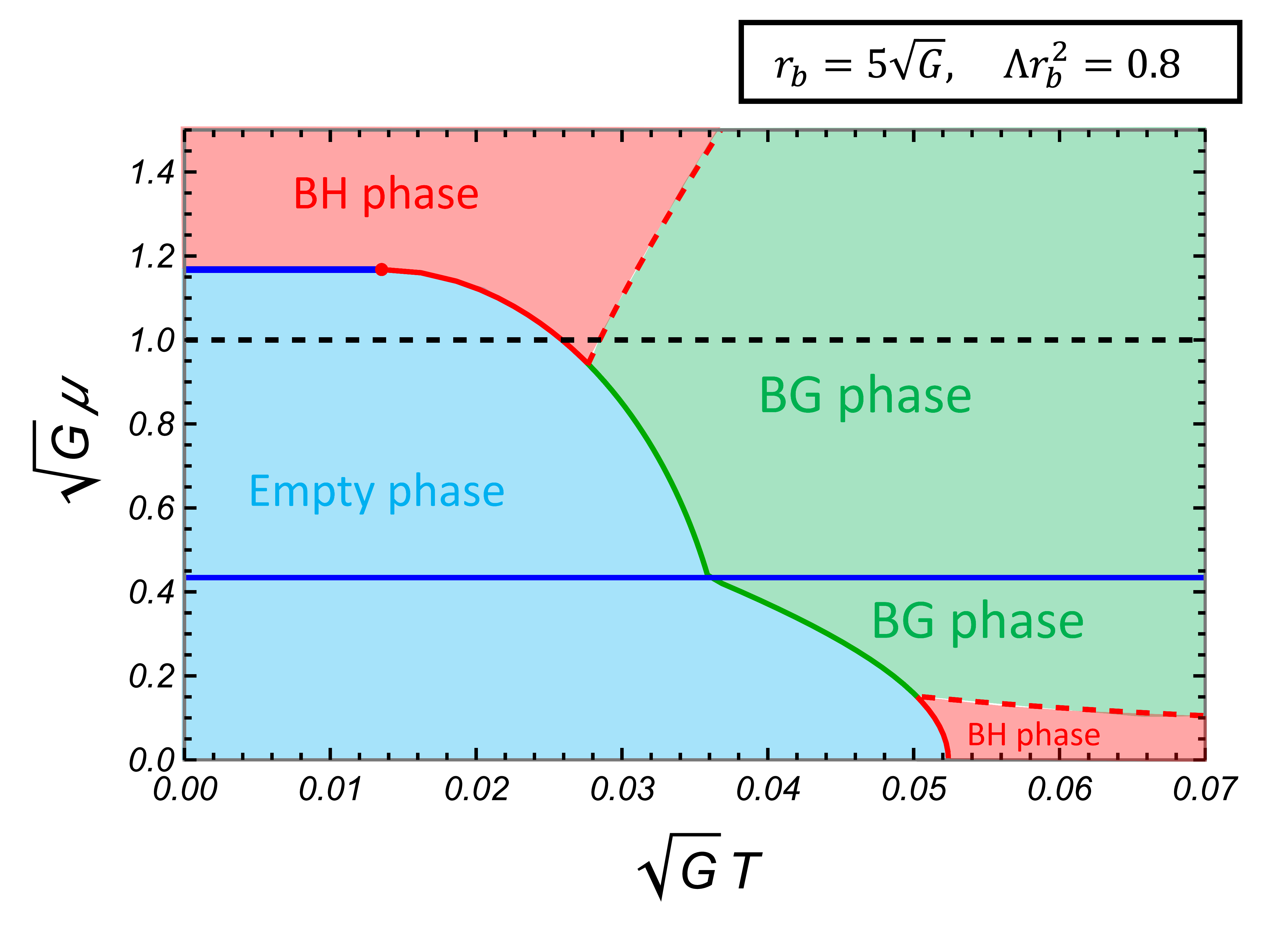} 
	\includegraphics[width=5.3cm]{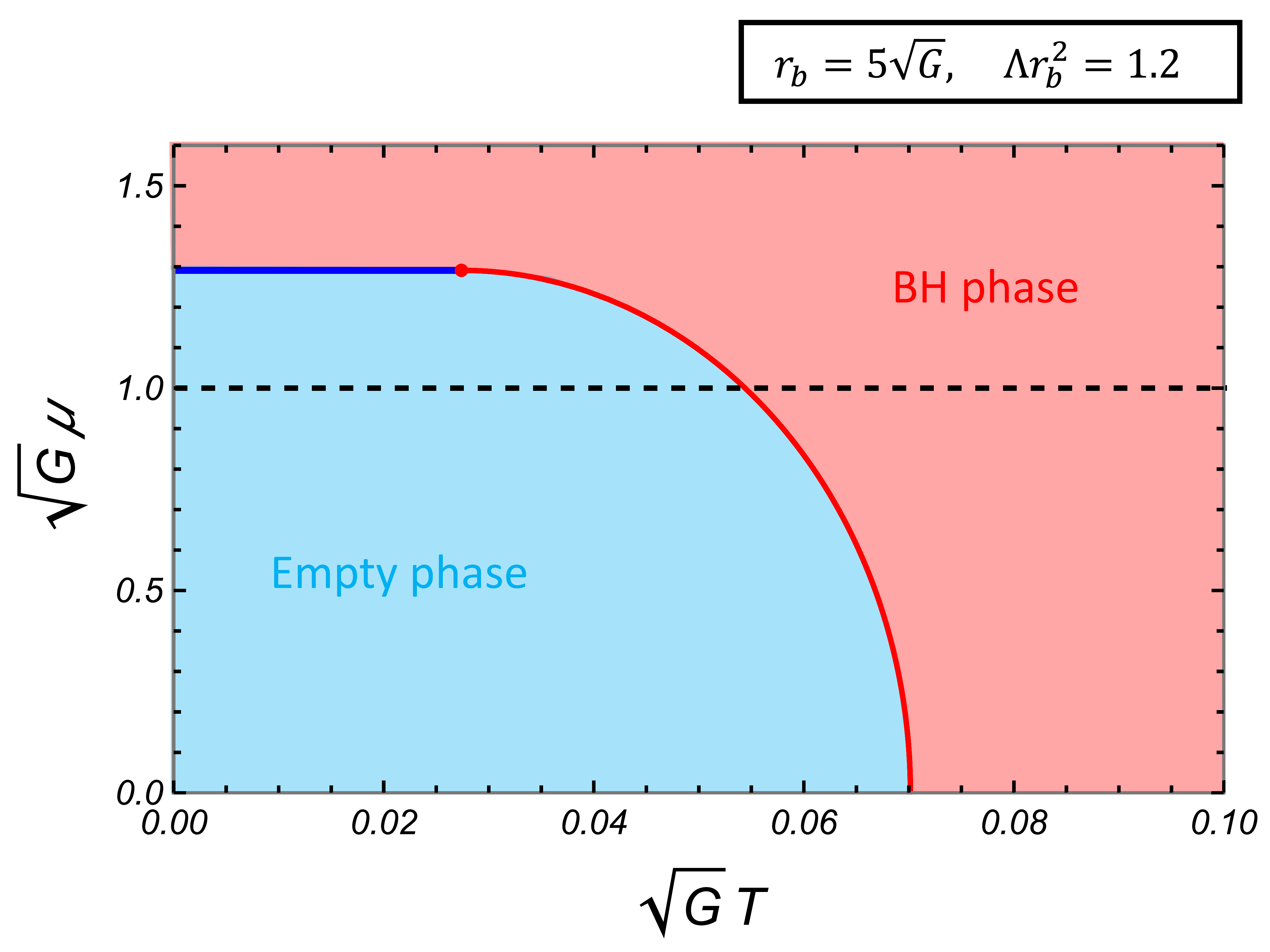} 

	\caption{Examples of phase diagrams. These are the extensions of the examples in Fig. \ref{15}. The black dashed lines represent $\sqrt{G}\mu=1$. (Left) An example of the case $0<\Lambda r_{b}^2 < \frac{1}{2}$. The blue line represents $\sqrt{G} \mu=\sqrt{ \frac{3(2-\Lambda r_{b}^2) -\sqrt{3\Lambda r_{b}^2 (4- \Lambda r_{b}^2) } }{2(3-\Lambda r_{b}^2)} }$, and on this line, the empty saddles are dominant as shown in Fig. \ref{B2}.  (Middle) An example of the case $\frac{1}{2} \leq \Lambda r_{b}^2<1 $. (Right) An example of the case $1 \leq \Lambda r_{b}^2<3 $. In this case, the BG phase does not exist. }
\label{B1}
\end{center}
\end{figure}
\fi
                                                 %
The blue line represents $\sqrt{G} \mu=\sqrt{ \frac{3(2-\Lambda r_{b}^2) -\sqrt{3\Lambda r_{b}^2 (4- \Lambda r_{b}^2) } }{2(3-\Lambda r_{b}^2)} }$, and on this line, the empty saddles dominate. As shown in Fig. \ref{B2}, for fixed $r_{b}$ and $\Lambda$, as $\sqrt{G}\mu$ approaches the value $\sqrt{ \frac{3(2-\Lambda r_{b}^2) -\sqrt{3\Lambda r_{b}^2 (4- \Lambda r_{b}^2) } }{2(3-\Lambda r_{b}^2)} }$, the BH/{\it good} BG branch and {\it bad} BG branch get closer to each other. When $\sqrt{G} \mu=\sqrt{ \frac{3(2-\Lambda r_{b}^2) -\sqrt{3\Lambda r_{b}^2 (4- \Lambda r_{b}^2) } }{2(3-\Lambda r_{b}^2)} }$, the two branches merge to form an unstable and subdominant branch. 
\footnote{
Although this branch is subdominant, it may lead to an instability similar to the instability of {\it hot flat space} \cite{GrossPerryYaffe}. Therefore, I regard this branch as {\it bad} BG and exclude it from the path integral.
}
                                                 %
\iffigure
\begin{figure}[h]
\begin{center}
	\includegraphics[width=5.3cm]{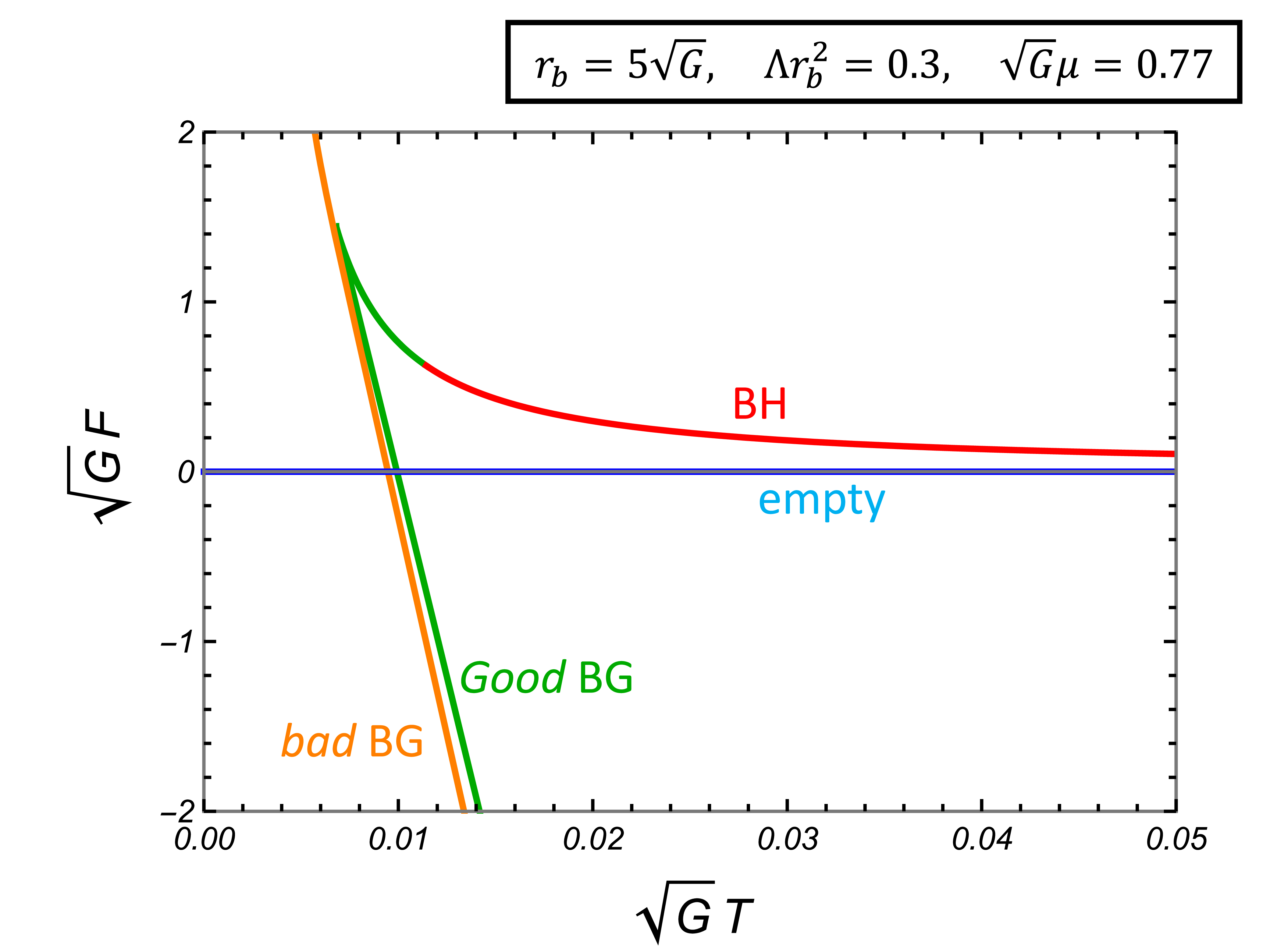} 
	\includegraphics[width=5.3cm]{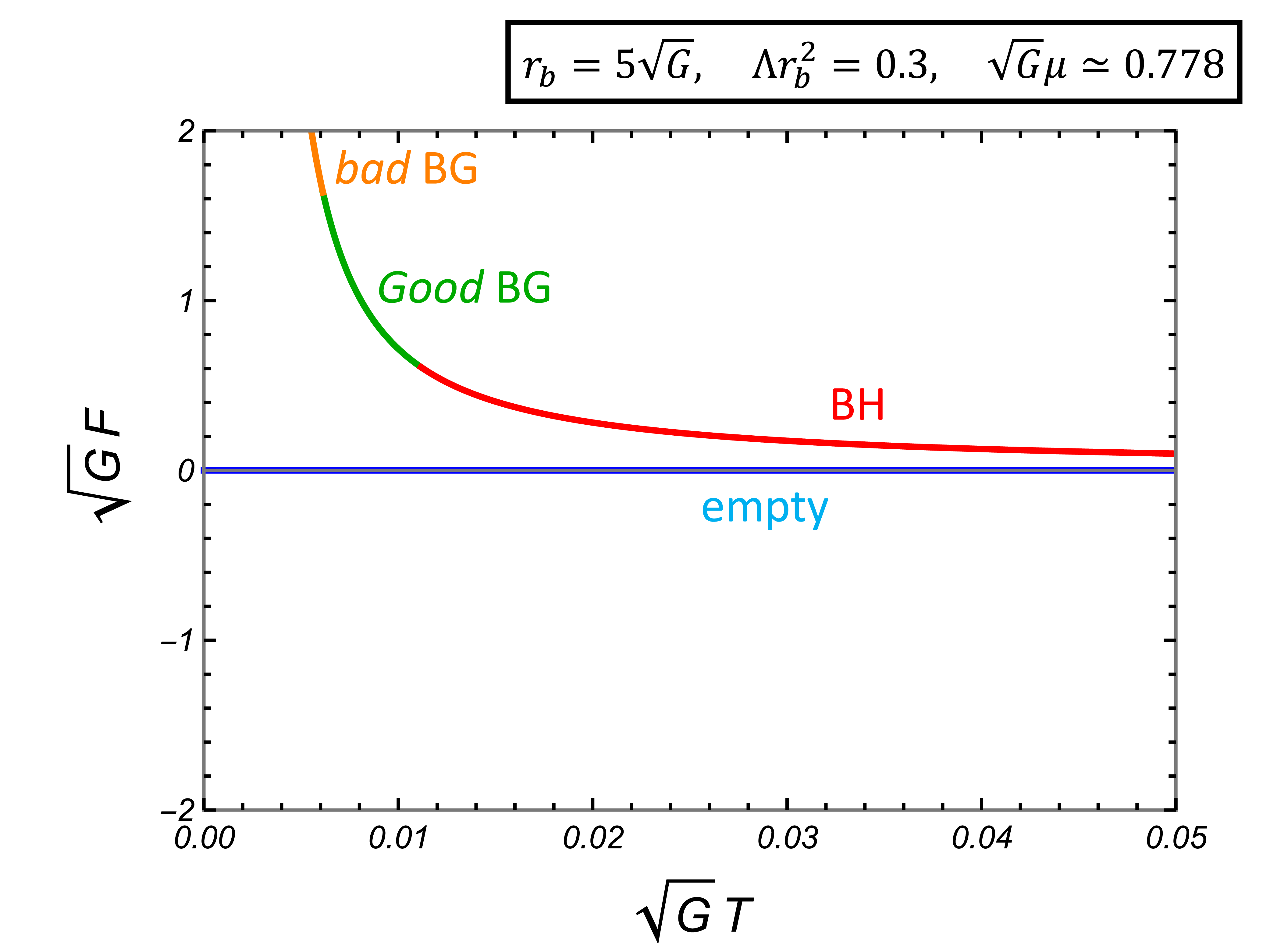} 
	\includegraphics[width=5.3cm]{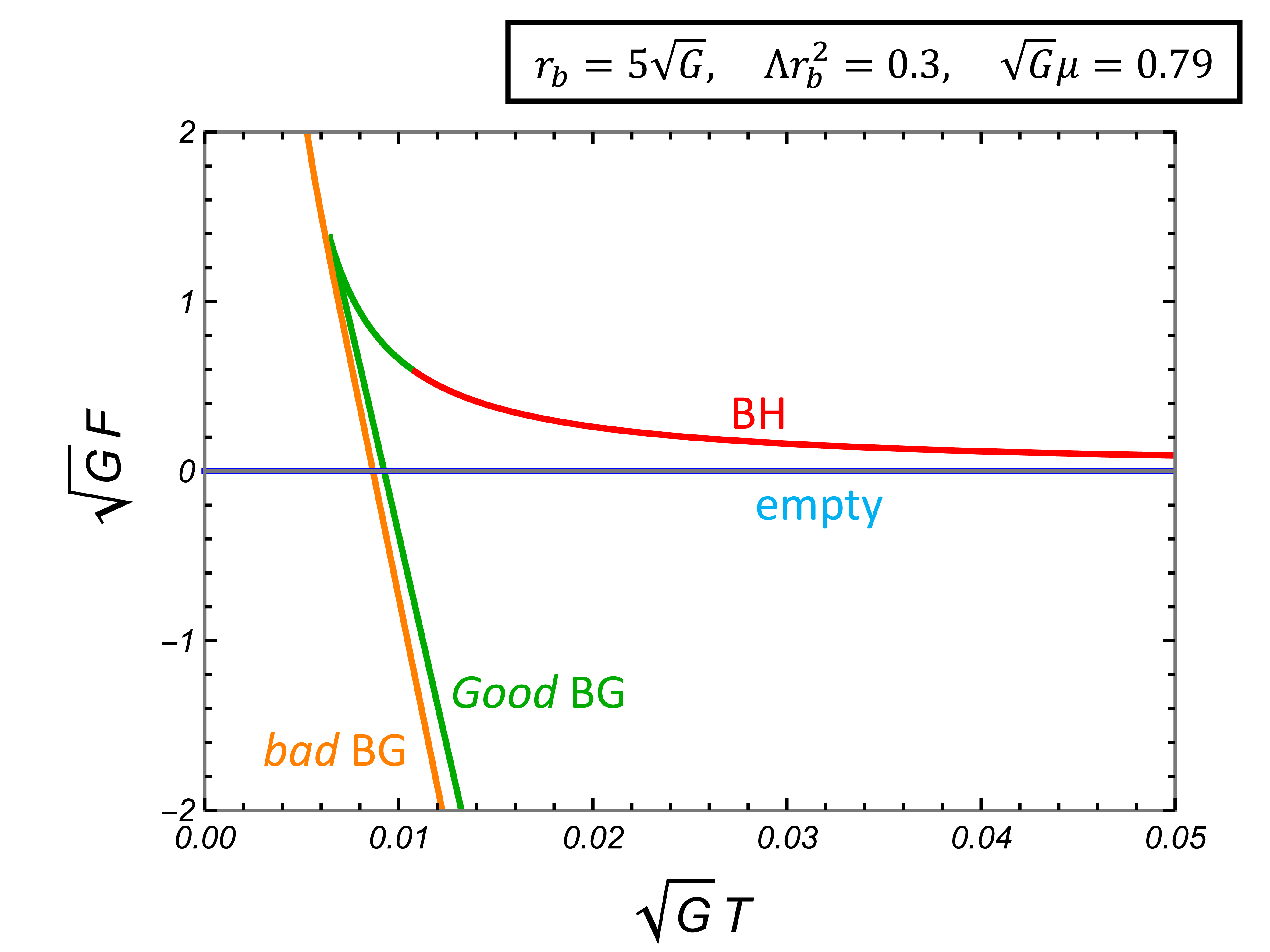} 

	\caption{Behaviors of the free energies when $\sqrt{G}\mu$ is close to, or equal to the critical value $\sqrt{ \frac{3(2-\Lambda r_{b}^2) -\sqrt{3\Lambda r_{b}^2 (4- \Lambda r_{b}^2) } }{2(3-\Lambda r_{b}^2)} }$. (Left) When $\sqrt{G}\mu$ is slightly below the critical value. (Middle) When $\sqrt{G}\mu$ is equal to the critical value. In other words, this is the case when the system is on the boarder between {\it \{good, bad$\}_{I}$} and {\it \{good, bad$\}_{II}$} in Fig. \ref{13}.  (Right) When $\sqrt{G}\mu$ is slightly above the critical value.}
\label{B2}
\end{center}
\end{figure}
\fi
                                                 %
Naively, we can think of these as follows. When $\Lambda>0$ there are two regions in the phase diagram. One is below the line and the other is above the line. The former consists of the phases continuously deformed from those of $\Lambda<0$. The latter consists of the new phases that appear when $\Lambda>0$. As we increase $\Lambda r_{b}^2$, the former region shrinks and vanishes at $\Lambda r_{b}^2=1$. When $\Lambda r_{b}^2 \geq 1$, the entire phase diagram becomes the latter region.

\subsection{canonical ensembles without {\it bad} BG}
For the canonical ensembles, we simply forget about the BG phase in Fig. \ref{A3}. 

For the microcanonical ensembles, we do not need to exclude them even if it somehow turns out  that the alternative assumption is true for grandcanonical and canonical ensembles.

\section{Bertotti-Robinson Geometry}
Bertotti-Robinson(BR) geometry takes a form of $AdS_{2} \times S^{2}$. And it is a family of solutions in the Einstein-Maxwell system. Under Dirichlet type boundary conditions, besides the parameter of the theory $\Lambda$, it is parametrized by
\bea
r_{b} : {\rm radius ~ of ~}S^2 \hspace{1.6cm} \notag \\
T_{H} : {\rm Hawking ~ temperature} \notag 
\ena
The metric and gauge field can be written as
\bea
\BS{g}= -\left[ 4\pi T_{H} r + \left( \frac{1}{r_{b}^2} - 2\Lambda \right) r^2 \right] dt^2 + \frac{1}{\left[  4\pi T_{H} r + \left(  \frac{1}{r_{b}^2} - 2\Lambda \right) r^2 \right]}dr^2 + r_{b}^2 d\Omega^2 \label{eqBR} \\
\BS{A}= \pm \frac{1}{r_{b}}\sqrt{\frac{1-\Lambda r_{b}^2}{G}} r \hspace{9.3cm} \\
r\in(0, 1] \notag 
\ena
Without loss of generality, we can set the boundary at $r=1$.
\footnote{
If we consider a geometry of $r_{b}, T_{H}$ but whose coordinate range of $r$ is $r\in(0, a]$ for some positive number $a$, we can easily check, by a suitable coordinate change,  that it is equivalent to that of $r_{b}, \frac{T_{H}}{a}$.
}
\footnote{
If we take the coordinate range $(0, \infty)$ for $r$, it still solves the Einstein-Maxwell equation. However, since (only) the $tt$ component of the metric diverges, it cannot be a solution under either Dirichlet type boundary conditions or conformal Dirichlet type (or asymptotically AdS) boundary conditions. Under appropriate types of boundary conditions, it will be a solution and $T_{H}$ may be interpreted as the temperature measured at infinity. 
}
Physical quantities are represented by
\bea
T=  \frac{T_{H}}{ \sqrt{ 4\pi T_{H}+ \left( \frac{1}{r_{b}^2} - 2\Lambda \right) } } \hspace{1.cm} \\
E= \frac{r_{b}}{G}\sqrt{1- \frac{\Lambda}{3}r_{b}^2}  \hspace{2.5cm} \\
Q= \pm r_{b}\sqrt{\frac{1-\Lambda r_{b}^2 }{G} } \hspace{2.3cm}  \\
\mu = \pm \frac{1}{\sqrt{G}} \sqrt{ \frac{ 1-\Lambda r_{b}^2 }{ 4\pi T_{H}r_{b}^2 + 1 - 2\Lambda r_{b}^2 } } \label{BRchemi}
\ena

\subsection*{Standard near horizon and near extremal limit of Reissner-Nordstr$\ddot{o}$m BH}
The BR geometry often appears in the context of the near horizon geometry of the RN BH. Firstly, let us review how BR geometry appears using the standard near horizon and near extremal limit. The metric and gauge field of the RN (A)dS BH is written as
\bea
&& \BS{g}= f(r)dt^2 + \frac{1}{f(r)}dr^2 + r^2 d\Omega^2 \\
&& ~~~~~~~~ f(r)=  1 - \frac{2G\ti{M}}{r}+ \frac{G\ti{Q}^2}{r^2} - \frac{\Lambda}{3}r^2  \\
&&  A_{t}(r)= \ti{Q} \left( \frac{1}{r_{H}}-\frac{1}{r} \right) 
\ena
Besides the theory parameter $\Lambda$, this class of geometries are controlled by $\ti{Q}$ and $r_{H}$. Let's fix $r_{H}$ by $r_{H}=r_{b}$. (Here, $r_{b}$ is some positive constant and is identified with $r_{b}$ in the equation (\ref{eqBR}).) The first and second derivatives of $f(r)$ at $r=r_{b}$ are 
\bea
f'(r_{b})= \frac{1}{r_{b}}- \frac{G \ti{Q}^2}{r_{b}^3} -\Lambda r_{b} \label{fderi}\\
f''(r_{b})= - \frac{2}{r_{b}^2} + \frac{4 G \ti{Q^2} }{r_{b}^4} \label{sderi}
\ena
Instead of $\ti{Q}$, let's parameterize this geometry by two parameters $T_{H}$ and $\varepsilon$ (although it seems that one of them is fictitious). They are defined by the relation
\bea
4\pi T_{H} \varepsilon = \frac{1}{r_{b}}- \frac{G \ti{Q}^2}{r_{b}^3} -\Lambda r_{b}
\ena
Then, (\ref{fderi}) and (\ref{sderi}) can be rewritten as
\bea
f'(r_{b})= 4\pi T_{H} \varepsilon \hspace{2.1cm} \\
f''(r_{b})= \frac{2}{r_{b}^2} - 4\Lambda - \frac{16\pi T_{H} \varepsilon}{r_{b}}
\ena
Therefore, by changing the coordinate $r=r_{b}+\varepsilon \ti{r},  t= \frac{\ti{t}}{\varepsilon}$, the metric becomes
\bea
\BS{g} = -\left[ \varepsilon f'(r_{b})\ti{r} + \varepsilon^2 \frac{1}{2}f^{\p\p}(r_{b}) \ti{r}^2 + \cdots \right] \frac{d\ti{t}^2}{\varepsilon^2} + \frac{\varepsilon^2}{\left[ \varepsilon f'(r_{b})\ti{r} + \varepsilon^2 \frac{1}{2}f^{\p\p}(r_{b}) \ti{r}^2 + \cdots \right]} d\ti{r}^2 + (r_{b}^2 + O(\varepsilon))d\Omega^2 \hspace{0.7cm} \notag \\
= -\left[ 4\pi T_{H} \ti{r} + \left( \frac{1}{r_{b}^2} - 2\Lambda \right) \ti{r}^2 + O(\varepsilon) \right] d\ti{t}^2 + \frac{1}{\left[ 4\pi T_{H} \ti{r} + \left( \frac{1}{r_{b}^2} - 2\Lambda \right) \ti{r}^2 + O(\varepsilon) \right]} d\ti{r}^2 + (r_{b}^2 + O(\varepsilon))d\Omega^2 \notag \\
\ena
Similary,
\bea
\BS{A}= \frac{\ti{Q} \ti{r}}{r_{b}(r_{b}+\varepsilon)}d\ti{t}
\ena
Then, by taking $\varepsilon\to 0$ limit, the metric becomes a BR geometry. Let me emphasize that taking this limit means not only a near horizon limit (i.e. zooming up some region of some geometry) but also a near extremal limit (i.e. changing the geometry itself). Only in the case of $T_{H}=0$, the limit becomes a near horizon limit of an extremal RN BH.

\subsection*{$T\to T_{end}$ limit in grandcanonical ensemble}

In the main text, I claimed that, when the temperature is equal to $T_{end}$ (\ref{negativeend}), the BR geometry appears in the grand canonical ensemble. Essentially, the $T\to T_{end}$ limit is a near horizon and near extremal limit similar to the one above. However, in the grand canonical ensemble, we do not fix the horizon radius $r_{H}$ and $Q$ is not a control parameter. Therefore, for the sake of clarity, I will briefly describe how the BR geometry appears in the $T\to T_{end}$ limit.

In grand canonical ensemble, the control parameters are $\mu, r_{b}$, and $T$. Let's fix $\mu$ and $r_{b}$. And for simplicity, I use the fact that, near $T_{end}$, $r_{H}$ is a monotonically increasing function of $T$ and approaches $r_{b}$. Therefore I use $r_{H}$ instead of $T$ as a control parameter. By introducing a parameter $\varepsilon$ as a control parameter by
\bea
r_{H}= r_{b} -\varepsilon
\ena
the first and second derivatives of $f(r)$ at $r=r_{H}=(1-\varepsilon) r_{b}$ are 
\bea
f'(r_{H})= \frac{1-G\mu^2 - \Lambda r_{b}^2 (1-2G\mu^2) }{G\mu^2 r_{b}^2} \varepsilon +O(\varepsilon^2) \\ 
f''(r_{H})= \left(- \frac{2}{r_{b}^2} + 4\Lambda \right) + O(\varepsilon) \hspace{2.5cm}
\ena
After changing the coordinate by $r=r_{b}+\varepsilon \ti{r},  t= \frac{\ti{t}}{\varepsilon}$, the metric becomes
\bea
\BS{g} = -\left[ 4\pi T_{H} \ti{r} + \left( \frac{1}{r_{b}^2} - 2\Lambda \right) \ti{r}^2 + O(\varepsilon) \right] d\ti{t}^2 + \frac{1}{\left[ 4\pi T_{H} \ti{r} + \left( \frac{1}{r_{b}^2} - 2\Lambda \right) \ti{r}^2 + O(\varepsilon) \right]} d\ti{r}^2 + (r_{b}^2 + O(\varepsilon))d\Omega^2 \notag \\
~\\
\ti{r}\in(0, 1] \hspace{2cm} \notag 
\ena
where $T_{H}$ is given by
\bea
4\pi T_{H} = \frac{1-G\mu^2 - \Lambda r_{b}^2 (1-2G\mu^2) }{G\mu^2 r_{b}^2}
\ena
which is the same relation as (\ref{BRchemi}). By taking $\varepsilon \to 0$, it becomes a BR geometry.

\appendix

\end{document}